\documentclass[camera,letterpaper,nomarginnotes,nonarrowgutter]{jpaper}

\usepackage{graphicx}
\usepackage{textcomp}
\usepackage{enumitem}

\usepackage{rotating}
\usepackage{pdflscape}

\usepackage{amssymb}
\usepackage{siunitx}
\usepackage{multirow}
\usepackage{glossaries}
\usepackage{adjustbox}
\usepackage{tablefootnote}
\usepackage{subcaption}
\usepackage{datetime}

\usepackage{fvextra}

\usepackage{lipsum}
\usepackage{makecell}

\usepackage{marginnote, soul}
\usepackage{xspace}
\usepackage{tabularx}

\usepackage[bottom,hang,flushmargin]{footmisc}

\usepackage[most]{tcolorbox}
\usepackage{pifont}

\usepackage[sort,compress]{cite}
\usepackage{booktabs} 

\usepackage{algorithm}
\usepackage{algpseudocode}

\usepackage{soul}

\usepackage{mathtools}

\usepackage{xurl}

\definecolor{citeblue}{HTML}{1A56DB}    
\definecolor{refmaroon}{HTML}{800000}   
\definecolor{urlmagenta}{HTML}{A21CAF}  
\usepackage[bookmarks=true,breaklinks=true,colorlinks=true,
  citecolor=citeblue,linkcolor=refmaroon,urlcolor=urlmagenta]{hyperref}

\usepackage[colorinlistoftodos,prependcaption,textsize=scriptsize]{todonotes}

\usepackage{fancyhdr}
\fancypagestyle{firstpage}{%
  \fancyhf{}%
  \pagenumbering{arabic}%
  \fancyfoot[C]{\thepage}%
}
\usepackage{tikz}
\definecolor{darkspringgreen}{rgb}{0.09, 0.45, 0.27}
\definecolor{denim}{rgb}{0.08, 0.38, 0.74}
\definecolor{darkolivegreen}{rgb}{0.33, 0.42, 0.18}
\definecolor{tangerine}{rgb}{0.95, 0.52, 0.0}
\definecolor{mahogany}{rgb}{0.75, 0.25, 0.0}
\definecolor{coolblack}{rgb}{0.0, 0.18, 0.39}
\definecolor{yellow}{rgb}{1.1, 0.85, 0.1}
\definecolor{darkblue}{rgb}{0.0, 0.0, 0.7} 
\definecolor{lightmahogany}{rgb}{0.87, 0.45, 0.33}
\definecolor{cerulean}{rgb}{0.0, 0.48, 0.65}
\definecolor{violet}{rgb}{0.56, 0.0, 1.0}
\definecolor{lightorange}{rgb}{1.0, 0.9, 0.8}
\definecolor{seagreen}{rgb}{0.7, 0.9, 0.85}

\definecolor{darkspringgreen}{rgb}{0.09,0.45,0.27}
\definecolor{denim}{rgb}{0.08,0.38,0.74}
\definecolor{darkolivegreen}{rgb}{0.33,0.42,0.18}
\definecolor{tangerine}{rgb}{0.95,0.52,0.00}
\definecolor{mahogany}{rgb}{0.75,0.25,0.00}
\definecolor{uglyyellow}{rgb}{0.99,0.93,0.00}
\definecolor{nbs}{rgb}{0.8, 0.25, 0.33}
\definecolor{CheckGreen}{HTML}{8CB369}
\definecolor{ArithUnits}{HTML}{DDD6FE}
\definecolor{csprng}{HTML}{BFDBFE}
\definecolor{lightorange}{HTML}{FED7AA}
\definecolor{lightblue}{HTML}{A5F3FC}
\definecolor{lightgreen}{HTML}{BBF7D0}
\definecolor{cuteyellow}{HTML}{FEF08A}
\definecolor{darkbrown}{HTML}{6a4c93}
\definecolor{darkpink}{HTML}{dd2d4a}
\definecolor{goodgreen}{HTML}{006d77}
\definecolor{cadmiumgreen}{rgb}{0.0, 0.42, 0.24}
\definecolor{chocolate}{rgb}{0.92, 0.41, 0.12}
\definecolor{burgundy}{rgb}{0.5, 0.0, 0.13}
\definecolor{darkmagenta}{rgb}{0.55, 0.0, 0.55}
\definecolor{darkblue}{rgb}{0.0, 0.5, 1.0}
\definecolor{lavender}{RGB}{150, 120, 190}
\definecolor{lightred}{HTML}{FFCED0}

\newcommand{\mkspkon}[1]{\textcolor{black}{#1}}

\newif\ifshowcomments
\showcommentsfalse 

\ifshowcomments
  \newcommand{\agycomment}[1]{%
    \todo[size=\scriptsize,
      linecolor=black,
      bordercolor=black,
      backgroundcolor=white]{\scriptsize\textcolor{red}{\textbf{@gy:} #1}}}
\else
  \newcommand{\agycomment}[1]{}
\fi

\ifshowcomments
  \newcommand{\mkcomment}[1]{%
    \todo[size=\scriptsize,
      linecolor=black,
      bordercolor=black,
      backgroundcolor=white]{\scriptsize\textcolor{blue}{\textbf{@mk:} #1}}}
\else
  \newcommand{\mkcomment}[1]{}
\fi

\newcommand{\cqisca}[1]{\textcolor{black}{#1}}

\newcommand{\mkar}[1]{\textcolor{black}{#1}}

\newcommand{\mkisca}[1]{\textcolor{black}{#1}}

\newcommand{\pw}[1]{\textcolor{black}{#1}}
\newcommand{\nb}[1]{\textcolor{black}{#1}}

\newcommand{\mksp}[1]{{\color{black}#1}}

\newcommand{\hgsp}[1]{{\color{black}#1}}

\newcommand{\hga}[1]{\textcolor{black}{#1}}
\newcommand{\agy}[1]{\textcolor{black}{#1}}

\definecolor{lightpink}{rgb}{1.0, 0.9, 0.95}
\definecolor{lightgray}{rgb}{0.83, 0.83, 0.83}
\definecolor{lightgreen}{rgb}{0.8, 1.0, 0.8}
\definecolor{lightred}{rgb}{1,0.85,0.85}
\definecolor{lightblue}{rgb}{0.85, 0.92, 1}

\renewcommand{\footnoterule}{%
  \kern-3pt
  \hrule width 0.4\columnwidth height 0.4pt
  \kern 2.6pt
}

\makeatletter
\@ifundefined{algrenewcommand}{}{%
  \algrenewcommand{\algorithmiccomment}[1]{\hfill\textcolor{darkgreen}{\# #1}}%
}
\makeatother

\newcommand*\circled[1]{\tikz[baseline=(char.base)]{\node[shape=circle,fill,inner sep=1pt] (char) {\textcolor{white}{#1}};}}

\newcommand{\mkvvv}[1]{\textcolor{black}{#1}}

\newcommand{\harshita}[1]{\textcolor{black}{#1}}

\newcommand{\mayank}[1]{\textcolor{black}{#1}}
\newcommand{\mayankk}[1]{\textcolor{black}{#1}}

\newcommand{\mkrev}[1]{\textcolor{black}{#1}}
\newcommand{\mkrevb}[1]{\textcolor{black}{#1}}  
\newcommand{\mktches}[1]{\textcolor{black}{#1}}  
\newcommand{\hgrev}[1]{\textcolor{black}{#1}}

\newcommand{\phillip}[1]{\textcolor{black}{#1}}

\definecolor{hs}{rgb}{0.529, 0.298, 0.384}

\sethlcolor{lightblue}

\usepackage{chngcntr}
\usepackage{caption}

\newcommand{\tklbl}{Key Takeaway}
\newcounter{take}
\newcommand\takeaway[1]{%
    \refstepcounter{take} 
    \noindent 
    \colorbox{cyan!20}{\textbf{\tklbl{} \thetake.}} 
    \textit{#1}\par\medskip 
}

\newcommand{\summ}{Summary}
\newcounter{sum}
\newcommand\summary[1]{%
    \refstepcounter{sum} 
    \noindent 
    \colorbox{tangerine!20}{\textbf{\summ{}.}} 
    \textit{#1}\par\medskip 
}

\newcounter{fea}
\newcommand{\wi}[1]{\textcolor{black}{#1}}

\definecolor{darkgreen}{rgb}{0.0, 0.7, 0.0} 

\newcommand{\titlename}{HULK\xspace}
\newcommand{\titlenamed}{HULK}

\newcolumntype{Y}{>{\centering\arraybackslash}X}
\usepackage{tikz}

\newcommand{\squishlist}{
 \begin{list}{$\circ$}
  { \setlength{\itemsep}{0pt}
     \setlength{\parsep}{0pt}
     \setlength{\topsep}{3pt}
     \setlength{\partopsep}{0pt}
     \setlength{\leftmargin}{1em}
     \setlength{\labelwidth}{1em}
     \setlength{\labelsep}{0.5em} } }

\newcommand{\squishend}{
  \end{list}  }

\usepackage{blindtext,graphicx}
\usepackage[absolute]{textpos}

\usepackage{datetime}

\providecommand{\keysGRAPH}{%
  li2020ordering, danisch2018listing, kaushik2021gretch, besta2021graphminesuite, yan2019alleviating,
  basak2019analysis, qiu2018real, harshvardhan.pact14, tian.vldb13, nguyen2013lightweight,
  beamer-SC-2012, low.vldb12, hong.asplos12, kyrola2012graphchi, gonzalez.osdi12,
  luo2010effective, malewicz.sigmod10, ligra, page1999pagerank, graphlab}
\providecommand{\citeGRAPH}{\expandafter\cite\expandafter{\keysGRAPH}}

\providecommand{\keysCKKS}{%
  ebel2025orion, kim2025anaheim, bae2024bootstrapping, kim2022bts, samardzic2022craterlake,
  kim2022ark, lu2021pegasus, feldmann2021f1, cheon2017homomorphic, ozcan2024heongpu,
  drucker2024bleach, jung2021over, mouchet2020lattigo, boura2020chimera, sealcrypto,
  ding2024apache, TensorFHEPlus25TC, fan2025fast, park2025fhendi, alexandru2025general,
  jayashankar2025cinnamon, WarpDrive25HPCA, Neo25ISCA, WPC25CCS, CHKS25SHIP,
  Grafting25CCS, PhantomFHE23, cheon2024dacapo, BKSS24DiscreteCKKS, Trinity24MICRO,
  Alchemist24DAC, KSS24BFVBootCKKS, NeuJeans23CRYPTO, agrawal2023fab, kim2023sharp,
  TensorFHE23HPCA, Shen23CARM, Poseidon23HPCA, MAD23, HERMES23CRYPTO,
  cheon2018bootstrapping, CHKKS18FullRNS, Bossuat21EUROCRYPT, lee2021high, Bae22MetaBTS,
  OpenFHE22}
\providecommand{\citeCKKS}{\expandafter\cite\expandafter{\keysCKKS}}

\providecommand{\keysTFHE}{%
  nam2025affinity, bon2024optimized, liu2023amortized, lu2021pegasus, chillotti2021improved,
  bourse2018fast, chillotti2017faster, chillotti2016faster, ducasfhew, aziz2024secure,
  stoian2023deep, chillotti2021programmable, chillotti2020tfhe, boura2020chimera, sanyal2018tapas,
  micciancio2018ring, TFHE-rs, ding2024apache, MNEMOS26ISCA, TFHEAutomorphBoot25ASIACRYPT,
  Trinity24MICRO, Alchemist24DAC, NTRU4ms24TCHES, LMK23EfficientFHEWBoot, KLUCZNIAK23FDFB,
  XZDDF23FastBlindRot, LMP22LargePrecSign, CGGI20TFHE, MP21BootstrappingFHEW, chillotti2020concrete}
\providecommand{\citeTFHE}{\expandafter\cite\expandafter{\keysTFHE}}

\providecommand{\keysDBANALYTICS}{%
  nambiar2006making, chan1998bitmap, wu-icde-1998, oneil-ideas-2007, shao2007bitmap,
  fastbit, hsieh2018focus, huang2020tidb, cao2019titant, ankerdb,
  goodwin-SIGIR-2017, lee2017parallel, htap-survey, batchdb, peloton,
  pelkonen2015gorilla, sql-htap, lahiri2015oracle, li-vldb-2014, wu2014q100,
  yustaring, kocberber2013meet, li-sigmod-2013, le2012memory, Teubner:2013:MHJ:2510649.2511320,
  wu2013navigating, ChungDL13, sap-hana-evolution, balkesen2013multi, stonebraker2013bigdata,
  RDD, labrinidis2012challenges, hyper, Abadi_2009, bloom1970space,
  boncz.1999, ailamaki1999dbmss, wu2009fastbit, idreos2012monetdb}
\providecommand{\citeDBANALYTICS}{\expandafter\cite\expandafter{\keysDBANALYTICS}}

\providecommand{\keysGENOMICS}{%
  alser2020sneakysnake, cali-micro-2020, ghiasi2026sage, firtina2024rawhash2, alonso2024bimsa,
  bingol2024gatekeeper, mutlu2023accelerating, lindegger2023scrooge, cali2022segram, alser2020accelerating,
  firtina2024aphmm, marco2023optimal, doblas2023gmx, alser2019shouji, alser-bioinformatics-2017,
  xin.oss.bioinformatics15, xin-bioinformatics-2015, firtina2024rawsamble, eris2025rawbench, lindegger2024rawalign,
  firtina2023rawhash, hanhan2022edam, marco2021fast, lou2020helix, chen2020parc,
  alser2022molecules, lindegger2022algorithmic, cavlak2024targetcall, singh2024rubicon, rautiainen2019bit,
  li2018minimap2, turakhia2018darwin, fujiki2018genax, li2009fast, altschul-jmb-1990,
  Needleman1970Ageneral, alser2021technology}
\providecommand{\citeGENOMICS}{\expandafter\cite\expandafter{\keysGENOMICS}}

\providecommand{\keysCRYPTO}{%
  kipnis1999cryptAnalysis, canteaut2005symmetric, daemen1999aes, kim-hpca-2018, devadas2022designing,
  hellman2002overview, paillier1999public, rivest1978method, barrett1987implementing, diffie2022new,
  elgamal1985public, DR02AESRijndael, fips197}
\providecommand{\citeCRYPTO}{\expandafter\cite\expandafter{\keysCRYPTO}}

\providecommand{\keysPIMDATAMOVE}{%
  mutlu2023dac, mutlu.nsfpim20, gholami2020ai, mutlu2019enabling, boroumand2018google,
  mutlu2025memory, mutlu2024memory, ghiasi2022alp, devic2022pim, mutlu2021intelligent,
  mutlu2022modern, tang.micro17, wang.micro2016, balasubramonian-micro-2014, weinberg2005quantifying,
  saulsbury1996missing, jino1978magnetic, wulf1995hitting, ghose2019processing, mutlu2019processing,
  ghose2022road, mutlu2015main, Kim2018HowMC}
\providecommand{\citePIMDATAMOVE}{\expandafter\cite\expandafter{\keysPIMDATAMOVE}}

\providecommand{\keysHEFOUND}{%
  cheon2017homomorphic, chillotti2016faster, ducasfhew, brakerski2012fully, lyubashevsky2010ideal,
  Regev:2005aln, albrecht2022homomorphic, chillotti2020tfhe, boura2020chimera, HomomorphicEncryptionSecurityStandard,
  brakerski2014leveled, fan2012somewhat, cryptoeprint:2024/1587, mankali2025glitchfhe, kim2021revisiting,
  cryptoeprint:2014/356, lyubashevsky2013toolkit, lyubashevsky2013ideal, GSW13, GHS12AES,
  gentry2011implementing, BV11LWEFHE, BV11RLWE, van2010fully, regev2009lattices,
  gentry2009fully, aharonov2005lattice, bossuat2024security, BGV12LeveledFHE, RAD78PrivacyHomo,
  Marcolla22Survey, acar2018survey, Peikert16Decade}
\providecommand{\citeHEFOUND}{\expandafter\cite\expandafter{\keysHEFOUND}}

\providecommand{\keysPIMPNM}{%
  gomez2022benchmarking, singh2021fpga, wang2024ndsearch, park2024lpddr, li2024stream, ham2024low,
  schwedock2024leviathan, tianndpbridge2024, seo2024ianus, wang2023affinity, zhou2023dimm,
  eggermann202316, orenes2023dalorex, boroum2019conda, ahn2015scalable, ahn2015pim,
  denzler2021casper, niu2022isscc, ke2021near, kim2022aquabolt, lee20221ynm,
  feng2022menda, lee2021hardware, kwon202125, asgarifafnir, park2021trim,
  sun2021abc, sadredini2021sunder, gu2020ipim, newtonpim, oliveira2021anew,
  rezaei2020nom, jang2019charon, schuiki2018scalable, huang2019active, zhang2018minnow,
  drumond-isca-2017, patterson-ieeemicro-1997, kogge1994, gokhale1995processing, devaux2019true,
  kwon2022system, kim2021aquabolt, impica, oskin1998active, gao2016hrl,
  kim2016isca, asghari2016chameleon}
\providecommand{\citePIMPNM}{\expandafter\cite\expandafter{\keysPIMPNM}}

\providecommand{\keysPIMPUM}{%
  tokuda2026pudghost, tokuda2026clutch, yuksel2024simultaneous, olgun2022pidram, mutlu2024memory,
  oliveira2024mimdram, yuksel2024functionally, hajinazar2021simdram, orosa2021codic, seshadri-micro-2017,
  GS-DRAM, seshadri-micro-2013, gao2022fracdram, rohbani-dac-2022, ferreira2021pluto,
  lee20223d, zahedi2022system, wan2022compute, jung2022crossbar, zhang2021max,
  truong-micro-2021, lin-vlsidat-2021, huang2021mixed, leitersdorf2021efficient, jain2020rxnn,
  kim2021colonnade, caminal2021cape, yuan2021forms, wen2020ckfo, lenjani_fulcrum_2020,
  xin-hpca-2020, ramanathan-micro-2020, wang-micro-2020, kim2020mvid, ankit2020panther,
  zha2020hyper, su202015, luo2020accelerating, luo2020benchmark, jiang2020c3sram,
  imani2020dual, jia2020programmable, challapalle2020gaas, angizi2019mrima, ankit2019puma,
  gao-micro-2019, fujiki-isca-2019, imani2019floatpim, deng-dac-2018, song2018graphr,
  fujiki-asplos-2018, eckert-isca-2018, angizi2018pima, li2018scope, li-micro-2017,
  aga-hpca-2017, song2017pipelayer, li-dac-2016, Shafiee-isca-2016, ping-isca-2016,
  gao2016draf, xin2019rocdrambased, borghetti2010memristive, kvatinsky2014imply, de2025proteus,
  leitersdorf2023aritpim, seshadri2017simple, seshadri2015fast, seshadri-ieeecal-2015, ma20232,
  garzon20234, seshadri2018rowclone, seshadri-arxiv-2019, stone-tc-1970, kautz-tc-1969,
  angizi2019redram}
\providecommand{\citePIMPUM}{\expandafter\cite\expandafter{\keysPIMPUM}}

\providecommand{\keysHEACCEL}{%
  Kabra2025CipherMatch, nam2025affinity, kim2025anaheim, gupta2023evaluating, kim2022bts,
  samardzic2022craterlake, kim2022ark, gupta2021accelerating, feldmann2021f1, zhou2025fhemem,
  ozcan2024heongpu, mwaisela2024evaluating, jung2021over, ding2024apache, Falcon26ASPLOS,
  MNEMOS26ISCA, HierarchicalGPUFHE26ISCA, NTPFPGA26HPCA, Libra26Security, TensorFHEPlus25TC,
  BitwiseNTT26TCAD, BarrettMult25TCAD, fan2025fast, park2025fhendi, jayashankar2025cinnamon,
  WarpDrive25HPCA, Neo25ISCA, HAWK25MICRO, AllMod25DAC, chen2024hmc,
  PhantomFHE23, Trinity24MICRO, Alchemist24DAC, agrawal2023fab, pal2023fully,
  kim2023sharp, yang2023poseidon, TensorFHE23HPCA, Shen23CARM, Poseidon23HPCA,
  FxHENN23HPCA, MAD23, BASALISC22, bpntt, park2023ntt,
  barik2026long, reagen2021cheetah, HEAX20, gupta2024memfhe, SoKFHEAccel24,
  dramaton, hexl, wang2012accelerating}
\providecommand{\citeHEACCEL}{\expandafter\cite\expandafter{\keysHEACCEL}}

\providecommand{\keysSTORISP}{%
  Park_MICRO2022, Chen_MICRO2024, Gao_MICRO2021, nadig2026conduit, ghiasi2026grains,
  Chen2025REIS, Ghiasi2024MegIS, Mansouri-Ghiasi_ASPLOS2022, Park_ASPLOS2025, kim2025crossbit,
  lee2025aif, mahapatra2025rag, pan2024instattention, wong2025anvil, lincoln-hpca,
  jang2024smart, gouk2024dockerssd, kang2024isp, wang2024beacongnn, chen2024search_tcad,
  leepresto2024, mahapatra2024storage, Duffy_PVLDB2023, Kim_HPCA2023, li2023ecssd,
  yang2023lambda, li2023optimizing, ghiasi2026enabling, Soysal2025MARS, Lee_ISCA2022,
  Liang_DAC2022, hu2022ice, kang-tc-2021, kang2021iceclave, kim-fast-2021,
  park2021sas, Matam_ISCA2019, mailthody-micro-2019, jun-isca-2018, koo-micro-2017,
  lee2017extrav, istvan2017caribou, gu2016biscuit, jo2016yoursql, jun2015bluedbm,
  seshadri-osdi-2014, do-sigmod-2013, tiwari-fast-2013, acharya-asplos-1998, keeton-sigmod-1998,
  riedel-vldb-1998, lee2020smartssd, ruan2019insider, barbalace2021computational}
\providecommand{\citeSTORISP}{\expandafter\cite\expandafter{\keysSTORISP}}

\providecommand{\keysBFV}{%
  Kabra2025CipherMatch, gupta2023evaluating, iliashenko2021faster, chen2018homomorphic, juvekar2018gazelle,
  chen2017fast, brakerski2012fully, ozcan2024heongpu, wibawa2022bfv, mouchet2020lattigo,
  boura2020chimera, halevi2019improved, bajard2016full, fan2012somewhat, huo2019note,
  sealcrypto, PhantomFHE23, RelaxedFuncBootBGV24, KSS24BFVBootCKKS, Shen23CARM,
  GIKV23PolyModBoot, kim2021revisiting, chen2019efficient, roy2019fpga, roy2018hepcloud,
  Badawi18GPUFV, SlotCoeffBGVBFV24, ozcan2023homomorphic, GV23BGVBootRevisited, BP23DemystifyBoot,
  OpenFHE22, SEAL}
\providecommand{\citeBFV}{\expandafter\cite\expandafter{\keysBFV}}

\providecommand{\keysPIMDATAMOVE}{mutlu2022modern,boroumand2018google,mutlu2019processing,ghiasi2022alp,mutlu2019enabling,alsop2024pim,devic2022pim,ghose2019processing,mckee2004reflections,mutlu2024memory,mutlu2025memory,wang.micro2016,asifuzzaman2023survey,balasubramonian-micro-2014,singh2018review,murphy2001characterization,mutlu2021intelligent,gomezluna2020.lecture,nair2015evolution,ca.fall2020.solution,ca.fall2020.challenges,mutlu2015main,pandiyan2014,jino1978magnetic,ghose2019arxiv,wilkes.can01,shaw1981non,loh2013processing,wulf1995hitting,zhang2013new,mutlu.msttalk17,mutlu.gwutalk19,mutlu.nsfpim20,wen2017rebooting,ivanov2021data,mutlu2023dac,boroumand2020practical,ghose2022road,weinberg2005quantifying,kestor.iiswc2013,ghose.bookchapter19,ghose.bookchapter19.arxiv,chishti2019memory,gholami2020ai,clapp.2015,tang.micro17,saulsbury1996missing,sites1996,Kim2018HowMC}
\providecommand{\citePIMDATAMOVE}{\expandafter\cite\expandafter{\keysPIMDATAMOVE}}

\providecommand{\keysPIMPUM}{ankit2019puma,de2025proteus,deng-dac-2018,gao-micro-2019,gao2022fracdram,hajinazar2021simdram,leitersdorf2023aritpim,lenjani_fulcrum_2020,li-dac-2016,li-micro-2017,oliveira2024mimdram,seshadri-arxiv-2016-pum,seshadri-arxiv-2019,seshadri-ieeecal-2015,seshadri-micro-2013,seshadri-micro-2017,song2018graphr,xin-hpca-2020,yuksel2024functionally,zhang2021max,aga-hpca-2017,fujiki-isca-2019,fujiki-asplos-2018,eckert-isca-2018,ramanathan-micro-2020,han-ieeetcas-2019,bayat-ieeetnnls-2017,kang-jproc-2020,angstadt-micro-2018,Shafiee-isca-2016,imani2019floatpim,seshadri-arxiv-2016,truong-micro-2021,ali-tcas-2019,wang-micro-2020,zhang-jssc-2017,lin-vlsidat-2021,alhawaj-iscas-2020,zhang-arxiv-2022,rohbani-dac-2022,ping-isca-2016,kautz-tc-1969,stone-tc-1970,deng2019lacc,gao2016draf,marinella2018multiscale,xi2020memory,zheng2016tcam,elliott1999computational,kim2020mvid,sun2020one,8645905,cho2020mcdram,shin2018mcdram,ferreira2021pluto,jain2018computing,feinberg2018enabling,prezioso2015training,alves2015reconfigurable,alves2015opportunities,kvatinsky2014imply,ankit2020panther,hasan2017fast,hamdioui2015memristor,xie2015fast,hamdioui2017myth,yu2018memristive,gaillardon2016plim,angizi2018cmp,elliott1992computational,GS-DRAM,saikia2019k,morad.taco15,kang.icassp14,zha2020hyper,ambrosi2018hardware,huang2021mixed,cheng2018time,jiang2019cimat,su202015,li20203d,luo2020accelerating,luo2020benchmark,levy.microelec14,seshadri2017simple,seshadri2015fast,truong2022adapting,leitersdorf2021efficient,li2018scope,sutradhar2021look,wang20211t2c,slesazeck20192tnc,ma20232,lee20223d,seshadri2018rowclone,sutradhar2020ppim,orosa2021codic,korgaonkar2019bitlet,si2019dual,long2018reram,bhattacharjee2017revamp,angizi2018pima,jain2020rxnn,liu2014reduction,charan2020accurate,long2019design,liu2015vortex,klachko2019improving,chen2017accelerator,zahedi2022system,seshadri.thesis16,jeloka201628,jiang2020c3sram,kim2021colonnade,agrawal2018x,kang2018multi,kvatinsky.tcasii14,lin2022all,jain2017computing,kang2017memory,kvatinsky.iccd11,zhou2022flexidram,burr2015large,chen2015optimized,kim2015reconfigurable,vadivel2020tdo,CASES_MVX,vieira2021compute,imani2020dual,xia2016switched,song2017pipelayer,yavits2014resistive,fan2016lowpower,siegl2016datacentric,angizi2017design,angizi2017energy,angizi2017rimpa,fan2017inmemory,he2017exploring,he2017leveraging,parveen2017hybrid,parveen2017lowpower,gao2018adesign,manning2018apparatuses,parveen2018hielm,parveen2018imcs2,rakin2018pimtgan,angizi2019redram,lea2019apparatuses,shin2019indram,xin2019rocdrambased,caminal2021cape,garzon20234,morad2016resistive,wu2022dramcam_generalpurpose,borghetti2010memristive,li2013memristor,holmes1993use,hu2012hardware,alibart2012high,jeong2017parasitic,lee2019exploring,ielmini2018memory,linn2012beyond,wan2022compute,jung2022crossbar,sebastian2020memory,cai2019fully,le202364,joshi2020accurate,li2019long,valavi201964,le2018mixed,yang2019sparse,jia2020programmable,tang2017binary,yuan2021forms,wen2020ckfo,wen2019memristor,chou2019cascade,feinberg2018making,challapalle2020gaas,angizi2019mrima,wang2018snrram,zhu2019configurable,yang2020retransformer,gao2016demonstration,tang2017aepe,xia2017fault,huang2017highly,cheng2017time,bojnordi2016memristive,chen2018regan,cai2018training,mao2018lergan,nag2018newton,yao2017face,hu2016dot,lehtonen2009stateful,kim2011field,lehtonen2012applications,mahmoudi2013implication,kim2019single,xie2017scouting}
\providecommand{\citePIMPUM}{\expandafter\cite\expandafter{\keysPIMPUM}}

\providecommand{\keysPIMPNM}{wang2024ndsearch,ahn2015scalable,ahn2015pim,doty1980magnetic,gao2016hrl,gokhale1995processing,kang1999flexram,nider2020processing,oskin1998active,patterson-ieeemicro-1997,pugsley2014ndc,singh2021fpga,drumond-isca-2017,farmahini-hpca-2015,tang-aspdac-2017,devaux2019true,niu2022isscc,ke2021near,lee2021hardware,kwon202125,draper2002architecture,farmahini2014drama,yitbarek2016exploring,kim2021aquabolt,kim2022aquabolt,kwon2022system,gu2020ipim,asgarifafnir,denzler2021casper,kim2016isca,park2021trim,nair2015active,gokhale2015rearr,lloyd2018dse,rodrigues2016scattergather,lloyd2015memory,zhu2013,akin2015,kogge1994,jeon2017hmc,kaxiras.1997,jeddeloh2012hybrid,Mai:2000:SMM:339647.339673,akin2014hamlet,hassan.memsys15,asghari2016chameleon,sun2021abc,newtonpim,lee20221ynm,kim1999assessing,gebis2000trends,landgraf2021combining,wang2023affinity,NIM,lim2017triple,HIVE,jang2019charon,smc_sim,santos2018processing,schuiki2018scalable,rezaei2020nom,impica,park2024lpddr,azarkhish2016logic,HughesA05,hsieh.tr16,ghiasi2022revamp3d,Asghari-Moghaddam_2016,huang2019active,kersey2017lightweight,sim2022computational,de2018design,guo2014wondp,eckert.wondp14,zhou2023dimm,sadredini2021sunder,kim2015understanding,zhang2018minnow,li2024stream,chen2024bridge,feng2022menda,hall1999mapping,ham2024low,schwedock2024leviathan,IRAM_WML_1997,NMP_2005,LiM_3D_FFT_MM,eggermann202316,orenes2023dalorex,kim2017heterogeneous,hong2016accelerating,baskaran2020decentralized,azarkhish2016design,akin2019case,nai2015instruction,tianndpbridge2024,seo2024ianus,fromm1997theenergy,pawlowski2011hybrid,agrawal2017amanycore,boroum2019conda,oliveira2021anew}
\providecommand{\citePIMPNM}{\expandafter\cite\expandafter{\keysPIMPNM}}

\providecommand{\keysPIMSYSTEMS}{ahmed2019compiler,boroumand2019conda,chen2022offload,hadidi2017cairo,hsieh.isca16,jiang20243,kim2017toward,maity2024coat,maity2025offload,maity2025unguided,olgun2022pidram,peng_chopper_2023,ramulator-pim,silentpim,singh2019napel,wei2022pimprof,wu2020tuning,yan2001codemap,boroumand-ieeecal-2017,zhang-hpdc-2014,santos-date-2017,gomez2021benchmarking,item2023transpimlibispass,gomez2022benchmarking,nider2021case,gomezluna2023repo,gomezluna2021.SLS,giannoula2021.SLS,hajinazar2021.SLS,cordeiro2021machine,gomezluna2021repo,jacob2016compiling,sura2015data,gokhale2015hmc,shelor2019reconfigurable,liu2017,pattnaik2016,lee.hpca01,fraguela.ppopp03,bruel2017generalize,falevoz2024energy,chmielewski2025upmem,son2025pimnet,hyun2024pathfinding,chen2023simplepim,jonatan2024scalability,oliveira2022heterogeneous,abumpimp2023,boroumand2016lazypim,gomezluna2021prim,oliveira2023dappa,oliveira2022methodologies,boroumand.arxiv17,wang2023infinity,mosanu2022pimulator,lenjani2021supporting,siddique2024architectural,zhao2024pim,lee2024pim,lockerman2020livia,sandia,oliveira2017generic,picorel2017near,sokulski2022spec,transpimlib2023repo,qu2024cim,farzaneh2024c4cam,baekpsyncpim2024,cilasunpim2024,kim2023virtual,noh2024pid,lopes2024pim,leitersdorf2024pypim,xu2019cuckoopim}
\providecommand{\citePIMSYSTEMS}{\expandafter\cite\expandafter{\keysPIMSYSTEMS}}

\providecommand{\keysPIMAPPS}{augusta2015jafar,besta2021sisa,dai2018graphh,fernandez2020natsa,gao.pact15,gomezluna2022isvlsi,gu2025pim,guz-ndp-2014,he2025papi,huang2020heterogeneous,kim-genomics-2018,medal2019,mirzadeh2015sort,nai2017graphpim,nejatollahi2020cryptopim,park2024attacc,perach-arxiv-2022,roy-2021-jetcas,singh2025scalable,vermij2017sorting,zhang2018graphp,zhuo2019graphq,angizi-glsvlsi-2019,giannoula-pomacs-2022,li-jcst-2021,kim-hpca-2021,boroumand-icde-2022,gomez2023evaluating,diab2023aim,lee2022improving,oliveira2022accelerating,shahroodi2023case,mao2022genpip,lim2023design,ahn2023retrospective,kwon2019tensordimm,diab2022hicomb,herruzo2021enabling,singh2021accelerating,boroumand2021icde,liu2018processing,gu2020dlux,park2021high,lee2021task,lloyd2017keyvalue,falahati2018origami,Zois2018,lavenier2020,lavenier2016,nag2019gencache,azarkhish2017neurostream,gao2017tetris,singh2020nero,ke2020recnmp,sun2020energy,rhyner2024pim,bssync,giannoula2022sigmetrics,chen2023uppipe,lavenier2016blast,bernhardt2023pimdb,baumstark2023adaptive,baumstark2023accelerating,das2022implementation,kim2024optimal,kang2023pim,Zarif_2023,dai2022dimmining,fernandez2024matsa,shahroodi2023swordfish,gomez2022experimental,fernandez2022exploiting,angizi2019dna,nider2022bulk,li2019pims,JAFAR,samsunghc23,wu2021sieve,wang2015association,wang2016sequential,sadredini2019eap,sadredini2020flexamata,zhou2021ultra,lenjani2022pulley,jaiyeoba2023acts,subramaniyan2017parallel,yun2023grande,zhou2022gnnear,chen2023metanmp,min2019neuralhmc,lee2016application,kim2017neurotrainer,huo2024pifs,glova2019near,asgari2020mahasim,pugsley2014comparing,xu2015scaling,xie2017processing,zhou2022transpim,abecassis2023gapim,angizi2019graphs,heo2024neupims,liuisca2024,yueisca2024,lee2024spid,li2024pim,kim2016genome,angizi2017imcenergyefficient,fan2017energy,fan2017leveraging,angizi2018dima,angizi2018imce,angizi2019deep,angizi2019parapim,angizi2020pimassembler,yang2020aflexible,caminal2022accelerating,garzon2022aida,yavits2023drama,khalifa2023clapim,jahshan2024majork,gogineni2024swiftrl,li2020timely,giannoula2024pygim}
\providecommand{\citePIMAPPS}{\expandafter\cite\expandafter{\keysPIMAPPS}}

\providecommand{\keysSTORNAND}{kang201913,maejima2018512gb,cho20221,cai2013program,cai2017error,cai2015data,cai2013threshold,micheloni2010inside,cai-date-2012,cai-dsn-2015,cai-insidessd-2018,cheong2018flash,grupp2009characterizing,luo-sigmetrics-2018,luo2015warm,luo2018heatwatch,tanakamaru-jssc-2013,cai-iccd-2012,cai-sigmetrics-2014,hatanaka2012nand,luo2016enabling,takeuchi2009novel,cai-inteltechj-2013,cai-hpca-2017,cai.bookchapter18.arxiv,mohan2013modeling,yucai-thesis,pekny2022isscc,lee20043,cho2001dual,park-dac-2016,kim-dac-2017,ha-ieeetcad-2015,kim-ieeejssc-2012,cha-etrij-2013,lee-ieeejssc-2002,suh-ieeejssc-1995,kim-glsvlsi-2018,kim-ieeejssc-2018,dong-ieeetcas-2010,Wang-ieeetcom-2016,feng-iccd-2017,lee-isscc-2002,crippa-springer-2010,dong-tcas-2010,zuolo-imw-2015,hu-fms-2012,jimenez-date-2013,cao-elecletters-2022,lu-jet-2018,cai2018reliability,park2000new,chen2019performance,yucai.bookchapter18,luo.thesis18,ravindra1992fowlernordheim,bez2003introduction,cooke2007theinconvenient,cho2024aero,koh2009nand,maislos2011anew,wise2014reverse,meza2015alargescale,park2015technology,park2015threedimensional,z2015amathematical,courbon2016reverse,schroeder2016flash,fukami2017improving,kang2017256gb,z2017biterrors}
\providecommand{\citeSTORNAND}{\expandafter\cite\expandafter{\keysSTORNAND}}

\providecommand{\keysSTORSSD}{chen2011caftl,chung2009survey,goodson2010design,park2009design,lee2007log,meijer2019self,flashvault,lightnvm,picoli2020open,znssds,das2013cpu,jaliminche2023enabling,liu2021self,wang2018oc,kwon2020dc,min2023multi,tavakkol2018flin,kim2018utilitarian,huang2017flashblox,agrawal2008design,eshghi2012ssd,hu2012exploring,hu2011performance,gal2005mapping,ma2014survey,kim2002space,bang2011memory,xie2014adaptive,yang2014garbage,min2021isolating,sun2023leaftl,yang2011efficient,hu2010achieving,chen2011essential,zhao2013ldpc,qi2017cdf,jin2017kaml,luo2013s,gonzalez2017multi,gupta2009dftl,li2018femu,tavakkol2018mqsim,kim2023nvmevirt,kwak2018cosmos,choi2018parallelizing,dirik2009performance,jung2012taking,kim2022networked,lim2010faster,ma_lazyftl_2011,park2006high,shahidi2016exploring,shin2009ftl,wang2024learnedftl,wu2016gcar,zhou2015efficient,park-asplos-2021,nadig2023venice,gal2005algorithms,kim2015ops,kim2020evanesco,chang2007wear,zheng2013power,ajdari-hpca-2019,cho2015design,do2019improving,harris2020ultra,kim2021decoupled,lee2013preemptible,lee2015external,liao2012multi,micheloni2013inside,myung2020efficient,park2025selective,stoica2019understanding,wang-eurosys-2019,wu2012reducing,prabhakaran2008txflash,liu2017vst,wei2011reliably,tavakkol2014design,li-iccad-2019,hong-nvmsa-2019,wu-dac-2018,cox-fms-2018,jeong-fast-2014,shim-micro-2019,coughlin2011solid,park2018improving,li2018access,VSSIM,cai2011fpga,2012cs3140,lu2013lighttx,narayanan2016ssdfailures,samsung860pro}
\providecommand{\citeSTORSSD}{\expandafter\cite\expandafter{\keysSTORSSD}}

\providecommand{\keysSTORISP}{Soysal2025MARS,Ghiasi2024MegIS,Mansouri-Ghiasi_ASPLOS2022,Li_ATC2021,Matam_ISCA2019,Lee_ISCA2022,Liang_DAC2022,Duffy_PVLDB2023,Park_ASPLOS2025,Bisson_IPCCC2018,Chen2025REIS,Kim_HPCA2023,Chen_MICRO2024,Park_MICRO2022,Gao_MICRO2021,gu2016biscuit,wang2016ssd,park2016storage,ruan2019insider,ghiasi2026enabling,nadig2026conduit,ghiasi2026grains,kang2021iceclave,heydarigorji2022storage,jang2024smart,gouk2024dockerssd,picoli2019ox,lee2020smartssd,acharya-asplos-1998,bae-cikm-2013,boboila-msst-2012,chen2024search,do-sigmod-2013,hu2022ice,jun-isca-2018,kang-msst-2013,keeton-sigmod-1998,liang-atc-2019,pei-tos-2019,riedel-computer-2001,riedel-vldb-1998,tiwari-fast-2013,torabzadehkashi-pdp-2019,wang2016ssd1,chun2022pif,jo2016yoursql,kang-tc-2021,kang2024isp,kim-fast-2021,kim2025crossbit,koo-micro-2017,lee2017extrav,lee2025aif,li2023ecssd,mahapatra2025rag,mailthody-micro-2019,pan2024instattention,seshadri-osdi-2014,shim2022gp3d,wang2024beacongnn,wong2024tcam,wong2025anvil,yang2023lambda,barbalace2021computational,barbalace_blockndp_2020,chen2024search_tcad,cho-wondp-2013,choi2015energy,choi2020flash,fakhry2023review,istvan2017caribou,jaliminchecs,jang2025inf,jeong-tpds-2019,jun-hpec-2016,jun2014scalable,jun2015bluedbm,kim-infosci-2016,kim-sigops-2020,lee2016activesort,li2018cisc,li2023optimizing,liang-fpl-2019,lim-icce-2021,lincoln-hpca,okafor2023fusing,quero2015self,salamat2021nascent,salamat2022nascent2,tiwari-hotpower-2012,torabzadehkashi-ipdpsw-2018,wang2018three,wu2015data,yavits2021giraf,zaman2025cord,tseng-iedm-2020,lue-iedm-2019,wang-edtm-2022,han-aisy-2021,park2021sas,kim2011fast,park2014query,ghiasi2022genstore,jun2016bluedbm,Cho_2013,De_2013,leepresto2024,mahapatra2024storage,kaplan2018prins}
\providecommand{\citeSTORISP}{\expandafter\cite\expandafter{\keysSTORISP}}

\providecommand{\keysSTORSYSTEMS}{nadig2026harmonia,lerner2020networking,sun2025fleetio,mao2012cache,narasimhamurthy2019sage,odun2017overview,lee-atc-2009,gray-arxiv-2007,jacob2010journaled,yang2023end,bongiovanni1980magnetic,memcached,giampaolo1998practical,fitzpatrick2004distributed,singh2022sibyl,park2022deepsketch,AtikogluXFJP12,LimMSRW13,ghemawat.sosp03,rodeh.tos13,frank1969analysis,teorey1972acomparative,abbott1988offtrack,jacobson1991disk,chang1999automatic,cardarilli2000development,fan2000summary,chang2001using,jiang2003crosstrack,bairavasundaram2007ananalysis,pinheiro2007failure,schroeder2007disk,schroeder2007understandingdisk,bairavasundaram2008ananalysis,tang2008understanding,wood2009thefeasibility,vasudevan2012using}
\providecommand{\citeSTORSYSTEMS}{\expandafter\cite\expandafter{\keysSTORSYSTEMS}}

\providecommand{\keysFORMAL}{coq,chlipala2013certified,huet1997coq,qiao2019formal,chang2020determinizing,bodenmuller2021flashix,sel4,compcert,certikos,goguen1982security,ileri2018disksec,chajed2019argosy,chajed2019perennial,murray2013sel4,chen2015fscq,chen2017dfscq,sigurbjarnarson2016yggdrasil,reif1995kiv,amani2016cogent,tripathy2023formal,chajed2021gojournal,hance2020storage,schierl2009ubifs,schellhorn2014verified,ernst2015inside,joshi2007mini,butterfield2009mechanising,kang2008alloy,damchoom2009eventb,taverne2009raffs,pfahler2013erase,kim2008flashdriver,tripathy2019nand,prasad2005asurvey,demoura2008z3an,gomes2008satisfiability,brummayer2009boolector,sulflow2009wolfram,cimatti2010satisfiability,dillig2012minimum,bjorner2015nuzan}
\providecommand{\citeFORMAL}{\expandafter\cite\expandafter{\keysFORMAL}}

\providecommand{\keysMEMDRAM}{chang-hpca-2016,chang.hpca14,chargecache,ghose2018your,ghose2019demystifying,kim-hpca-2019,lee.hpca13,lee.hpca15,lee.sigmetrics17,lee201425,lee2016simultaneous,mutlu.superfri15,oh20143,olgun2021quactrng,salp,yuksel2025dram,pandey2016power,feng2013system,desrochers2016validation,olgun2024sectored,kim2014hbm,ca.fall2020.refresh,kim2014memory,fujun2020stacked,weber2005current,peng2015design,kwak2017refresh,kwon2017wearable,hasan2019crow,koppula2019eden,donghyuk-ddma,yoongu-thesis,chang.sigmetrics16,chang.phdthesis2017,facchini2009system,Ani2010,loh2008stacked,mandelman.ibmjrd02,2t-1c-1,migration-cell,shim201816gb,bostanci2022dr,luo2020clr,kim2018solar,wang2018cal,mutlu2018recent,das.dac18,mukundan.isca13,lee.thesis16,chang.sigmetrics17,keeth2007dram,zhang2000permutation,rockicki1996indexing,yauglikcci2022hira,lee2021greendimm,o2021energy,lee2016reducing,son2013reducing,choi2015multiple,sato1998fast,olgun2022sectored,lee2017partial,ha2016improving,zhang2014half,cooper2010fine,keeth2001dram,kim20121,embedDRAM,chatterjee2017architecting,oconnor2017fine,son2014microbank,zheng2008minirank,yoon2012dynamic,yoon2011adaptive,ahn2009multicore,alawneh2021dynamic,zhang2017enabling,kim1998dram,itoh1997limitations,radulovic2015another,kim2013memory,patel2024rethinkingieee,seongil2014row,itoh2013vlsi,itoh1977semiconductor,chatterjee1979leakage,smith1981laser,frailong1985xorschemes,nakagome1988theimpact,yanagisawa1988semiconductor,yoshihara1988atwisted,banerjee1989twotransistor,konishi1989analysis,lu1989advanced,min1990wordline,henkels1991a4mb,tanabe1992a30ns,patel1994dram_power,horiguchi1997redundancy,hamamoto1998onthe,hiraiwa1998localfieldenhancement,lieneweg1998assessment,ohsawa1998optimizing,huang1999aprogrammable,katayama1999faulttolerant,kimura199964mb,naritake1999a12ns,vandegoor1999industrial,davis2000modern,kim2000dynamic,rosar2000anew,saino2000impact,song2000method,yamaguchi2000theoretical,amrutur2001fast,hsu2001dram_array,kim2001blockbased,lee2001anew,lee2001novel,lin2001reducing,mori2001anew,zhang2001cached,batson2002dram_cam,redeker2002aninvestigation,sekiguchi2002alowimpedance,gu2003challenges,ho2003method,kim2003blockbased,lin2003method,alars2004effects,jin2004modeling,lazar2004dram_with,sim2004chargetransferred,wada2004postpackaging,wicht2004yield,yi2004offstate,jin2005prediction,kim2005selfrefresh,mori2005theorigin,mueller2005challenges,patel2005energyefficient,weber2005data,yi2005numerical,ahn2006adaptive,cho2006analysis,kim2006an8gbspin,ohyu2006quantitative,patel2006dram_cam,venkatesan2006retentionaware,yu2006leakage,ditali2007xray,sim2007circuit,takemura2007longretentiontime,emma2008rethinking,kong2008analysis,schloesser20086f2buried,isen2009eskimo,kim2009rtslike,chia2010newdram,kim2010random,lee2010amechanism,stuecheli2010elastic,vogelsang2010understanding,cha2011dram_and,horiguchi2011nanoscale,kim2011characterization,kim2011study,lee2011simultaneous,li2011dram_yield,liu2011flikker,liu2011hardwaresoftware,oh2011characterization,chatterjee2012leveraging,desai2012process,lee2012dram_data,lim2012a12v,lin2012secret,malladi2012towards,baek2013refresh,balasubramonian2013adram,bhati2013coordinated,hou2013anfpgabased,nair2013acase,agrawal2014mosaic,arahmati2014refreshing,bacchini2014characterization,bacchini2014total,bhati2014scalable,cui2014dtail,han2014dataaware,jung2014optimized,kumar2014detection,lee2014a12v,nair2014refresh,wang2014proactivedram,zhang2014cream,bains2015method,bhati2015flexible,jung2015omitting,mutlu2015themain,parkin2015memory,shin2015dramlatency_optimization,weis2015thermal,zhang2015exploiting,zhang2015hifi,bauer2016lest,bhati2016dram_refresh,duganapalli2016modelling,lo2016methods,cha2017defect,ito2017apparatus,jones2017holistic,jun2017hbmhigh,kotra2017hardwaresoftware,mathew2017using,seyedzadeh2017mitigating,sharifi2017online,tannu2017cryogenicdram,tovletoglou2017accessaware,weis2017dramspec_a,yang2017scanning,zhang2017addressing,zhang2017onthe,cho2018aninnovative,chun2018a16gb,gong2018memory,kraft2018improving,liu2018getout,nguyen2018anapproximate,nguyen2018nonblocking,seyedzadeh2018mitigating,talukder2018ldpuf,tu2018rana,bahartalukder2019prelatpuf,balasubramonian2019innovations,hur2019adaptive,lee2019anevaluation,lee2019exploiting,pan2019hiding,pan2019thecolored,talukder2019exploiting,william2019memory,jafri2020refresh,kim2020chargeaware,wang2020dramdig_a,gurumurthi2021hbm3,lee2021cryoguard,mathew2021thermoelectric,nabavilarimi2021understanding,nguyen2021zemzerocycle,bai2022alowcost,safari2022selfmanaging,kaur2022workinprogress,shi2022exploiting,nam2023xray,hassan2022acase}
\providecommand{\citeMEMDRAM}{\expandafter\cite\expandafter{\keysMEMDRAM}}

\providecommand{\keysMEMRELIAB}{yuksel2024simultaneous,yuksel2025columndisturb,yuksel2025pudhammer,girard2020survey,yauglikcci2021blockhammer,binarystar,kim2009new,chou2015reducing,luo2014characterizing,micron2017ecc,qiao2016new,bhattacharya2016curious,jang2017sgx,aga2017good,tatar2018defeating,cojocar19exploiting,zhang2018triggering,barenghi2018software,bhattacharya2018advanced,patel2019understanding,patel2020bit,patel2017reaper,kim2020revisiting,raidr,mutlu.isca2013talk,khan.sigmetrics14,khan.dsn16,kim-isca2014,mutlu2017rowhammer,sridharan.asplos15,meza.dsn15,khan.cal16,seaborn.2015,vanderveen.2016,orosa2021deeper,hassan2021uncovering,patel2021harp,throwhammer,nethammer,cloudflops,anotherflip,cojocar2020susceptible,frigo2020trr,mutlu2020retrospective,mutlu2023retrospective,olgun2024read,qureshi.dsn15,qureshi2024impress,kim.arxiv16,schroeder.sigmetrics09,giray-thesis,luo.arxiv17,khan.micro17,luo2023rowpress,gruss.2015,razavi.2016,rowhammer-topinhes18,mineshphd,kim2020improving,binder1975satellite,pickel1978cosmic,may1979alphaparticleinduced,bossen1980asystem,nibby1985remap,yaney1987ametastable,wyns1988temperature,kirton1989noise,dekker1990arealistic,kalter1990a50ns,saleh1990reliability,mashiko1991semiconductor,vandegoor1991testing,restle1992dram_variable,mazumder1993design,agrawal1994aproposed,ogorman1994theeffect,hamamoto1995well,hiraiwa1996statistical,lantz1996soft,baker1997shmoo,dell1997awhite,croswell2000amodel,johnston2000scaling,locklear2000chipkill,cheng2002neighborhood,kim2002dram_reliability,vandegoor2002address,maiz2003characterization,messer2004susceptibility,mukherjee2004cache,alars2005dram_fault,baumann2005radiationinduced,chen2005modeling,radaelli2005investigation,alars2006dramspecific_space,slayman2006impact,tang2006assessment,bajura2007models,cha2007high,kim2007alow,li2007amemory,seifert2007onthe,borucki2008comparison,dell2008system,lavery2008discriminating,richter2008newlinear,autran2009altitude,chao2009fault,wang2009soft,li2010arealistic,mrozek2010analysis,seong2010safer,yoon2010virtualized,nightingale2011cycles,white2011astudy,huang2012alternate,hwang2012cosmic,tol2023don,rakin2019bit,sridharan2012astudy,udipi2012lotecc,chen2013e3cc,jian2013adaptive,jian2013lowpower,ladbury2013useof,nair2013archshield,siddiqua2013analysis,sridharan2013feng,chen2014memguard,gold2014providing,greenfield2014method,harris2014flipping,kwon2014understanding,park2014activeprecharge,shirley2014copula,chen2015eecc,ghasempour2015armor,gong2015cleanecc,gorman2015memory,kim2015avert,kim2015bamboo,kim2015frugal,micheloni2015apparatus,wang2015radar,wang2015theory,weis2015retention,zhang2015vlsi,baseman2016improving,bautistagomez2016unprotected,chen2016rattecc,chen2016using,edri2016siliconproven,fiala2016flipsphere,kim2016allinclusive,kim2016eccaspirin,kim2016relaxfault,lim2016active,liu2016compilerdirected,lu2016voltage,memory2016imecc,nair2016xedexposing,park2016experiments,sfikas2016testing,cho2017anovel,duan20172dtcad,fieback2017dram_reliability,gong2017dram_scaling,halbert2017ondie,jung2017aplatform,kline2017sustainable,lin2017handling,malek2017oddecc,reed2017method,ryu2017overcoming,tsern2017dram_retention,wang2017architecting,chen2018configurableecc,gong2018duoexposing,hong2018eareccaided,kolditz2018ahead,levy2018lessons,meza2018large,pfaffli2018tcad,sankaranarayanan2018system,wang2018content,wang2018dram_retention,yeleswarapu2018sscmsdsinglesymbol,zhang2018exploring,chen2019care,goiffon2019radiationinduced,mrozek2019multirun,mukhanov2019workloadaware,nguyen2019stdrc,wang2019detect,zhang2019boosting,2020dynamic,bains2020read,boixaderas2020costaware,choi2020reducing,criss2020improving,jeong2020pair,kline2020flower,kwong2020rambleed,memory2020imecc,mukhanov2020dstress,yao2020deephammer,2021badpage,2021utrr,alam2021lightweight,chen2021care,safari2021blockhammer,jiang2021trrscope,longofono2021predicting,manzhosov2021muse,nguyen2021obet,pae2021minimal,rahman2021utilizing,sullivan2021characterizing,bepary2022dram_retention,kogler2022halfdouble,liu2022generating,loughlin2022moesiprime,manzhosov2022revisiting,patel2022acase,rakin2022deepsteal,roohi2022efficient,staudigl2022neurohammer,tomita2022extracting,yang2022sociallyaware,kwon2023temperature,marazzi2023protrr,zhou2023dnndefender,yaglikci2024spatial,brasser2017cant,vanderveen2018guardion,ji2019pinpoint,weissman2020jackhammer,zhang2020pthammer,deridder2021smash,jattke2022blacksmith,tol2022toward,zhang2022implicit,cohen2022hammerscope,fahrjr2022when,tobah2022spechammer,aydin2022cyber,mus2022jolt,wang2022research,lefforge2023reverse,fahr2022theeffects,france2022modeling,lenovo2015rowhammer,greenfield2012throttling,kim2014architectural,bains2016rowhammer,bains2016distributed,aichinger2015ddrmemory,aweke2016anvil,gomez2016dram_rowhammer,yang2016suppression,son2017making,seyedzadeh2017counterbased,you2019mrloc,lee2019twice,park2020graphene,yaglikci2021security,kang2020cattwo,qureshi2022hydra,saileshwar2022randomized,konoth2018zebram,vig2018rapid,gautam2019rowhammering,kim2022mithril,zhang2022softtrr,joardar2022learning,juffinger2023csirowhammercryptographic,saxena2022aqua,naseredini2022alarm,joardar2022machine,zhang2020leveraging,loughlin2021stop,olgun2024abacus,mutlu2019rowhammer,luo2024rowpress,canpolat2024understanding,canpolat2024breakhammer,canpolat2025chronus,bostanci2024comet,saxena2023ptguard,bennett2021panopticon,france2022reducing,guha2022criticality,devaux2021method,han2021surround,fakhrzadehgan2022safeguard,saroiu2022theprice,saroiu2022howto,zhou2022ltpim,hong2023dsac,mutlu2023fundamentally,marazzi2022rega,didio2023copyonflip,sharma2022areview,woo2023scalable,park2022rowhammer_reduction,wi2023shadow,kim2023a11v,guderamarao2023defending,park2016statistical,lim2018study,yun2018study,yang2019trapassisted,walker2021ondram,jiang2021quantifying,khan2018analysis,agarwal2018rowhammer_for,he2023whistleblower,baeg2022estimation,olgun2023anexperimental,zhou2023doublesided,kurmus2017from,orosa2024spyhammer,BitFlipEncNN25ICPADS,CorrosionHammer25CF,CorrosionHammer25Cybersec}
\providecommand{\citeMEMRELIAB}{\expandafter\cite\expandafter{\keysMEMRELIAB}}

\providecommand{\keysMEMCTRL}{gupta2024relief,mutlu2007stall,rixner2000memory,rixner2004memory,weiner2022tmo,zuravleff1997controller,kim2010atlas,muralidhara2011reducing,david2011memdvfs,deng2011memscale,kim2016bounding,kim2014bounding,kim2010thread,pa-micro08,Ipek2008,subramanian-hpca2013,usui-taco2016,moscibroda-usenix2007,hashemi2016accelerating,lee-tc2011,kang.memoryforum14,lavanya-thesis,rachata-thesis,807426,subramanian-micro2015,mutlu-isca2008,lee2009improving,jeong2012balancing,hur2004adaptive,ebrahimi2011parallel,moscibroda2008distributed,xie2014improving,wang2014memory,das2013application,Khalifa_HMC,shingari2015characterization,ghose.isca13,bird1991using,rau1991pseudorandomly,sohi1991highbandwidth,farkas1997memorysystem,carter1999impulse,hong1999access,lebeck2000power,lee2000eager,mckee2000dynamic,tremaine2001pinnacle,zhang2001theimpulse,suh2002anew,lin2003qualityaware,briggs2004microarchitecture,sprangle2004method,lee2005anefficient,huan2006processor,nesbit2006fair,conway2007theamd,diniz2007limiting,ghosh2007smart,rafique2007effective,shao2007aburst,aggarwal2008powerefficient,bitirgen2008coordinated,hur2008acomprehensive,owen2008northbridge,blagodurov2010contentionaware,cheng2010memory,stuecheli2010thevirtual,zhuravlev2010addressing,kaseridis2011minimalist,mutlu2011memory,tang2011theimpact,udipi2011combining,bojnordi2012pardis,liu2012asoftware,goossens2013conservative,suzuki2013coordinated,ding2014compiler,hussain2014pmss,shin2014nuat,wang2015xchange}
\providecommand{\citeMEMCTRL}{\expandafter\cite\expandafter{\keysMEMCTRL}}

\providecommand{\keysMEMCACHE}{ayers2018memory,liang2025ariadne,li-asplos19,vijaykumar2018xmem,vijaykumar2018locality,hajinazar2020virtual,qureshi2006case,khan2014improving,seshadri2012evicted,ganapathy1998general,pekhimenko2013lcp,mosaic-osr,seshadri-taco2015,ebrahimi-isca2011,basu2013efficient,jaleel2010high,ebrahimi2009,roth1998,seshadri2014dirty,sathaye2009structure,power.hpca14,bhattacharjee.micro13,bhattacharjee.asplos10,pichai.asplos14,bhattacharjee.hpca10,srikantaiah.micro10,lustig.taco13,goodman.isca83,wang2021stream,schwedock2023optimizing,wang2022near,wang2019bit,simon2020blade,abali2001mxt,pekhimenko2012lcp,daley1968virtual,bensoussan1972multics,qureshi-hpca07,samira2014improving,gena-thesis,burger1995declining,karakostas2014performance,LipastiSKR95,LukM96,JosephG97,RothS99,YangL00,KarlssonDS00,CollinsWTHLLS01,SolihinTL02,Wu02,CollinsSCT02,CookseyJG02,DBLP:conf/hpca/HuMK03,SrinathMKP07,ebrahimi.micro09,DBLP:conf/micro/YuHSD15,kanellopoulos2023utopia,kanellopoulos2023victima,qureshi.isca07,ishii2009access,fu1992stride,bakhshalipour2018domino,bera2019dspatch,zhang2000frequent,hallnor2005unified,islam2009zero,yang2000frequent,kadjo2014b,dusser2009zero,pekhimenko2012base,gonzalez1997speculative,cao1995study,tyson1995modified,hashemi2018learning,kim2002adaptive,sembrant2014direct,memik2003just,johnson1997run,johnson1999run,kharbutli2008counter,gupta2013adaptive,charney1995,sridharan2016discrete,tsai2019compress,tsai2017jenga,gaur2016base,pekhimenko2015exploiting,pekhimenkoenergy,pekhimenko2015toggle,pekhimenko2016case,zhang2007accelerating,annavaram2001data,talatiprodigy,ekman2005robust,arelakis2014sc,charneyphd,schwedock2022tako,SiNUCA_Thesis_2014,embedDRAM2,zhong2009program,gu2009component,ayers2020classifying,koo2017access,van2002memory,azarkhish2016memory,santos2016exploring,tsai2018rethinking,miguel2015doppelganger,tuck2006scalable,palacharla1994evaluating,wilkes1965slave,gindele1977buffer,smith1978sequential,denning1980working,kroft1981lockupfree,puzak1985analysis,chang1987cache,lee1987data,baer1988onthe,so1988cache,kessler1989inexpensive,mcfarling1989program,pomerene1989prefetching,gupta1990improving,hidaka1990thecache,alex1991anarchitecture,baer1991aneffective,mcfarling1991cache,mcfarling1991procedure,chen1992anefficient,dubnicki1992adjustable,kessler1992page,mowry1992design,smith1992prefetching,wada1992ananalytical,agarwal1993columnassociative,chen1993theeffect,golden1993hardware,gosmann1993code,hsu1993instruction,seznec1993acase,williamyuweichen1993data,wolfe1993twoported,alex1994dynamic,bershad1994avoiding,johnson1994expansion,jouppi1994tradeoffs,mendlson1994compile,wilton1994anenhanced,austin1995streamlining,calder1995next,dahlgren1995sequential,lee1995instruction,mcfarl1995cmos,mekhiel1995multilevel,seznec1995dasc,calder1996predictive,juan1996thedifferencebit,mehrotra1996examination,pierce1996wrongpath,rivers1996reducing,wilson1996increasing,xia1996instruction,charney1997prefetching,chen1997instruction,hashemi1997efficient,hashemi1997procedure,ki1997adaptive,racunas1997achieving,wilson1997designing,kumar1998exploiting,luk1998cooperative,srinivasan1999astatic,wilson1999thecase,zhang2000hardwareonly,zhang2000highlyassociative,jain2001controlling,lai2001deadblock,cooksey2002contentsensitive,karkhanis2002aday,peir2002bloom,wang2002using,balasubramonian2003adynamically,zhuang2003ahardwarebased,alameldeen2004frequent,iacobovici2004effective,iyer2004cqos,jnesbit2004acdc,kim2004fair,hsu2006communist,hur2006memory,rafique2006architectural,iyer2007qospolicies,nesbit2007virtual,jaleel2008adaptive,lin2008gaining,wilkerson2008trading,herdrich2009ratebased,xie2009pipp,zhao2009variationtolerant,lee2010dramaware_lastlevel,wilkerson2010reducing,alameldeen2011energyefficient,agrawal2013refrint,kang2015amnesic,seshadri2015page,son2015cidra,danger2018ccficache,hong2018attaché,wolff2018word,bera2021pythia,enomoto2022efficient}
\providecommand{\citeMEMCACHE}{\expandafter\cite\expandafter{\keysMEMCACHE}}

\providecommand{\keysMEMNVM}{bender2015memsys,song2020improving,zhang2009exploring,ramos2011page,yu2017banshee,meza2012enabling,li2017utility,yoon2012row,yoon-taco2014,tsai:micro:2018:ams,su2015hpmc,lee-isca2009,mikolajick2001feram,scott2007applications,bondurant1990ferroelectronic,scott1989ferroelectric,DBLP:conf/hpca/MeswaniBRSIL15,che2016challenges,qureshi.isca09,kultursay.ispass13,wong.procieee10,naeimi.itj13,lee.ieeemicro10,zhou.isca09,lee.cacm10,meza.weed13,zhao.micro14,ren.micro15,salkhordeh2019analytical,phadke2011mlp,patel2023xfm,nalli2017analysis,Liu:2017:HCC:3079079.3079089,deoliveira2021optane,pirovano2004reliability,kim2005reliability,lee2005lowenergy,kang2006a01um,huai2008spintransfer,kim2008future,raoux2008phasechange,condit2009better,lee2013clock,raybuck2021hemem,wang2019panthera,dhiman2009pdram,dong2009leveraging,qureshi2009enhancing,raychowdhury2009design,chen2010advances,ipek2010dynamically,ishigaki2010amultilevelcell,kong2010improving,schechter2010useecp,atwood2011current,coburn2011nvheaps,park2011power,qureshi2011payasyougo,volos2011mnemosyne,yoon2011freep,awasthi2012efficient,meza2012acase,narayanan2012wholesystem,zhang2012multilevel,apalkov2013spintransfer,wen2013cdecc,bishnoi2014read,li2014write,lu2014looseordering,meena2014overview,ren2015dualgranularity,vatajelu2015sttmrambased,zhang2015optimizating,chauhan2016nvmove,guo2017sanitizer,hamdioui2017test,tavana2017remap,zhang2017dphsim,zhang2017dynamic,khan2018test,ni2018write,vatajelu2018state,nair2019sudoku,nishi2019advances,pal2019design,vonthun2020qualification,wang2020characterizing,yavits2020wolfram,genssler2022onthe,zhang2011stt,chen2014rram,atwood2018pcm,bock2011analyzing,burr2008overview,du2013bit,ferreira2010increasing,jiang2012fpb,jiang2013hardware,kannan2016energy,qureshi2010improving,qureshi2010morphable,sebastian2017temporal,wang2015exploit,yue2013accelerating,zhou2012writeback,zhou2013writeback,song2019enabling,yoon2013techniques,song2021aging,akinaga2010resistive,wang2013low,diao2007spin,hosomi2005novel}
\providecommand{\citeMEMNVM}{\expandafter\cite\expandafter{\keysMEMNVM}}

\providecommand{\keysARCHCORE}{halpern2016mobile,janapa2010web,reddi2011mobile,wang2013architectural,wang2017xpro,zhu2013high,suggs2020amd,harris_harris,yuffe2011,christy2020,singh2017,saini1993design,jaggar1997arm,waterman2016design,flynn1966very,gustafson1988reevaluating,amdahl1976validity,burtonsmith1978,smith1982architecture,thornton1970,henessy.patterson.2012.fgmt,ddca.spring2020.fgmt,haj2020sysscale,haj2020techniques,friedrich2014power8,lotfi-kamran.2012,mesi1984,lim-isca09,smith1982decoupled,rogers1995supporting,glew1998mlp,black3dstack,Turbo,SCD,Coarse-grain,HSC,SCT,streamRegister,multi-grain,phillips2014m7,barroso2009datacenter,lim2008understanding,sanchez.micro07,dmp,bulksc,censier.tc78,arch-os,os-hardware,burks.1946,hashemi2016continuous,gillespie2014steam,howard201048core,jowani2010x8664,kilburn1962one,fotheringham1961dynamic,kung1982systolic,kumar.isscc2009,sriraman2019softsku,ferdman.asplos12,MutluKP05,ZillesS01,Luk01,DBLP:conf/hpca/MutluSWP03,DBLP:conf/isca/CragoP11,DBLP:conf/isca/KurianHC92,DBLP:journals/micro/MutluSWP03,DBLP:conf/isca/MutluKP05,DBLP:journals/tc/MutluKP06,DBLP:journals/micro/MutluKP06,tcc,smith.tc86,smith.computer89,joao.isca09,tomasulo.ibmjrd67,patt.hps.micro85,patt.critical.micro85,muzahid.isca09,pokam.micro09,lamport.tc79,lopes2015shrink,lopes2013isa,arafa2019cascade,jacob2009memory,hammerstrom1977information,kirman2005checkpointed,ceze2006cava,austin1995zero,lipasti1996exceeding,goeman2001differential,endo2017interactions,eickemeyer1993load,yazdanbakhsh2016rfvp,gabbay1996speculative,tuck2005multithreaded,tune2002quantifying,tullsen1999storageless,sazeides1997predictability,sato2002low,nakra1999global,burtscher1999exploring,lipasti1996value,yazdanbakhsh2016mitigating,orosa2018avpp,patterson2016computer,farrens1994study,mutlu2004cache,mutlu2005using,fallin2011chipper,moscibroda2009case,das2010aergia,dally2004principles,besta2018slim,benini2002networks,grot2011kilo,nychis2012chip,das2009application,tyson1994effects,luo2001balancing,michaud2012demystifying,snavely2000symbiotic,kozyrakis2010server,ausavarungnirun2014design,yeh1991two,yuan2023rambda,shooman1960parallel,batcher1982bit,Kumar_2014,kuck1980structure,padua1986advanced,allen1987automatic,thornton1964parallel,trifunovic2009polyhedral,pohl2018cost,patsilaras2012efficiently,everman2007memory,ramirez2010efficient,van2009mlp,collins2001dynamic,jacob_book_2008,AutoVectorization,kumar2015dasx,yuan2019halo,srinivasan2004continual,brewer2010instruction,instruction_scheduling,lukefahr2012composite,kim2007composable,suleman2009accelerating,chaudhry2009simultaneous,joao2012bottleneck,pohl2020vectorization,lattner2008llvm,pohl2019portable,padmanabha2017mirage,kumar2003single,alipour2019fiforder,kumar2019freeway,alipour2020delay,tarjan2008federation,joao2013utility,suleman2010data,ipek2007core,villavieja2014yoga,petrica2013flicker,fallin2014heterogeneous,wang1997highly,calder1999selective,oliveira2018employing,chou2004microarchitecture,avizienis1961signeddigit,riseman1972theinhibition,jayashree1976onbinary,keller1976lookahead,russel1978thecray1,connors1979theibm,dennis1980data,johnson1980adigital,fisher1981trace,smith1981astudy,harbison1982anarchitectural,allen1983conversion,lee1984branch,weiss1984instruction,young1984asimulation,gurd1985themanchester,smith1985implementation,aho1986compilers,hsu1986highly,mcfarling1986reducing,ferrante1987theprogram,hwu1987checkpointproc,melvin1987aclarification,smith1987thezs1,colwell1988avliw,ebciouglu1988some,melvin1988hardware,smith1988implementing,alex1989runtime,briggs1989coloring,dehnert1989overlapped,hwu1989comparing,jouppi1989available,nakatani1989``combining,rau1989thecydra,smith1989limits,weber1989exploring,arvind1990executing,butler1990anareaefficient,chow1990theprioritybased,colwell1990architecture,davidson1990reducing,golumbic1990instruction,grohoski1990machine,hswarren1990instruction,love1990aninvestigation,moon1990hardware,pettis1990profile,smith1990boosting,tirumalai1990parallelization,bernstein1991global,butler1991single,chang1991comparing,chang1991impact,johnson1991superscalar,kaeli1991branch,melvin1991exploiting,nemirovsky1991disc,papadopoulos1991multithreading,popescu1991themetaflow,stone1991computer,wall1991limits,austin1992dynamic,chen1992branch,fisher1992predicting,franklin1992register,franklin1992theexpandable,granlund1992eliminating,hirata1992anelementary,hwu1992efficient,lam1992limits,mahlke1992effective,mahlke1992sentinel,moon1992anefficient,mueller1992avoiding,pan1992improving,richardson1992caching,yeh1992acomprehensive,yeh1992alternative,abraham1993predictability,ball1993branch,butler1993aggressive,drach1993midee,franklin1993themultiscalar,hank1993superblock,hwu1993thesuperblock,mcfarling1993combining,moudgill1993register,perleberg1993branch,rau1993dynamically,rau1993instructionlevel,schlansker1993acceleration,silberman1993anarchitectural,warter1993reverse,windheiser1993ksr1,yeh1993acomparison,yeh1993branch,yeh1993increasing,calder1994fast,calder1994reducing,franklin1994afillunit,gwennap1994digital_leads,hao1994theeffect,hsu1994design,kathail1994hplplaydoh,krall1994improving,mahlke1994characterizing,pnevmatikatos1994guarded,schlansker1994critical,schlansker1994height,smith1994powerpc,sprangle1994facilitating,vassiliadis1994interlock,vassiliadis1994scisma,young1994improving,chang1995alternative,chang1995profileguided,conte1995optimization,dutta1995control,farkas1995howuseful,gwennap1995intels,jourdan1995exploring,nair1995dynamic,sechrest1995therole,tullsen1995simultaneous,uhlig1995instruction,uht1995disjoint,young1995acomparative,bondi1996integrating,chang1996improving,chen1996analysis,clift1996floating,evers1996using,franklin1996arba,gwennap1996digital_21264,gwennap1996intels,hao1996increasing,heil1996selective,jacobsen1996assigning,jourdan1996theeffects,keller1996the21264,nagendra1996areatimepower,rotenberg1996trace,sechrest1996correlation,seznec1996multipleblock,soltis1996inside,tullsen1996exploiting,yeager1996themips,yung1996design,bugnion1997disco,chang1997predicting,dean1997profileme,driesen1997accurate,friendly1997alternative,gieseke1997a600mhz,jacobson1997control,jacobson1997pathbased,jourdan1997recovery,lee1997thebimode,menezes1997path,michaud1997trading,moshovos1997dynamic,moshovos1997streamlining,nair1997exploiting,palacharla1997complexityeffective,patel1997critical,patt1997onebillion,rotenberg1997trace,sodani1997dynamic,sprangle1997theagree,vajapeyam1997improving,wallace1997multiple,akkary1998adynamic,chrysos1998memory,colwell1998maintaining,evers1998ananalysis,farrell1998issue,friendly1998putting,gonzalez1998virtualphysical,gowan1998power,grohoski1998reining,hao1998increasing,lynch1998lowload,manne1998pipeline,onder1998superscalar,patel1998improving,skadron1998improving,sodani1998understanding,stark1998variable,austin1999diva,barnes1999a500mhz,bekerman1999correlated,bekerman1999early,chappell1999simultaneous,govil1999cellular,jacobson1999instruction,kessler1999thealpha,pai1999code,patel1999trace,reinman1999ascalable,shen1999replenishing,barroso2000piranha,canal2000alowcomplexity,henry2000circuits,jimenez2000theimpact,moon2000scalable,stark2000onpipelining,balasubramonian2001dynamically,dally2001route,folegnani2001energyeffective,goshima2001ahighspeed,hinton2001themicroarchitecture,jimenez2001dynamic,leenstra2001a18ghz,michaud2001dataflow,morancho2001recovery,patel2001replay,peh2001adelay,roth2001speculative,sager2001a018$mu$m,tendler2001power4,aragon2002dual,briggs2002intel,lebeck2002alarge,martinez2002cherry,snavely2002symbiotic,akkary2003checkpoint,akkary2003recycling,balakrishnan2003exploiting,barnes2003beating,cristal2003acase,raasch2003theimpact,towles2003throughputcentric,wang2003ybranches,zhou2003enhancing,bera2022hermes,armstrong2004wrong,cazorla2004dynamically,ceze2004cava,cristal2004outoforder,millberg2004guaranteed,mullins2004lowlatency,thottethodi2004exploiting,annavaram2005mitigating,borkar2005designing,chou2005store,kim2005alow,kim2005wish,kongetira2005niagara,mutlu2005onreusing,qureshi2005microarchitecturebased,reis2005swift,spracklen2005chip,gabor2006fairness,gratz2006implementation,gschwind2006synergistic,kim20062dprofiling,kim2006divergemerge,kim2006wish,laudon2006ultrasparc,wagner2006shielding,borkar2007thousand,choi2007modeling,constantinides2007softwarebased,hoskote2007a5ghz,hwu2007implicitly,joao2007dynamic,kim2007divergemerge,kim2007flattened,kim2007profileassisted,kim2007vpcprediction,kumar2007express,li2007applicationlevel,michelogiannakis2007approaching,neelakantam2007hardware,owens2007research,sarangi2007patching,shye2007using,wentzlaff2007onchip,cai2008meeting,cho2008interaction,constantinides2008online,grot2008regional,hill2008amdahls,lee2008globallysynchronized,nesbit2008multicore,shye2008plra,wagner2008using,constantinides2009aflexible,grot2009express,grot2009preemptive,kim2009virtual,li2009operating,saltzer2009principles,eyerman2010modeling,grot2010topologyaware,li2010concurrent,mutlu2010asymmetry,mutlu2010rethinking,neelakantam2010areal,nychis2010next,ware2010architecting,watanabe2010widget,borkar2011thefuture,das2011aergia,hennessy2011computer,martinezalvarez2011compilerdirected,suleman2011data,warnock2011a52ghz,chang2012hatheterogeneous,elsayed2012temperature,fallin2012minbd,nachiappan2012applicationaware,dubois2013criticality,dubois2013perthread,mishra2013aheterogeneous,martinez2014xchange,fattah2015alowoverhead,weste2015cmos,zhu2015microarchitectural,ausavarungnirun2016acase,balasubramonian2017cacti,gustafson2017beating,sun2017study,tian2017approxlut,yahya2022darkgates}
\providecommand{\citeARCHCORE}{\expandafter\cite\expandafter{\keysARCHCORE}}

\providecommand{\keysARCHGPUHPC}{a100,ampere,baghsorkhi2016flexvec,choquette2020nvidia,khadem2023vector,olivier2019hexo,cheng2014professional,firasta2008intel,huang2019gpu,xie2017cumf_sgd,rocki2020fast,rapids,catanzaro2014decomposition,van2013simulation,gomez2013atomics,dotsenko2008fast,sengupta2008efficient,yan2013streamscan,hwukirk2016.scan,degonzalo2019automatic,harris2007optimizing,rabenseifner2004optimization,gomez2013optimized,blackford2002updated,saad2003iterative.parallel,saad2003iterative.sparsematrices,liu2015csr5,rauber2013parallel,henessy.patterson.2012.simd,hager2010introduction.scaling,ddca.spring2020.gpu,ddca.spring2020.simd,gomez2016matrix,sung2014matrix,sung2012dl,bell2012thrust,merrill2015cuda,gomezluna2015ds,boyer2013,blelloch1989,sadi2019efficient,kanellopoulos2019smash,keckler2011gpus,li2019evaluating,werner2025towards,pattnaik2019opportunistic,rhu2013locality,chen2014adaptive,sadrosadati2019itap,davis2012spmv,wahib2014scalable,sadrosadati2018ltrf,mask,mosaic,jog.asplos2013,jog.isca2013,veynu.2011,nandita.2016,nandita.2015,jog.2016,sadrosadati2021highly,olabi2022compiler,pan2021exploring,darabi2022morpheus,chen2013guided,paul.2015,medic,sinclair_micro2015,kim2018coda,nvidia2009nvidia,lindholm2008nvidia,im1999optimizing,sengupta2007scan,cole2004bringing,cole1989algorithmic,seitz1985thecosmic,gaffar2004unifying,liang2006bluegenel,bryant2007dataintensive,schroeder2007understanding,seiler2008larrabee,vangal2008an80tile,kelm2009rigel,chung2010singlechip,lan2010astudy,xiao2010interblock,brodtkorb2013graphics,cavus2014gpubased,hestness2014acomparative,kayiran2014managing,villa2014scaling,dally2015thepath,fang2017letgo,salami2018comprehensive}
\providecommand{\citeARCHGPUHPC}{\expandafter\cite\expandafter{\keysARCHGPUHPC}}

\providecommand{\keysARCHMLACCEL}{Yu2024CambriconLLM,boroumand2021google,sheng2023flexgen,mahajan2016tabla,de2017understanding,abts2020think,jouppi2017datacenter,jouppi2021ten,chetlur2014cudnn,boroumand2021mitigating,hwukirk2016.nn,gondimalla2019sparten,hegde2019extensor,zhu2019sparse,cho2019fa3c,wang2019accelerating,chung2018serving,umuroglu2017finn,peccerillo2022survey,nori2021reduct,han2016eie,gupta2020architectural,zhao2022understanding,tangram,hauswald2015sirius,han2017ese,dnpu,zhang2021sara,kara2017fpga,Farabet2011,Du2015,eyeriss,sze2017efficient,chen2018tvm,wang2020neural,wang2020systematic,wang2019benchmarking,kim2018deeptrain,li2016optimizing,rhu2016vdnn,jain2018architectural,peemen2013memory,eyerissv2,scnn,esmaeilzadeh2012neural,wang2016anoverview,garofalo2020pulpnn}
\providecommand{\citeARCHMLACCEL}{\expandafter\cite\expandafter{\keysARCHMLACCEL}}

\providecommand{\keysARCHTOOLS}{oliveira2021damov,che2009rodinia,hara2008chstone,kanev.isca15,kim-cal-2016,luo2023ramulator2,noauthor_patmos_hlsbenchmarks_nodate,olgun2023dram,pouchet2012polybench,qemu,ramulator-github,khan2018rapl,hahnel2012measuring,gomezluna2017chai,advisor,roofline,rapl,smi,matrixmarket,lattner2004llvm,llvm-builtin,upmem-explorer,mccalpin1995,gaekegups,luszczek_hpcc2006,yasin2014,hassan2017softmc,wang.hpca14,council2005transaction,GEM5,gutierrez2014sources,DRAMSim2,zilles2001benchmark,Perf,OProfile,ramulator.github,softmc.github,zsim,ajayi2019openroad,shao2013isa,MemoryBo20,stengel2015quantifying,LiASBTJ13,gem5-gpu,gpgpusim,conte1995advances,conte1991brief,Conte91,flexus,marss,yazdanbakhsh2016axbench,Venkata_iiswc2009,zsimplusplus,llvmprofiler,stratton2012parboil,woo_isca1995,muralimanohar2007optimizing,pisa,narancic2014evaluating,corda2019memory,bienia2008parsec,limaye2018workload,hsia2020cross,lottarini2018vbench,wang2022characterizing,richins2020missing,sriraman2020accelerometer,gan2018architectural,jia2016understanding,kasture2016tailbench,hmc_sim,spec2006,spec2017,cashmc,SiNUCA_HPCC,Thomas_CortexSuite_IISWC_2014,hadidi2017demystifying,cacti,mcpat_bad,panda2001systemc,mathew2017bank,adolf2016fathom,energy-roofline-model,chdrampower_opensource,drampower_source,einsim,safariplutogithub,safariplutozenodo,safariramulator,harp,rambuspowermodel,ptmpredictive,nagel1973spice,srivastava1994atom,vladimirescu1994thespice,burger1996evaluating,moudgill1998anapproach,bhargava1999accurately,austin2002simplescalar,kleinosowski2002minnespec,sherwood2002automatically,skadron2003challenges,hamerly2005simpoint,luk2005pinbuilding,yi2005improving,bhansali2006framework,kanellopoulos2024virtuoso,nagpurkar2008workload,fritts2009mediabench,li2009mcpat,muralimanohar2009cacti,li2010design,carlson2011sniper,gutierrez2011fullsystem,chatterjee2012usimm,jeong2012drsim,poremba2012nvmain,ahn2013mcsima+,huang2014moby,parasyris2014gemfi,roberts2014faultsim,jung2015dramsys_a,khairy2020accel,safari2017softmc,francismezger2018araspberry,mutlu2018lecture,software2019memtest86,patel2021harpartifacts,safari2023ramulator2}
\providecommand{\citeARCHTOOLS}{\expandafter\cite\expandafter{\keysARCHTOOLS}}

\providecommand{\keysENERGYTECH}{mutlu.imw13,hong2010memory,strukov2008,wong2012,dennard1974design,rich2020nano,sabry2015abundant,denning.2016,dally.2015,itrs.2009,gonzalez1996energy,rotem2011power,haj2020flexwatts,wu2005voltage,mallik2006user,kim2008system,schmidt1998wafer,kagawa2016novel,dennard1968field,gopireddy2019m3d,mitra2018vlse,hwang2018cmos,mitra2015nano,lau2023recent,kim2023van,ebrahimi2014monolithic,sabry2019n3xt,chua.tct71,lefurgy.2003,ware.2010,fong2015spin,TSV,sevilla2022compute,garello2014ultrafast,yu2016emerging,gallagher2019recent,schaller1997moore,sakurai1993closedform,1994thenational,mizuno1994experimental,1998international,frank2001device,roy2003leakage,wang2003electrical,chang2005stable,davis2005demystifying,friedberg2005modeling,wann2005sram,leung2006systeminpackage,zhao2006newgeneration,association2007process,chen2008thenext,pavlov2008cmos,guo2009largescale,iniewski2011nanosemiconductors,kim2011variationaware,integration2012predictive,karim2012onchip,itrs2013international,yoon2013flash,neuberger2014protecting,itrs2015more,lee2016technology,yang2013memristive}
\providecommand{\citeENERGYTECH}{\expandafter\cite\expandafter{\keysENERGYTECH}}

\providecommand{\keysGENOMICS}{ghiasi2026sage,alkan.naturegenetics09,alser-bioinformatics-2017,altschul-jmb-1990,cali-micro-2020,eris2025rawbench,firtina2023rawhash,firtina2024rawhash2,firtina2024rawsamble,lander-nature-2001,lindegger2024rawalign,loving-bioinformatics-2014,papageorgiou2018genomic,xin-bioinformatics-2015,rautiainen2019bit,li2013aligning,bhukya2011exact,dandass2008accelerating,alser2020accelerating,alser2022molecules,zook2016extensive,sayers2021database,goodwin2016coming,glenn2011field,reuter2015high,lappalainen2019genomic,mount2001bioinformatics,Needleman1970Ageneral,li2018minimap2,turakhia2018darwin,fujiki2018genax,li2009fast,alser2019shouji,bitwise-alignment,gotoh1982improved,paten2017genome,pevzner2001eulerian,ameur2019goodbye,rakocevic2019fast,kaye2021genome,eizenga2020pangenome,cali2022segram,alonso2024bimsa,marco2023optimal,marco2021fast,alser2020sneakysnake,cali2018nano,alser2021technology,mutlu2023accelerating,lindegger2023scrooge,lindegger2022algorithmic,xin.bmcgenomics13,xin.oss.bioinformatics15,lee.methods14,bingol2024gatekeeper,cavlak2024targetcall,firtina2024aphmm,singh2024rubicon,wood2019improved,lapierre2020metalign,karasikov2020metagraph,chaisson2012mapping,liu20173d,ahmed2016comparison,appuswamy_ipdpsw2018,doblas2023gmx,treangen2012repetitive,kaplan2017aresistive,angizi2020exploring,angizi2020pimaligner,hanhan2022edam,bonito,lou2020helix,chen2020parc,hhrlich2011metahit,sunagawa2015structure,fierer2017embracing,FindeR19PACT}
\providecommand{\citeGENOMICS}{\expandafter\cite\expandafter{\keysGENOMICS}}

\providecommand{\keysDBANALYTICS}{bloom1970space,chan1998bitmap,elmasri2007fundamentals,fastbit,goodwin-SIGIR-2017,graf2020xor,graf2022binary,idreos2012monetdb,jia2017understanding,kocberber2013meet,li-sigmod-2013,li-vldb-2014,myers-jacm-1999,oneil-ideas-2007,panda2014data,stonebraker2013voltdb,wu-icde-1998,wu2014q100,nambiar2006making,hsu2001characteristics,olson2020approach,shao2007bitmap,wu2009fastbit,pelkonen2015gorilla,koudas2004flexible,gadepally2016storage,erdem2016tree,le2012memory,ahmed2014hardware,tsai2015big,runkler2020data,ceri1985translating,zhu2018matrix,baeza1992new,wu1992fast,yeh2016matrix,mysql,yustaring,Teubner:2013:MHJ:2510649.2511320,RDD,kallman2008h,OLTP,boncz.1999,ailamaki1999dbmss,wu2013navigating,bruni2012optimizing,harizopoulos2008oltp,lehman1981efficient,carey1994shoring,bicompression,silberschatz1997database,ChungDL13,cao2019titant,sql-htap,lee2017parallel,snappy-data,sahay2008real,ta2016big,wiser,huang2020tidb,sap-hana-evolution,htap,htap-gartner,peloton,htap-survey,hyper,ankerdb,batchdb,lahiri2015oracle,awan.2015,awan.2016,yasin.2014,hsieh2018focus,balkesen_TKDE2015,yoo_iiswc2009,sirin_damon2017,balkesen2013multi,stonebraker2013bigdata,Copeland_1985,Manegold_2000,boncz_2005,boncz_2013,russomp_2012,Abadi_2009,dhindsap_2012,Russom_2013,chen2014big,labrinidis2012challenges,fang1998computing,cohen2003spectral,karp2003asimple,dean2004mapreduce,wu2005fastbit,bonomi2006animproved,ranger2007evaluating,vandurme2009probabilistic,cooper2010benchmarking,li2012compression}
\providecommand{\citeDBANALYTICS}{\expandafter\cite\expandafter{\keysDBANALYTICS}}

\providecommand{\keysGRAPH}{beamer-SC-2012,bundy1984breadth,low.vldb12,danisch2018listing,li2020ordering,gianinazzi2021parallel,jabbour-kcsl-2018,salihoglu2013gps,shi2018graph,luo2010effective,hwukirk2016.bfs,rmat,graphgowalla,page1999pagerank,ligra,Seraph,snap,harshvardhan.pact14,hong.asplos12,malewicz.sigmod10,tian.vldb13,kyrola2012graphchi,nguyen2013lightweight,qiu2018real,graphlab,gonzalez.osdi12,umuroglu.2015,xu.2014,bulucc2011combinatorial,yan2019alleviating,basak2019analysis,ainsworth2016graph,kaushik2021gretch,nilakant2014prefedge,besta2021graphminesuite,west2015hybrid,zheng2020distdgl,graph_mining1,graph_mining2,graph_clustering,bron-kerbosch}
\providecommand{\citeGRAPH}{\expandafter\cite\expandafter{\keysGRAPH}}

\providecommand{\keysMLALGO}{Karpathy,LSTM,qlearning_ML_1992,reinforce1992,touvron2023llama,wang2020survey,aiolli2013efficient,rastegari2016xnor,zhou2016dorefa,hubara2016binarized,minaee2021image,maulud2020review,sarangdhar2017machine,akbar2017predictive,martin2022iot,sivasankari2022detection,dedeturk2020spam,chang2008partitioned,wijaya2016hybrid,han1995influence,gholamisurvey,zmora2021quantization,bo2019cluster,breiman2001random,suthaharan2016decision,geron2019,alpaydin2020,goodfellow2016,mohri2018,shalev2014,raschka2019,sutton_2018,pedregosa2011scikit,hinton1987learning,abadi2016tensorflow,watkins1989learning,rummery1994line,sutton1999policy,baird1995residual,liang2016deep,mnih2013playing,silver2016mastering,silver2017mastering,geurts2006extremely,Lloyd82leastsquares,calinski1974dendrite,hubert1985comparing,baldi2014searching,susy,skin,Dua:2019,elkan2003using,vassilvitskii2006k,dunner2018snap,he2016deep,lecun1998gradient,stich2018sparsified,bottou2010large,jiang2021towards,epoch2023trendsinthedollartrainingcostofmachinelearningsystems,zhou2020towards,zinkevich2010parallelized,mcdonald2010distributed,dekel2012optimal,li2014communication,MAL-016_Boyd_ADMM,yu2019parallel,wang2017memory,chin2015fast,chin2015learning,geman1992neural,zhang2019mllib,liu2020distributed,lian2017can,lan2020communication,wang2023cocktailsgd,xu2020compressed,alistarh2017qsgd,strom2015scalable,dettmers2022gpt3,xi2023training,paszke2019pytorch,xue2024repeat,robbins1951stochastic,bottou2012stochastic,li2014efficient,boser1992training,cortes1995support,hastie2009elements,huang2005using,sirb2016consensus,naumov2019deep,desa2018high,hinton2015distilling,zhou2017kunpeng,quah2008real,ivakhnenko1965,rosenblatt1958perceptron,yegnanarayana2009artificial,varmedja2019credit,kannan2016smart,oliveira2016computer,lecun.cognitiva1985,gers.icann1999,graves2013generating,google-translation,lrcn,rcnn-google,karpathy2015deep,gru,kanai2017preventing,graves.icmlworkshop2012,transducer3,he.icassp2019,pinheiro.icml2014,liang.cvpr2015,cho2014properties,vaswani2017attention,xingjian2015convolutional,xu2015show,sutskever2014sequence,srivastava2015unsupervised,ranzato2014video,cho2014learning,lstm-google,graves2013hybrid,soltau2016neural,rcnn-caption,crnn2,crnn,rcnn-weather,transducer4,greff.tnnls2017,hochreiter.neco1997,szegedy2015going,zeiler2014visualizing,russakovsky2015imagenet,lecun1989handwritten,gu2018recent,li2015convolutional,qin2016joint,jiang2017face,simonyan2015very,fukushima.biologicalcybernetics1980,alexnet2012,squeezenet,ghosh2016contextual,he2020sparse,redmon2016yolo9000,zhang2016colorful,karpathy2015unreasonable,rumelhart.nature1986,vanhoucke2011improving,xiang2017binary,qiao2020stbnn,qin2020bipointnet,chen2021bnn,lin2017towards,deng2012mnist,krizhevsky2010convolutional,ren2015exploring,bojarski2016end,chen2015deepdriving,bag2017deep,mobilenet,zhao.ietits2017,devlin2018bert,raffel2020exploring,achiam2023gpt,milyaev2013image,denton2014exploiting,giurgiu2017predicting,hubara2017quantized,khoram2018adaptive,hong2019terminal,nguyen2020anapproximate,li2020sample,1606.01540,wang2014energy,kipf2016semi,xu2018powerful,hamilton2017inductive,kingma2013auto,goodfellow2014generative,liu2019fault}
\providecommand{\citeMLALGO}{\expandafter\cite\expandafter{\keysMLALGO}}

\providecommand{\keysHDC}{karunaratne-nature-2020,kanerva-congcomp-2009,imani-hpca-2017,kanerva-1992,lee-fnins-2020,ge2020classification,kang2015energy,roy2019towards,sharad2013ultra,fan2015hierarchical}
\providecommand{\citeHDC}{\expandafter\cite\expandafter{\keysHDC}}

\providecommand{\keysHEFOUND}{brakerski2012fully,brakerski2014leveled,fauzi2022ind,marcedone2014obfuscation,cryptoeprint:2024/1587,lopez2017multikey,lee2019security,regev2009lattices,aharonov2005lattice,arora1997hardness,HomomorphicEncryptionSecurityStandard,cheon2017homomorphic,albrecht2022homomorphic,fan2012somewhat,fontaine-eurasipjis-2007,bossuat2024security,alaya2020homomorphic,lyubashevsky2010ideal,gentry2009fully,moore2014practical,naehrig2011can,Regev:2005aln,regev2010learning,chillotti2016faster,acar2018survey,kim2021revisiting,pedrouzo2021revisiting,bonnoron2017somewhat,silverberg2013fully,van2010fully,gomathisankaran2011horns,gentry2011implementing,tourky2016homomorphic,yi2014homomorphic,chase2017security,cryptoeprint:2014/356,ogburn2013homomorphic,chaudhary2019Analysis,lyubashevsky2013toolkit,chillotti2020tfhe,shihab2018design,lyubashevsky2013ideal,cheon2020remark,ducasfhew,boura2020chimera,RAD78PrivacyHomo,Gentry09Thesis,BV11LWEFHE,BGV12LeveledFHE,GSW13,Peikert16Decade,BV11RLWE,CGGI20TFHE,Marcolla22Survey,smart2010fully,mu2026vulnerability,mankali2025glitchfhe,GHS12AES,SlotCoeffBGVBFV24,CKLS21LWEConvert}
\providecommand{\citeHEFOUND}{\expandafter\cite\expandafter{\keysHEFOUND}}

\providecommand{\keysHEALGO}{lu2021pegasus,huo2019note,chen2018homomorphic,li2018lazy,chen2019efficient,yasuda2015new,bon2024optimized,hu2013improving,gentry2012better,gentry2012fully,chillotti2021improved,chillotti2021programmable,chillotti2017improving,halevi2019improved,bajard2016full,yasuda2014practical,chillotti2017faster,lee2021high,kim2022approximate,gai2017blend,qin2019addition,castryck2018homomorphic,smart2014fully,mono2022finding,cheon2018bootstrapping,micciancio2018ring,liu2023amortized,bae2024bootstrapping,drucker2024bleach,iliashenko2021faster,CHKKS18FullRNS,CCS19ImprovedBootstrap,HS15HElibBoot,MP21BootstrappingFHEW,HS14AlgorithmsHElib,HS18FasterLin,Bossuat21EUROCRYPT,satriawan2023conceptual,alexandru2025general,li2024evalcomp,satriawan2024complete,han2020better,brakerski2013packed,Cheon19ChebyshevACrypt,Bae22MetaBTS,cheon2024dacapo,Kim22EvalRound,BKSS24DiscreteCKKS,AP13QuasilinearBoot,GV23BGVBootRevisited,GIKV23PolyModBoot,RelaxedFuncBootBGV24,KSS24BFVBootCKKS,HHC19ImprovedDFTBoot,LLKK22HighPrecErrVar,BTP22BootNegligibleFail,CHKS25SHIP,OverModRaise25CIC,BP23DemystifyBoot,NeuJeans23CRYPTO,HERMES23CRYPTO,Grafting25CCS,LMP22LargePrecSign,LMK23EfficientFHEWBoot,KLUCZNIAK23FDFB,DKMS23FasterFHEWAutomorph,NTRU4ms24TCHES,BlockBinaryTFHE23ASIACCS,TFHEAutomorphBoot25ASIACRYPT,JW22LibTFHE,XZDDF23FastBlindRot,BootSurveyCybersec25,BootOptSurveyCybersec26}
\providecommand{\citeHEALGO}{\expandafter\cite\expandafter{\keysHEALGO}}

\providecommand{\keysHESW}{mouchet2020lattigo,ebel2025orion,TFHE-rs,jiang2022fhebench,sealcrypto,chillotti2020concrete,polyakov2017palisade,dathathri2019chet,dathathri2020eva,HS20HElibDesign,OpenFHE22,SEAL,ChenLainePlayer17SEAL,HElibRepo,TFHErs,Lattigo,TenSEAL21,IbarrondoViand21Pyfhel,PALISADE,HEIR,nGraphHE19,nGraphHE2_19,Cingulata15,Porcupine21PLDI,HECO23,Fhelipe24PLDI,ViandJattkeHithnawi21SoKCompilers,Marble18,ConcreteML,HEProfiler24,singh2025efficient,GPUFHEFramework26ASPLOS,FHEFusion26CGO,FHECompilersSurvey25Cybersec}
\providecommand{\citeHESW}{\expandafter\cite\expandafter{\keysHESW}}

\providecommand{\keysHEAPPS}{pulido2021privacy,celsystems2021he,verma2022efficient,cruz2017secure,kim2021efficient,titus2018sig,bed2016privately,park2019secure,mr2016homomorphic,mahato2023comparative,bonte2020homomorphic,kumar2022homomorphic,wang2015generalized,hariss2017fully,aziz2024secure,petrean2024homomorphic,kim2016efficiency,cheon2015search,kim2017private,yasuda2014privacy,feer2024privacy,barni2010privacy,ren2016secure,lu2014confidentiality,dawoud2014privacy,wood2020homomorphic,kim2015private,ren2022heda,boneh2013private,el2015challenges,bian2023he3db,zhang2024arcedb,wibawa2022bfv,sebert2023combining,ding2017encrypted,fiore2014efficiently,yaji2018privacy,catak2020practical,doroz2016homomorphic,al2020towards,pradel2021privacy,alexandru2020cloud,raj2020web,chatterjee2019fully,stoian2023deep,sanyal2018tapas,lagesse2021you,brutzkus2019low,juvekar2018gazelle,xie2014crypto,bourse2018fast,yasuda2013secure,Delphi20,LeeMPCNN22,bian2025engorgio,SealPIR18,OnionPIR21,Spiral22,SimplePIR23,KimEtAl21Genotype,GraepelLauterNaehrig12,CryptoGCN22,SavPoseidon21,TawoseHE23,cheon2019polynomial,cryptoeprint:2024/136,lee2022privacy,gilad2016cryptonets,Lee22PPMLCACM,Potey16HECloud,WPC25CCS,CipherPrune25ICLR}
\providecommand{\citeHEAPPS}{\expandafter\cite\expandafter{\keysHEAPPS}}

\providecommand{\keysHEACCEL}{Kabra2025CipherMatch,ozcan2024heongpu,suzuki2023designing,reagen2021cheetah,gupta2023evaluating,ding2024apache,reis2020computing,gupta2021accelerating,zhou2025fhemem,mwaisela2024evaluating,nam2025affinity,fan2025fast,dai2016cuhe,dai2014accelerating,cousins2014fpga,su2022highly,kim2022bts,al2020multi,cousins2016designing,ozturk2016custom,agrawal2023fab,meftah2022towards,mert2019design,syafalni2022efficient,de2021does,jayet2015polynomial,samardzic2022craterlake,feldmann2021f1,jung2021over,wang2014accelerating,chen2024hmc,wang2012accelerating,wang2013exploring,nabeel2023cofhee,wang2023he,morshed2020cpu,cao2014high,cao2015optimised,cao2013accelerating,roy2018hepcloud,roy2019fpga,kim2022ark,pal2023fully,riaziheax,turan2020heaws,kim2023sharp,gupta2024memfhe,chen2014high,wang2013fpga,yang2023poseidon,park2025fhendi,kim2025anaheim,PhantomFHE23,NuFHE,Jung21CKKS100x,TensorFHE23HPCA,Shen23CARM,Badawi18GPUFV,HEAX20,TuranRoy20HEAWS,Poseidon23HPCA,FxHENN23HPCA,FAST25FPGA,MAD23,BASALISC22,SoKFHEAccel24,CiMHE19,wang2026fheins,barik2026long,hexl,cuFHE,heaan,bpntt,dramaton,park2023ntt,ozcan2023homomorphic,wang2024chameleon,xiang2023fast,jayashankar2025cinnamon,sun2025sliding,Trinity24MICRO,Alchemist24DAC,WarpDrive25HPCA,Neo25ISCA,HAWK25MICRO,Falcon26ASPLOS,MNEMOS26ISCA,HierarchicalGPUFHE26ISCA,NTPFPGA26HPCA,LPHenn25Cybersec,FHEMultiChiplet25CASES,Libra26Security,UnderstandingGPUFHE25Cybersec,TensorFHEPlus25TC,BitwiseNTT26TCAD,BarrettMult25TCAD,FlexModMult26ISCAS,AllMod25DAC,BleakFutureStorageFHE25APPT}
\providecommand{\citeHEACCEL}{\expandafter\cite\expandafter{\keysHEACCEL}}

\providecommand{\keysPETOTHER}{ccetin2016blind,ziegeldorf2017bloom,kales2019mobile,ion2017private,pu2019two,giraud2017practical,poh2017searchable,sedghi2010searching,wei2022secure,verma2019hybrid,moataz2018substring,grubbs2017leakage,faber2015rich,leontiadis2018storage,chase2015substring,guo2024gridse,goldreich2019play,goldreich1987towards,song2000practical,chen2018labeled,chor1998private,yaogenerate,essex2019secure,shahandashti2012private,yao,mainardi2019privacy,hahn2018practical,kerschbaum2015frequency,snyder2014yao,moataz2013boolean,tan2021cryptgpu,damgaard2012multiparty,de2013secure,hazay2010computationally,baron20135pm,zarezadeh2020efficient,pinkas2019efficient,chandran2022circuit,chen2017fast,PipeZK21ISCA,SecureML17,ABY3_18,VIPBench21,wang2009privacy,franklin1996design,melzi2024overview,aziz2022privacy,DSZ15ABY,ABY2_21,MOTION22,Keller20MPSPDZ,BGI15FSS,BGI16FSSImproved,CGRRS19EzPC,CrypTen21,HLHD22Cheetah,BCGI18PCG,GI14VOLE,Groth16SNARK,GWC19PLONK,CBBZ23HyperPlonk,AST24Jolt,STW24Lasso,DP23Binius,BBHR18STARK,GLSTW23Brakedown,AHIV17Ligero,KST22Nova,KS24HyperNova,BBBPWM18Bulletproofs,LWYCFCWC23cuZK,Sppark,Xavier22PipeMSM,DRG24SZKP,DMABKGR25zkSpeed,SHEAGDA24NoCap,Goldilocks22,BabyBear23,Mersenne31_24,Circom23,Noir23,Cairo21,Halo2_22,ZoKrates18,arkworks20,libsnark20,DMNS06DP,Mironov17RDP,Abadi16DPSGD,USCensusDP22,AppleDP17,ErlingssonPK14RAPPOR,McMahan17FedAvg,McMahan18DPLM,Bonawitz17SecAgg,Buterin21Rollups,zkSync20,StarkNet21,Scroll23,ANTACE24,RingORAM21HPCA}
\providecommand{\citePETOTHER}{\expandafter\cite\expandafter{\keysPETOTHER}}

\providecommand{\keysCRYPTO}{fips197,fips180-4,chang2011workload,daemen1999aes,daemen2002data,devadas2022designing,diffie2022new,han-spie-1999,kim-hpca-2018,manavski-spcom-2007,tuyls-springer-2005,canteaut2005symmetric,fuller2003analysis,hellman2002overview,abdullah2017advanced,grabbe2010algorithm,han1996improved,amounas2012ecc,banerjee2020ecc,biryukov-asiacrypt-2009,liestyowati2020public,knudsen2013selected,mohsen2018performance,akleylek2016efficiency,bisheh2021high,fritzmann2019efficient,kipnis1999cryptAnalysis,tan2019ring,xor2,rivest1978method,elgamal1985public,paillier1999public,bosselaers1993comparison,barrett1987implementing,1802.shs,rukhin2001statistical,enc1,wolkerstorfer2002anasic,arnold2004theibm,zoltak2004vmpc,bernstein2005salsa20,fiskiran2005onchip,koc2009about,paul2011rc4stream,taylor2011behind,yang2014163a,liu2015experimental,tehranipoor2015dram_based,yan2015anovel,ye2015opuf,zhang2015highly,anagnostopoulos2016lowtemperature,danger2016pufs,sutar2016dpuf,tehranipoor2016drambased_intrinsic,tehranipoor2016robust,vatajelu2016sttmrambased,xiong2016runtime,zeitouni2016remanence,aysu2017anew,eckert2017drng,liu2017acropuf,pang2017optimization,sutar2017memorybased,tang2017adram,tehranipoor2017investigation,wang2017current,anagnostopoulos2018anoverview,schaller2018decaybased,sutar2018dpuf,talukder2020towards,yue2020drambased_authentication,islam2022signature,li2022cyberradar,li2023fphammer,DR02AESRijndael}
\providecommand{\citeCRYPTO}{\expandafter\cite\expandafter{\keysCRYPTO}}

\providecommand{\keysSECHWSYS}{zaddach2013backdoor,huang2017flashguard,baek2018ssdinsider,hajiabadi2025cassandra,lee2005architecture,lemay2021cryptographic,wang2022eager,chowdhuryy2023dshield,wertenbroek2024pandora,ahn2020sgxssd,juffinger2025secret,juffinger2025hmb,reidys2022rssd,shaikh2016attacks,duncan2015overview,moghaddam2013comparative,kumar2019review,gruschka2010attack,bella2022study,li2019leveraging,fathullah2022review,apple2015about,pessl2016drama,fournaris2017exploiting,poddebniak2018attacking,carre2018openssl,bostanci2025revisiting,kanellopoulos2024amplifying,glitch-vu,gu2016leveraging,shapiro.sosp99,shapiro.usenixatc02,bosman.2016,standaert2009introduction,lofstrom2000icidentification,gutmann2001data,gassend2002silicon,pappu2002physical,govindavajhala2003using,suh2003efficient,lee2004atechnique,lim2005extracting,duc2006cryptopage,tuyls2006readproof,guajardo2007fpga,guajardo2007physical,holcomb2007initial,su2007a16pjbit,suh2007physical,safari2008tcgplatform,halderman2008lest,hammouri2008unclonable,kumar2008thebutterfly,maes2008intrinsic,majzoobi2008testing,mcgregor2008braving,ozturk2008physical,ozturk2008towards,halderman2009lest,helinski2009aphysical,holcomb2009powerup,pyo2009dram_as,ruhrmair2009onthe,torrance2009thestateoftheart,huang2010ivec,james2010silicon,maes2010physically,majzoobi2010fpga,muller2010aesse,pabel2010frozencache,vanderleest2010hardware,chhabra2011invmm,lee2011correcting,maiti2011theimpact,simmons2011security,bhargava2012reliability,katzenbeisser2012pufs,wang2012flash,cortez2013adapting,gruhn2013onthe,koeberl2013memristor,rose2013foundations,ruhrmair2013pufmodeling,yin2013design,zheng2013resp,che2014anonvolatile,henson2014memory,herder2014physical,hilgers2014postmortem,iyengar2014dwmpuf,keller2014dynamic,liu2014atrustworthy,xiao2014bitselection,bacha2015authenticache,chen2015exploiting,das2015mram,gao2015memristive,gu2015ultracompact,guan2015protecting,hashemian2015arobust,lindenlauf2015cold,burleson2016invited,costan2016intel,irazoqui2016mascat,rahmati2016probable,yitbarek2017cold,bu2018srasa,kocher2018spectre,lipp2018meltdown,saileshwar2018morphable,saileshwar2018synergy,schilling2018pointing,west2018dynamic,cheng2019cattmew,seol2019amnesiac,vila2019theory,villanuevapolanco2019cold,ainsworth2020theguardian,hastings2020anew,arikan2022processor,cai2022onthe,zheng2022trojvit,witchel2005mondrix,hajyahya2021ichannels,ajorpaz2022evax,KPW16AMDSEV,PinSeshia19TrustZone,IntelTDX23,AMDSEVSNP20,ARMCCA21,NVIDIAH100CC23,AWSNitroEnclaves20,Foreshadow18,SGXStepBulck17,Shaikh11CloudAttacks,Duncan12Insider}
\providecommand{\citeSECHWSYS}{\expandafter\cite\expandafter{\keysSECHWSYS}}

\providecommand{\keysSTANDARDSWEB}{nvmenamespaces,aws_ec2_instances,azure_lsv_vms,nvme_base_spec,snai_sdc2021,cortex_r4,daisyplus_openssd,onfi,snia_cs_architecture,nvme_computational_programs,Oracle,adatasu630,arm-cortexR5,arm-cortexR8,armssdcontroller,clang,ddr4sheet,guide2016intel,inteloptane,intelp4610,intelqlc,intels4510,jedec2021lpddr4,llvm,m-profile,marvellsc5bravera,microchip-16-channel-controller,mobile-tensorflow,mqsim-github,noauthor_hexopsfastfilter_2024,nvme,pcie-2017,redis-bitmaps,samsung-980pro,samsung2017znand,samsungmlc,sata,standard-jedec-2020,xeon5118,tcg_key_per_io,nvme_nvm_command_set,IntelCor20,GeForceR28,jedecddr6,NVIDIAMa14,intelskylake,980PROCo56,cernea-uspatent-2006,micron-flyer-2016,leong-uspatent-2008,macronix-technote-2013,micron-datasheet-2009,samsung-datasheet-2009,toshiba-datasheet-2012,onfi-2020,jedec-electrically-2009,jedec-stress-2010,upmem,gf22,sata_2,NVMe2021,nvme_2,samsung-ssd,i5-8250U,upmemsdk,AMDCPU,cho-patent-2022,weka_gpu,runai_gpu,mkl,cublas,titanv,xeon-e3-1225,xeon-4215,xeon-4110,jedec2012ddr4,cudasamples,comm-,avx2011intel,guide2011intel,seal2001arm,volta,hopper,mcginnis2017pci,upmem-guide,upmem-sdk,upmem2018,jedec2020ddr5,lee2014green,micron2016ddr4,drammer-github,drammer-app-github,CortexA57,linarogem5,filippo2012technology,CortexR4,wideio,wideio2,jedec.hbm.spec,hmc.spec.1.1,hmc.spec.2.0,xeonw3550,SAP,MemSQL,sanfilippo2009redis,whitedb,safari.github,impica.github,chrome,rowhammer.github,samsungssd,upmemtechpaper2022,upmemproductsheet2022,criteo1tb2014,upmem_sdk,LIBSVMData2023,zenodo-artifact,pimopt-artifact-github,amdepyc2019,isvlsi2022.pim,isvlsi2022,PIM.Spring2022.upmem1.lecture,pimtutorialhpca,hipeacinterview24,hipeacinterview18,pimtutorialasplos,pimtutorialisca,pimtutorialmicro,pimtutorialmicro24,pimtutorialppopp25,pimworkshop,pimtutorialisca24,pimtutorialheart24,PIM.Spring2022.upmem2.lecture,PIM.Spring2022.fimdram.lecture,PIM.Spring2022.aim.lecture,PIM.Spring2022.axdimm.lecture,PIM.Spring2022.hbpnm.lecture,youtube,axeleraai,skype,google-hangouts,giannoula2022repo,diab2023repo,openCAPI,van2019hoti,coughlin2021higher,PIM_fall2021,PIM_spring2022,PIM_spring2023,PIM_fall2022,Seminar_fall2021,Seminar_fall2022,Seminar_fall2023,Seminar_fall2024,CompArch_fall2021,CompArch_fall2022,CompArch_fall2023,CompArch_fall2024,damov2021_repo,olgun2021pidram_repo,gomezluna2021repo_2,intel_gold_cpu,nvidia_2080ti,nvidia_3080ti,nvidia_p100,AlveoU2590,zynq_ultrascale_zcu102,xilinxzcu102,standard2012jesd79,jetson,movidius,gtx745,vp9-specification,ampere_2,edge-tpu,dagum1998openmp,systime,mpi,synopsysdc,mythic,jedec2012ddr3,mkl_2,jedec2012jedec,micron2014ddr4,jedec2017jedec,amatref,lecturedrambender,oooexec,devicethree,coral,rldram,redmon_darknet2013,artracfd,openaccbench,hardwareeffects,libvpx,OpenACCA1:online,dimacs,codexl,nvprof,gprof,vtune_HT,tensorflow-lite,quant-aware-training,HMC2,DiRAM4,tpch,Intel_XeonPhi,arm-a5,TSVpatent,MicronDDR3,safari-drambender,TitanX,i7-6700K,E5-2650,E3-1240,UseAutom2:online,lopes2014getting,sarda2015llvm,avx512doc,writingpass,reddy2008neon,lomont2011introduction,AutoVect65:online,Autovect67:online,IntelManual,lpddr4,edge-tpu-compiler,manual2010arm,adexelecddr4sodv1,advancedvalidationlabsavltesting,alsopdram_manufacturers,amazon,amds,arduino,bapco,bdkg,bee2,beer,bittware_xupvvh,bittware_xusp3s,choworkshop,consortiumhybrid,core,corpltspice,corsaircmk16gx4m2b3200c16,corsairskucmv4gx4m1a2133c15,crucialct4g4dfs824a,crucialct4g4dfs8266,demouraz3py,dram_datasheet,hynixh5an8g8ndjrxnc,hynixh5anag8najrxn,hynixh5anag8ncjrxn,intel3rdgen,amdepyc,amdsecure,micronddr4partnumbering,xilinxultrascale,xilinxu200,xilinxu50,intel,issint5ad256m16e4jr,jedecjc42,jujetelectronicscojet5612a,lapedusdram_scaling,maxwellft20x,memoryadvanced,memorynethmaa4gu6ajr8nxn,memorynethmaa4gu7cjr8nxn,memtest86,memtest86+,micronddr4mta18asf2g72pz,micronmt40a1g16kd062e,micronmt40a1g16rc062eb,micronmt40a2g4we083eb,micronmt40a4g4jc062ee,micronmta18asf2g72pz2g3b1qk,micronmta36asf8g72pz2g9e1ti,micronmta4atf1g64hz3g2b2,micronmta4atf1g64hz3g2e1,microsoft,mscproducts,rdma,samsungk4a4g085wfbctd,samsungk4a8g045wcbctd,samsungk4a8g085wbbctd,samsungm393a1k43bb1ctd,samsungm393a2k40cb2ctd,samsungm471a5143eb0cpb,samsungsamsung,semiconductorsqoriq,sunutility,technologies16mb,technologies34465a,thegnu,theultrasparc$^tm$,thinkpad,timesten,transaction,ttiplplp,volano,welcome,wikichipcascade,wikichipcore,wolframresearchwolframalpha,ziff,intel19701103,mostek1971mk4006,mostek1973mk4096,mostek1973mk4096kn61611,corporation1975intel,mostek1976mk4027,mostek1976mk4116p,fairchild1978f4116,intel19792118,mitsubishi1981m5k4164anp,mostek1982mk4564,instruments1984tms4416,nte1984nte4164,mitsubishi1985m5m4256p,instruments1987tms4464,samsung1987km41464a,instruments1989smj4464,intel199021256,mitsubishi1991m5m44100,sites1992alpha,1993pentium,mitsubishi1993m5m44256b,vitelic1993v404j8,jedec1994synchronous,mesh-transformer-jax,mipstechnologies1994r10000,jedec1995eiajesd511,price1995mips,siemens1995hyb511000bj,1997digital_semiconductor,1997digital_unix,1997mpc750,corporation1997cost,gwennap1997klamath,hyundai1997hy5118164b,samsung1997km48s2020ct,diefendorff1998jalapeno,diefendorff1998k7challenges,diefendorff1998xeon,gwennap1998alpha,henry1998winchip,intel1998pcsdram,meyer1998amdk7$^tm$,micron1998mt48lc4m4a1,pollack1998description,siemens1998hyb39s16400,vitelic1998v53c808h,micron1999tn0430,samsung1999k4s160822d,samsung1999k4s561632a,amic2000a416316b,rambus2000direct,issi2000is41lv32256,samsung2000k4d623237a5,yu2000client,200119th,2001ia32,alliance2001as4c256k16e0,alliance2001as4c4m4foq,issi2001is41c4400x,jedec2001arrhenius,micron2001mt46v2m32,micron2001mt46v32m4,tech2001t224160b,vitelic2001v58c265164s,micron2002tn0008,2003amdathlon$^tm$,micron2003mt46v64m4,samsung2003k4s560432e,semiconductor2004soft,amic2005a43l0632,elite2006m52s32321a,issi2006is42s81600b,sasmung2006k4t51043qc,20071gbddr2,micron2007mt47h256m4,isaac2008theremarkable,jedec2008ddr3,jedec2008jesd79f,2009rdma,amd2009bkdg,corporation2009intel,cavicchi2010patent,commission2010competition,flamm2010theimpact,jedec2010jesd218,jedec2010jesd219,kang2010astudy,samsung2010k4b4g0446a,hewlettpackarddevelopmentcompany2011whybuy,hynix2011h5tq4g43mfr,nanya2011nt5cb512m,specification2002pci,sata_3,xilinx2011virtex6,2012fcram,2012hybrid,jedec2012jesd793d,association2012jedec,corporation2012kingston,henderson2012ibmpower7,intel2012intel,jedec2012double,jedec2012jep122g,jedec2012lowpower,jedec2012ddr3spd,jedec2012jesd793f,ltd2012corelink,ross2012migrating,xilinx2012ml605,2013top500,amd2013bkdg,dramexchange2013trendforce,howard2013dram_scores,hynix2013ddr3l,intel20132ndgeneration,intel20134thgeneration,jedec2013standard,lee2013astrategic,samsung2013m471b5173qh0,xilinx2013virtex6,consortium2014hybrid,micron2014tn4003,micron2014using,instruments2014rs485,jedec2014lowpower,memory2014reliability,technology2014ddr4,corporation2015mobile,enterprise2015hpmoonshot,field2015microsoft,hynix2015366ball,jedec2015annex,micron2015mt40a2g,semiconductors2015itrs,software2015memtest86,synopsys2015reliability,wang2015backward,2016ct8g4sfs8,goodin2016cuttingedge,google2016google,greenberg2016forget,hynix2016h5an4g4nafr,jedec2016graphics,jedec2016jep122h,kaplan2016amdmemory,nasa2016nasa,nvidia2016p100,samsung2016k4a8g165wc,2017aptget,2017toppicks,adata2017adata,jedec2017jesd794b,crucial2017mt41k512m8da107p,hynix2017h5an8g4nafr,iso2017iso148822017,issi2017quality,jedec2017jesd2094b,nvidia2017nvidia,openssh2017openssh,samsung2017288pin,technologies2017smart,zhao2017apparatuses,2018128mx8,2018ddr5,jedec2024jesd795c,2018sentaurus,2018terminology,amd2018amdopteron,boost2018boost,electronics2018memory,electronics2018mobile,electronics2018stringent,micron2018mt40a,microntechnology2018mobile,newman2018aningenious,s2018every,samsung2018260pin,samsung2018288pin,samsungelectronicsco2018288pin,zawodny2018apparatuses,corporation2019intel,hynix2019h5an8g4ncjr,makosiej2019cammemory,markoff2019ibms,research2019global,rooney2019micron,samsung2019k4aag045wa,20202gb4gb8gb,corporation2020intel,corporation2020platform,hynix202016gb,micron20208gbx4,micron2020tn4040,intel2020intel,issi20208gbx16,jedec2020jesd2095a,jedec2020jesd235b,kingston2020ksm32rd816hdr,s2020fabriation,samsung2020euvthe,samsung2020k4aag085wa,technology20202020,xilinx2020vitis,xilinx2020vivado,2021dram_the,2021rldram,2021rugged,corporation2021kingstons,gskill2021f42400c17s8gnt,hynix2021hmcg66meb,xilinx2021adaptable,xilinx2021fpgas,jedec2021graphics,jedec2021jesd235d,kingston2021kvr24n17s88,micron2021mt60b1g16,samsung2021k4a4g085wfbctd,2022intel,2022quarterly,advantest2022t5833t5833es,corporation2022intel,corporation2022radhard,safari2022placeholder,infineon2022radiation,jedec2022high,jedec2022jedec,mathworks2022textttimbinarize,statista2022dram_manufacturers,wang2022dueto,xilinx2022axihigh,alsop2023dram_manufacturers,jedec2016jesd232a,jedec2021jesd250c,datasheetlpdddr5x,IntelXe74:online}
\providecommand{\citeSTANDARDSWEB}{\expandafter\cite\expandafter{\keysSTANDARDSWEB}}

\providecommand{\keysOTHER}{wolski2008using,farhat1991lagrange,cook2023complexity,chen2017psi,paterson1973number,harrison1964classification,jain1997identity,smith2010fast,schick2016fast,yanowitz1989new,chen2013face,organick2020probing,goldman2013towards,gonzales2002digital,arrhenius-zpc-1889,bruce-yuv-2000,su2012linear,larson2008analysis,davidian1987variance,jacquier2003geometric,zhao2019approximating,eyupoglu2015performance,sontag1985real,zippel1993effective,moenck1976practical,harvey2017faster,chan2010optimal,harvey2022polynomial,bookstein2002generalized,antonopoulos2010cloud,voorsluys2011introduction,furht2010handbook,xiao2022real,weisstein2004taylor,hosmer2013applied,yan2009linear,freedman2009statistical,polyak1987,boyd2004convex,cayley1858ii,knuth1971optimum,hungerford1997abstract,vanhateren1998,kuhn2012structure,dean2013tail,jones.gcbook1996,wilson1992uniprocessor,waldvogel1997,buell2013methodology,btt-knuth,hacker-delight,weissteinFrobenius,manuel2008efficient,lattigo_2,amato2016yfcc100m,yfcc100m,aysal2009broadcast,boyd2005gossip,carli2010gossip,cooley1965algorithm,walther1971unified,volder1959cordic,churin1988camac,Abril07,naylor1990,chisholm2014adopting,alaee2021time,berndt1994using,esling2012time,poletto1999,al2011artificial,mccool2012structured,johnson1973near,knowlton1966programmer,John98,mo2019mesoscale,firstfit,ho2017efficient,delimitrou2018amdahl,CHO1998161,garey1972worst,azmat2015self,kessler2015elon,harris2015documents,burns2013sustainable,tarjan1976edge,cormen2022introduction,yan2015characterizing,pooch1973survey,bjorck2024numerical,hartigan1979algorithm,carnegie,collaborative,craconference,craw,etude,grand,research,surf,surg,winece,wright1921correlation,bhattacharyya1943onameasure,hamming1950error,smith1956various,bellman1957dynamic,hocquenghem1959codes,bose1960onaclass,reed1960polynomial,berger1961anote,gallager1963lowdensity,forney1965concatenated,knowlton1965afast,coffman1973operating,kleinrock1976queuing,macwilliams1977thetheory,tukey1977exploratory,morris1978counting,carter1979universal,garey1979computers,bird1980tabulation,james1980monte,taylor1980redundancy,klove1981error,boyer1982afast,costello1982error,fischer1982finding,misra1982finding,rousseeuw1987silhouettes,williamson1989thebox,baker1990opening,efron1992bootstrap,kitchen1992game,flamm1993measurement,cover1994elements,lamport1994latex,waldspurger1994lottery,wicker1995error,inacio1996thedsp,biran1998thecambridge,matocha1998extended,ch2000surplus,ch2001deadline,jones2001scipy,oconnell2002soul,huffman2003fundamentals,jha2003testing,rai2003analysis,moon2005error,qin2005rxtreating,2006integrative,nyirongo2006concatenated,roth2006introduction,mansour2007efficient,mcneely2007lowpower,richardson2008modern,aven2009identification,mutlu2009modernizing,navi2009twonew,aalst2010process,baek2010green,ieee2010isoiecieee,sun2010apipelined,wang2010design,hovinen2011getting,lastname12011yetanother,wilcox2011mesa,wilcox2011tool,faber2012statistics,clarkjr2013errorcorrection,alkaldy2014anovel,lastname12015another,lastname12016avery,gong2017asurvey,hofmann2017lettervalue,kobayashi2017highlyreliable,suresh2017compiletime,besnard2019aframework,pinto2021amethodology,archives2022summary,apple2022final,silberschatz2018operating,kirchhoff1859uber,kirchhoff1978verhaltnis,kirchhoff1860relation,mishchenko2006dag}
\providecommand{\citeOTHER}{\expandafter\cite\expandafter{\keysOTHER}}

\providecommand{\keysBFV}{Kabra2025CipherMatch,brakerski2012fully,mouchet2020lattigo,huo2019note,chen2018homomorphic,ozcan2024heongpu,chen2019efficient,gupta2023evaluating,reis2020computing,fan2012somewhat,al2020multi,mert2019design,syafalni2022efficient,jiang2022fhebench,sealcrypto,halevi2019improved,wibawa2022bfv,kim2021revisiting,bajard2016full,roy2018hepcloud,roy2019fpga,turan2020heaws,polyakov2017palisade,juvekar2018gazelle,iliashenko2021faster,boura2020chimera,chen2017fast,OpenFHE22,SEAL,ChenLainePlayer17SEAL,Lattigo,TenSEAL21,IbarrondoViand21Pyfhel,PhantomFHE23,PALISADE,HEIR,Cingulata15,Porcupine21PLDI,Shen23CARM,Badawi18GPUFV,TuranRoy20HEAWS,lattigo_2,hexl,ozcan2023homomorphic,GV23BGVBootRevisited,GIKV23PolyModBoot,RelaxedFuncBootBGV24,SlotCoeffBGVBFV24,KSS24BFVBootCKKS,BP23DemystifyBoot}
\providecommand{\citeBFV}{\expandafter\cite\expandafter{\keysBFV}}

\providecommand{\keysCKKS}{lu2021pegasus,mouchet2020lattigo,ozcan2024heongpu,chen2019efficient,ebel2025orion,cheon2017homomorphic,ding2024apache,fan2025fast,kim2022bts,agrawal2023fab,de2021does,samardzic2022craterlake,feldmann2021f1,jiang2022fhebench,jung2021over,sealcrypto,kim2022ark,riaziheax,kim2023sharp,lee2021high,kim2022approximate,cheon2020remark,dathathri2019chet,dathathri2020eva,park2025fhendi,kim2025anaheim,cheon2018bootstrapping,bae2024bootstrapping,drucker2024bleach,boura2020chimera,HS20HElibDesign,CHKKS18FullRNS,CCS19ImprovedBootstrap,OpenFHE22,SEAL,HElibRepo,Lattigo,TenSEAL21,IbarrondoViand21Pyfhel,PhantomFHE23,PALISADE,HEIR,nGraphHE19,nGraphHE2_19,Fhelipe24PLDI,LeeMPCNN22,CryptoGCN22,SavPoseidon21,Bossuat21EUROCRYPT,Jung21CKKS100x,TensorFHE23HPCA,Shen23CARM,HEAX20,Poseidon23HPCA,FAST25FPGA,MAD23,HEProfiler24,alexandru2025general,li2024evalcomp,cryptoeprint:2024/136,lattigo_2,hexl,heaan,han2020better,wang2024chameleon,xiang2023fast,jayashankar2025cinnamon,Lee22PPMLCACM,Cheon19ChebyshevACrypt,Bae22MetaBTS,cheon2024dacapo,Kim22EvalRound,BKSS24DiscreteCKKS,Trinity24MICRO,Alchemist24DAC,WarpDrive25HPCA,Neo25ISCA,FHEFusion26CGO,TensorFHEPlus25TC,WPC25CCS,KSS24BFVBootCKKS,HHC19ImprovedDFTBoot,LLKK22HighPrecErrVar,BTP22BootNegligibleFail,CHKS25SHIP,OverModRaise25CIC,BP23DemystifyBoot,NeuJeans23CRYPTO,HERMES23CRYPTO,Grafting25CCS,BootSurveyCybersec25}
\providecommand{\citeCKKS}{\expandafter\cite\expandafter{\keysCKKS}}

\providecommand{\keysTFHE}{lu2021pegasus,bon2024optimized,TFHE-rs,aziz2024secure,petrean2024homomorphic,ding2024apache,nam2025affinity,jiang2022fhebench,chillotti2016faster,chillotti2021improved,chillotti2021programmable,chillotti2017improving,morshed2020cpu,pradel2021privacy,gupta2024memfhe,chillotti2020concrete,chillotti2017faster,chillotti2020tfhe,stoian2023deep,sanyal2018tapas,bourse2018fast,ducasfhew,micciancio2018ring,liu2023amortized,boura2020chimera,CGGI20TFHE,MP21BootstrappingFHEW,OpenFHE22,TFHErs,PALISADE,HEIR,ConcreteML,NuFHE,cuFHE,wang2024chameleon,singh2025efficient,Trinity24MICRO,Alchemist24DAC,MNEMOS26ISCA,BP23DemystifyBoot,LMP22LargePrecSign,LMK23EfficientFHEWBoot,KLUCZNIAK23FDFB,DKMS23FasterFHEWAutomorph,NTRU4ms24TCHES,BlockBinaryTFHE23ASIACCS,TFHEAutomorphBoot25ASIACRYPT,JW22LibTFHE,XZDDF23FastBlindRot}
\providecommand{\citeTFHE}{\expandafter\cite\expandafter{\keysTFHE}}

\providecommand{\keysENCDB}{bian2023he3db,zhang2024arcedb,ren2022heda,bian2025engorgio,jung2026nshedb,boneh2013private,kim2016efficiency,kim2017private,cheon2015search,TawoseHE23,kim2021efficient,ren2016secure,dawoud2014privacy,raj2020web}
\providecommand{\citeENCDB}{\expandafter\cite\expandafter{\keysENCDB}}

\providecommand{\keysENCGENOMICS}{kim2015private,cruz2017secure,titus2018sig,park2019secure,bed2016privately,aziz2024secure,celsystems2021he,KimEtAl21Genotype,de2013secure,ziegeldorf2017bloom,wood2020homomorphic,aziz2022privacy}
\providecommand{\citeENCGENOMICS}{\expandafter\cite\expandafter{\keysENCGENOMICS}}

\providecommand{\keysENCGRAPH}{CryptoGCN22}
\providecommand{\citeENCGRAPH}{\expandafter\cite\expandafter{\keysENCGRAPH}}

\makeatletter
\providecommand*\bibalias[2]{\@namedef{bibalias@@#1}{#2}}
\def\bibalias@translate#1{%
  \let\bibalias@out\@empty
  \@for\bibalias@k:=#1\do{%
    \edef\bibalias@k{\expandafter\@firstofone\bibalias@k\@empty}%
    \@ifundefined{bibalias@@\bibalias@k}{}{\edef\bibalias@k{\csname bibalias@@\bibalias@k\endcsname}}%
    \ifx\bibalias@out\@empty\let\bibalias@out\bibalias@k
    \else\edef\bibalias@out{\unexpanded\expandafter{\bibalias@out},\unexpanded\expandafter{\bibalias@k}}\fi}}
\def\bibalias@call#1#2#3#4{%
  \edef\bibalias@tmp{\noexpand\bibalias@callB\unexpanded{{#1}{#2}{#3}{#4}}{\unexpanded\expandafter{\bibalias@out}}}%
  \bibalias@tmp}
\def\bibalias@callB#1#2#3#4#5{%
  \IfBooleanTF{#2}%
    {\IfValueTF{#4}{\csname bibalias@orig@#1\endcsname*[#3][#4]{#5}}%
      {\IfValueTF{#3}{\csname bibalias@orig@#1\endcsname*[#3]{#5}}{\csname bibalias@orig@#1\endcsname*{#5}}}}%
    {\IfValueTF{#4}{\csname bibalias@orig@#1\endcsname[#3][#4]{#5}}%
      {\IfValueTF{#3}{\csname bibalias@orig@#1\endcsname[#3]{#5}}{\csname bibalias@orig@#1\endcsname{#5}}}}}
\def\bibalias@wrap#1{%
  \@ifundefined{#1}{}{%
    \expandafter\NewCommandCopy\csname bibalias@orig@#1\expandafter\endcsname\csname #1\endcsname
    \expandafter\RenewDocumentCommand\csname #1\endcsname{s o o m}%
      {\bibalias@translate{##4}\bibalias@call{#1}{##1}{##2}{##3}}}}
\AtBeginDocument{%
  \bibalias@wrap{cite}\bibalias@wrap{nocite}\bibalias@wrap{citet}\bibalias@wrap{citep}%
  \bibalias@wrap{citealp}\bibalias@wrap{citealt}\bibalias@wrap{citeauthor}\bibalias@wrap{citeyear}}
\makeatother

\bibalias{10.1093/bioinformatics/btae631}{alonso2024bimsa}
\bibalias{10158230}{item2023transpimlibispass}
\bibalias{1997thenational}{1994thenational}
\bibalias{2021terminology}{2018terminology}
\bibalias{AAUC18Survey}{acar2018survey}
\bibalias{acharya.1998}{acharya-asplos-1998}
\bibalias{aga.hpca17}{aga-hpca-2017}
\bibalias{aga2017}{aga-hpca-2017}
\bibalias{aga2017compute}{aga-hpca-2017}
\bibalias{ahn-isca-2015}{ahn2015scalable}
\bibalias{ahn-isca-2015-2}{ahn2015pim}
\bibalias{ahn.pei.isca15}{ahn2015pim}
\bibalias{ahn.tesseract.isca15}{ahn2015scalable}
\bibalias{ahn.tesseract.isca15talk}{ahn2015scalable}
\bibalias{aho1987compilers}{aho1986compilers}
\bibalias{ajdari2019cidr}{ajdari-hpca-2019}
\bibalias{akin2015data}{akin2015}
\bibalias{AKP25GenFunctionalBoot}{alexandru2025general}
\bibalias{al2022openfhe}{OpenFHE22}
\bibalias{AlAziz24SecureGenomicSearch}{aziz2024secure}
\bibalias{ali2019memory}{ali-tcas-2019}
\bibalias{alser.bioinformatics17}{alser-bioinformatics-2017}
\bibalias{amdahl1967validity}{amdahl1976validity}
\bibalias{amiraliphd}{boroumand2020practical}
\bibalias{angizi2019graphide}{angizi-glsvlsi-2019}
\bibalias{angstadt2018aspen}{angstadt-micro-2018}
\bibalias{ArcEDB24}{zhang2024arcedb}
\bibalias{AresFlash24}{Chen_MICRO2024}
\bibalias{ARK22MICRO}{kim2022ark}
\bibalias{asgari2021fafnir}{asgarifafnir}
\bibalias{asghari-moghaddam.micro16}{asghari2016chameleon}
\bibalias{augusta-sigmod-2015}{augusta2015jafar}
\bibalias{azarkhish2018neurostream}{azarkhish2017neurostream}
\bibalias{bae2013intelligent}{bae-cikm-2013}
\bibalias{baek2014refresh}{baek2013refresh}
\bibalias{balasubramonian2014near}{balasubramonian-micro-2014}
\bibalias{Barik26LongNTT}{barik2026long}
\bibalias{BCKSS24BootBits}{bae2024bootstrapping}
\bibalias{bekerman2000early}{bekerman1999early}
\bibalias{Bella22CloudThreats}{bella2022study}
\bibalias{bensoussan.cacm72}{bensoussan1972multics}
\bibalias{besta-micro-2021}{besta2021sisa}
\bibalias{besta2021sisa_micro}{besta2021sisa}
\bibalias{besta2021sisaarxiv}{besta2021sisa}
\bibalias{BGGJ20CHIMERA}{boura2020chimera}
\bibalias{BI20HEStringSearch}{bonte2020homomorphic}
\bibalias{bitfunnel}{goodwin-SIGIR-2017}
\bibalias{bitweaving}{li-sigmod-2013}
\bibalias{Bloom:1970:STH:362686.362692}{bloom1970space}
\bibalias{bmidc}{oneil-ideas-2007}
\bibalias{bmide}{chan1998bitmap}
\bibalias{bon2023optimized}{bon2024optimized}
\bibalias{boroumand.asplos18}{boroumand2018google}
\bibalias{boroumand.asplos18talk}{boroumand2018google}
\bibalias{boroumand2016pim}{boroumand-ieeecal-2017}
\bibalias{boroumand2018}{boroumand2018google}
\bibalias{BP23DemystifyBootTutorial}{BP23DemystifyBoot}
\bibalias{Brakerski12ScaleInv}{brakerski2012fully}
\bibalias{brakerski2014efficient}{BV11LWEFHE}
\bibalias{BS11DNAStringMatching}{bhukya2011exact}
\bibalias{BTS22ISCA}{kim2022bts}
\bibalias{burleson2016whois}{burleson2016invited}
\bibalias{cai-hpca-2015}{cai2015data}
\bibalias{cai-iccd-2013}{cai2013program}
\bibalias{cai-procieee-2017}{cai2017error}
\bibalias{cai.procieee17}{cai2017error}
\bibalias{cai2012error}{cai-date-2012}
\bibalias{cai2012flash}{cai-iccd-2012}
\bibalias{cai2013error}{cai-inteltechj-2013}
\bibalias{cai2014neighbor}{cai-sigmetrics-2014}
\bibalias{cai2015read}{cai-dsn-2015}
\bibalias{cai2017vulnerabilities}{cai-hpca-2017}
\bibalias{cali2020genasm}{cali-micro-2020}
\bibalias{CD16IntelSGX}{costan2016intel}
\bibalias{CGGI16FasterBootstrap}{chillotti2016faster}
\bibalias{CGGI17CircuitBoot}{chillotti2017faster}
\bibalias{CH18HomLowerDigits}{chen2018homomorphic}
\bibalias{chan-signmod-1998}{chan1998bitmap}
\bibalias{chang.hpca16}{chang-hpca-2016}
\bibalias{chang.sigmetrics16talk}{chang.sigmetrics16}
\bibalias{chang2016low}{chang-hpca-2016}
\bibalias{chang2017thesis}{chang.phdthesis2017}
\bibalias{chang2017understanding}{chang.phdthesis2017}
\bibalias{che_iiswc2009}{che2009rodinia}
\bibalias{Chen13FaceRecognition}{chen2013face}
\bibalias{Chen2017}{eyeriss}
\bibalias{chen2024ares}{Chen_MICRO2024}
\bibalias{ChenLaineRindal17PSI}{chen2017fast}
\bibalias{CHET19PLDI}{dathathri2019chet}
\bibalias{chi2016prime}{ping-isca-2016}
\bibalias{ChillottiCGGI17Packed}{chillotti2017faster}
\bibalias{CHKKS18Bootstrap}{cheon2018bootstrapping}
\bibalias{cho2013xsd}{cho-wondp-2013}
\bibalias{choi-iscas-2020}{choi2020flash}
\bibalias{choquette2021nvidia}{ampere}
\bibalias{chun-hotstorage-2022}{chun2022pif}
\bibalias{CJP21PBSInferenceDNN}{chillotti2021programmable}
\bibalias{ckks}{cheon2017homomorphic}
\bibalias{CKKS17}{cheon2017homomorphic}
\bibalias{ckksbootstrapping}{cheon2018bootstrapping}
\bibalias{ckksparams}{Bossuat21EUROCRYPT}
\bibalias{CLOT21ImprovedPBS}{chillotti2021improved}
\bibalias{Concrete20}{chillotti2020concrete}
\bibalias{cont-runahead}{hashemi2016continuous}
\bibalias{CraterLake22ISCA}{samardzic2022craterlake}
\bibalias{cryptoeprint:2014/106}{cryptoeprint:2024/136}
\bibalias{cryptoeprint:2020/931}{bonte2020homomorphic}
\bibalias{CryptoPIM20}{nejatollahi2020cryptopim}
\bibalias{dai-tcad-2019}{dai2018graphh}
\bibalias{DaiSunar15cuHE}{dai2016cuhe}
\bibalias{danisch-procwwwc-2018}{danisch2018listing}
\bibalias{datasheetCMK16GX4M2B3200C16}{corsaircmk16gx4m2b3200c16}
\bibalias{datasheetCMV4GX4M1A2133C15}{corsairskucmv4gx4m1a2133c15}
\bibalias{datasheetCT4G4DFS824A}{crucialct4g4dfs824a}
\bibalias{datasheetCT4G4DFS8266}{crucialct4g4dfs8266}
\bibalias{datasheetH5AN8G8NDJR-XNC}{hynixh5an8g8ndjrxnc}
\bibalias{datasheetH5ANAG8NCJR-XN}{hynixh5anag8ncjrxn}
\bibalias{datasheetHMAA4GU6AJR8N-XN}{memorynethmaa4gu6ajr8nxn}
\bibalias{datasheetHMAA4GU7CJR8N-XN}{memorynethmaa4gu7cjr8nxn}
\bibalias{datasheetK4A4G085WF-BCTD}{samsungk4a4g085wfbctd}
\bibalias{datasheetK4A8G045WC-BCTD}{samsungk4a8g045wcbctd}
\bibalias{datasheetK4A8G085WB-BCTD}{samsungk4a8g085wbbctd}
\bibalias{datasheetKSM32RD8}{kingston2020ksm32rd816hdr}
\bibalias{datasheetKSM32RD8/16HDR}{kingston2020ksm32rd816hdr}
\bibalias{datasheetM378A1K43DB2}{samsung2018288pin}
\bibalias{datasheetM393A1K43BB1}{samsung2017288pin}
\bibalias{datasheetM393A1K43BB1-CTD}{samsungm393a1k43bb1ctd}
\bibalias{datasheetM393A2K40CB2-CTD}{samsungm393a2k40cb2ctd}
\bibalias{datasheetM471A5143EB0}{samsungm471a5143eb0cpb}
\bibalias{datasheetM471A5244CB0}{samsung2018260pin}
\bibalias{datasheetMT40A1G16KD-062E}{micronmt40a1g16kd062e}
\bibalias{datasheetMT40A1G16RC-062E:B}{micronmt40a1g16rc062eb}
\bibalias{datasheetMT40A2G4WE-083E:B}{micronmt40a2g4we083eb}
\bibalias{datasheetMT40A4G4JC-062E:E}{micronmt40a4g4jc062ee}
\bibalias{datasheetMTA18ASF2G72PZ}{micronddr4mta18asf2g72pz}
\bibalias{datasheetMTA18ASF2G72PZ-2G3B1QK}{micronmta18asf2g72pz2g3b1qk}
\bibalias{datasheetMTA36ASF8G72PZ-2G9E1TI}{micronmta36asf8g72pz2g9e1ti}
\bibalias{datasheetMTA4ATF1G64HZ-3G2B2}{micronmta4atf1g64hz3g2b2}
\bibalias{datasheetMTA4ATF1G64HZ-3G2E1}{micronmta4atf1g64hz3g2e1}
\bibalias{davis2001modern}{davis2000modern}
\bibalias{DBLP:conf/hpca/GaoK16}{gao2016hrl}
\bibalias{DBLP:conf/IEEEpact/LeeSK15}{bssync}
\bibalias{DBLP:conf/isca/AkinFH15}{akin2015}
\bibalias{DBLP:conf/sigmod/BabarinsaI15}{augusta2015jafar}
\bibalias{DCFT13SecureGenomic}{de2013secure}
\bibalias{de2019automatic}{degonzalo2019automatic}
\bibalias{deng.dac2018}{deng-dac-2018}
\bibalias{deng2018dracc}{deng-dac-2018}
\bibalias{denzler2023casper}{denzler2021casper}
\bibalias{deoliveira2021old}{oliveira2021damov}
\bibalias{devaux2019}{devaux2019true}
\bibalias{devlin2019bert}{devlin2018bert}
\bibalias{DGHV10}{van2010fully}
\bibalias{diab2022framework}{diab2023aim}
\bibalias{diab2022high}{diab2022hicomb}
\bibalias{diab2023framework}{diab2023aim}
\bibalias{DM15FHEW}{ducasfhew}
\bibalias{DMPS24BLEACH}{drucker2024bleach}
\bibalias{do2013query}{do-sigmod-2013}
\bibalias{DPSZ12SPDZ}{damgaard2012multiparty}
\bibalias{DRAMA_CAL_2014}{farmahini2014drama}
\bibalias{Draper:2002:ADP:514191.514197}{draper2002architecture}
\bibalias{drumond2017mondrian}{drumond-isca-2017}
\bibalias{ducas2015fhew}{ducasfhew}
\bibalias{dusser2009zerocontent}{dusser2009zero}
\bibalias{eckert2018neural}{eckert-isca-2018}
\bibalias{EGR25Orion}{ebel2025orion}
\bibalias{elliott.dt99}{elliott1999computational}
\bibalias{EVA20PLDI}{dathathri2020eva}
\bibalias{F121MICRO}{feldmann2021f1}
\bibalias{FAB23HPCA}{agrawal2023fab}
\bibalias{farmahini2015nda}{farmahini-hpca-2015}
\bibalias{FASTable25ISCA}{fan2025fast}
\bibalias{FastBitA9}{fastbit}
\bibalias{ferreira2022pluto}{ferreira2021pluto}
\bibalias{Firtina23RawHash}{firtina2023rawhash}
\bibalias{flashcosmos}{Park_MICRO2022}
\bibalias{FlashCosmos22}{Park_MICRO2022}
\bibalias{friedman2001elements}{hastie2009elements}
\bibalias{fujiki2018memory}{fujiki-asplos-2018}
\bibalias{fujiki2019duality}{fujiki-isca-2019}
\bibalias{FV12BFV}{fan2012somewhat}
\bibalias{gao-hpca-2016}{gao2016hrl}
\bibalias{gao-micro-2021}{Gao_MICRO2021}
\bibalias{gao2015practical}{gao.pact15}
\bibalias{gao2019computedram}{gao-micro-2019}
\bibalias{gao2020computedram}{gao-micro-2019}
\bibalias{gao2021parabit}{Gao_MICRO2021}
\bibalias{gassend2003physical}{gassend2002silicon}
\bibalias{Gazelle18}{juvekar2018gazelle}
\bibalias{Gentry09FHE}{gentry2009fully}
\bibalias{GH11Implementing}{gentry2011implementing}
\bibalias{ghiasi2024megis}{Ghiasi2024MegIS}
\bibalias{ghiasimegis2024}{Ghiasi2024MegIS}
\bibalias{gholami2020ai_and_memory_wall}{gholami2020ai}
\bibalias{ghose.sigmetrics20}{ghose2019demystifying}
\bibalias{ghose2018vampire}{ghose2018your}
\bibalias{ghoseibm2019}{ghose2019processing}
\bibalias{GHS12BetterBoot}{gentry2012better}
\bibalias{GHS12PolylogBoot}{gentry2012fully}
\bibalias{giannoula2021syncron}{giannoula2021.SLS}
\bibalias{giannoula2022sparsep}{giannoula-pomacs-2022}
\bibalias{giannoula2022towards}{giannoula2022sigmetrics}
\bibalias{gomez2021cut}{gomez2021benchmarking}
\bibalias{gomez2022machine}{gomezluna2022isvlsi}
\bibalias{gomez2023evaluatingispass}{gomez2023evaluating}
\bibalias{gomezluna2021benchmarking}{gomezluna2021prim}
\bibalias{gomezluna2021cut}{gomez2021benchmarking}
\bibalias{gomezluna2022ieeeaccess}{gomez2022benchmarking}
\bibalias{gomezluna_icpp2015}{gomezluna2015ds}
\bibalias{gomezluna_ispass2017}{gomezluna2017chai}
\bibalias{Goodwin17BitFunnel}{goodwin-SIGIR-2017}
\bibalias{graphpim}{nai2017graphpim}
\bibalias{gruss2016rowhammerjs}{gruss.2015}
\bibalias{gu-isca-2016}{gu2016biscuit}
\bibalias{gu.isca16}{gu2016biscuit}
\bibalias{guomachine}{gomez2023evaluating}
\bibalias{gupta2019architectural}{gupta2020architectural}
\bibalias{gupta2022memfhe}{gupta2024memfhe}
\bibalias{Gupta23HEPIM}{gupta2023evaluating}
\bibalias{gursoy2022privacy}{celsystems2021he}
\bibalias{guz2014realtime}{guz-ndp-2014}
\bibalias{hajinazar-asplos-2021}{hajinazar2021simdram}
\bibalias{hajinazarsimdram}{hajinazar2021simdram}
\bibalias{halevi2012algorithms}{HS14AlgorithmsHElib}
\bibalias{HalevPolyakovShoup19}{halevi2019improved}
\bibalias{han2019novel}{han-ieeetcas-2019}
\bibalias{han2021flash}{han-aisy-2021}
\bibalias{Hash:NME}{mirzadeh2015sort}
\bibalias{hashemi.isca16}{hashemi2016accelerating}
\bibalias{hassan2015near}{hassan.memsys15}
\bibalias{HBM}{lee201425}
\bibalias{he.cvpr2016}{he2016deep}
\bibalias{He16ResNet}{he2016deep}
\bibalias{he2020newton}{newtonpim}
\bibalias{he2025papiexploitingdynamicparallelism}{he2025papi}
\bibalias{HE3DB23}{bian2023he3db}
\bibalias{HEDA22}{ren2022heda}
\bibalias{HEStandard18}{HomomorphicEncryptionSecurityStandard}
\bibalias{HK20BetterBoot}{han2020better}
\bibalias{hsieh2016}{hsieh.isca16}
\bibalias{hsieh2016accelerating}{hashemi2016accelerating}
\bibalias{hsieh2016transparent}{hsieh.isca16}
\bibalias{hynix2013ddr3lHMT451S6BFR8A}{hynix2013ddr3l}
\bibalias{IliashenkoZucca21PoPETs}{iliashenko2021faster}
\bibalias{intel-skylake}{intelskylake}
\bibalias{ipek2008self}{Ipek2008}
\bibalias{isen2009eskimoenergy}{isen2009eskimo}
\bibalias{jafri2021refresh}{jafri2020refresh}
\bibalias{jeong2019react}{jeong-tpds-2019}
\bibalias{jo-vldb-2016}{jo2016yoursql}
\bibalias{jun-isca-2015}{jun2015bluedbm}
\bibalias{jun2016storage}{jun-hpec-2016}
\bibalias{jun2018grafboost}{jun-isca-2018}
\bibalias{kabra2025ciphermatch}{Kabra2025CipherMatch}
\bibalias{kanerva2009hyperdimensional}{kanerva-congcomp-2009}
\bibalias{kang-isscc-2019}{kang201913}
\bibalias{kang-micro-2021}{kang2021iceclave}
\bibalias{kang-procieee-2020}{kang-jproc-2020}
\bibalias{kang2012flexram}{kang1999flexram}
\bibalias{kang2013enabling}{kang-msst-2013}
\bibalias{kang2021s}{kang-tc-2021}
\bibalias{kannan.kdd2016}{kannan2016smart}
\bibalias{Kautz1969}{kautz-tc-1969}
\bibalias{kautz1969cellular}{kautz-tc-1969}
\bibalias{ke2019recnmp}{ke2020recnmp}
\bibalias{keeton.1998}{keeton-sigmod-1998}
\bibalias{kevinchang-thesis}{chang.phdthesis2017}
\bibalias{kilburn.iretec62}{kilburn1962one}
\bibalias{kim-asplos-2020}{kim2020evanesco}
\bibalias{kim-isca-2012}{salp}
\bibalias{kim.arxiv17}{kim-genomics-2018}
\bibalias{kim.bmc18}{kim-genomics-2018}
\bibalias{kim.hpca18}{kim-hpca-2018}
\bibalias{kim.hpca18talk}{kim-hpca-2018}
\bibalias{kim.hpca19}{kim-hpca-2019}
\bibalias{kim.hpca19talk}{kim-hpca-2019}
\bibalias{kim.isca16}{kim2016isca}
\bibalias{kim.isca2014talk}{kim-isca2014}
\bibalias{kim.sc17}{kim2017toward}
\bibalias{Kim14RowHammer}{kim-isca2014}
\bibalias{kim2016neurocube}{kim2016isca}
\bibalias{kim2016storage}{kim-infosci-2016}
\bibalias{kim2017grim}{kim-genomics-2018}
\bibalias{kim2017improving}{kim-dac-2017}
\bibalias{kim2021behemoth}{kim-fast-2021}
\bibalias{kim2021gradpim}{kim-hpca-2021}
\bibalias{Kim20PARA}{kim2020revisiting}
\bibalias{KimPolyakovZucca21}{kim2021revisiting}
\bibalias{kogge1994execube}{kogge1994}
\bibalias{koo2017summarizer}{koo-micro-2017}
\bibalias{Koudas04FlexibleStringMatch}{koudas2004flexible}
\bibalias{kvatinsky.tvlsi14}{kvatinsky2014imply}
\bibalias{kwak2020cosmos}{kwak2018cosmos}
\bibalias{LauterNaehrigVaikuntanathan11}{naehrig2011can}
\bibalias{lavenier2020variant}{lavenier2020}
\bibalias{lee-ieeecal-2020}{lee2020smartssd}
\bibalias{lee.taco16}{lee2016simultaneous}
\bibalias{lee2012correcting}{lee2011correcting}
\bibalias{lee2015bssync}{bssync}
\bibalias{lee2020neuromorphic}{lee-fnins-2020}
\bibalias{lee2022isscc}{lee20221ynm}
\bibalias{lee2023orion}{ebel2025orion}
\bibalias{Lee21HighPrecBoot}{lee2021high}
\bibalias{leitersdorf-arxiv-2022}{leitersdorf2023aritpim}
\bibalias{lenjani2020fulcrum}{lenjani_fulcrum_2020}
\bibalias{leong2008random}{leong-uspatent-2008}
\bibalias{LHHMQ21PEGASUS}{lu2021pegasus}
\bibalias{li-atc-2021}{Li_ATC2021}
\bibalias{li.dac16}{li-dac-2016}
\bibalias{li.micro17}{li-micro-2017}
\bibalias{li2016}{li-dac-2016}
\bibalias{Li24EvalComp}{li2024evalcomp}
\bibalias{liang2019ins}{liang-fpl-2019}
\bibalias{LiEvalComp24CKKSBoot}{li2024evalcomp}
\bibalias{lim2017active}{lim2016active}
\bibalias{lin2004error}{costello1982error}
\bibalias{liu-spaa17}{liu2017}
\bibalias{liu.isca13}{mutlu.isca2013talk}
\bibalias{liu2012flikker}{liu2011flikker}
\bibalias{liu2017concurrent}{liu2017}
\bibalias{LoLa19}{brutzkus2019low}
\bibalias{LPR10RLWE}{lyubashevsky2010ideal}
\bibalias{lue2019optimal}{lue-iedm-2019}
\bibalias{luo-jsac-2016}{luo2016enabling}
\bibalias{luo2018improving}{luo-sigmetrics-2018}
\bibalias{luo2023ramulator}{luo2023ramulator2}
\bibalias{mailthody2019deepstore}{mailthody-micro-2019}
\bibalias{mansouri-asplos-2022}{Mansouri-Ghiasi_ASPLOS2022}
\bibalias{mansouri2022genstore}{Mansouri-Ghiasi_ASPLOS2022}
\bibalias{mansouri2022genstore.arxiv}{ghiasi2022genstore}
\bibalias{Marazzi23REGA}{marazzi2022rega}
\bibalias{matam2019graphssd}{Matam_ISCA2019}
\bibalias{mcciede2024}{mutlu2024memory}
\bibalias{mckee.cf04}{mckee2004reflections}
\bibalias{Melzi24PETsBiometric}{melzi2024overview}
\bibalias{merrikh2017high}{bayat-ieeetnnls-2017}
\bibalias{micheloni-insidenand-2010}{micheloni2010inside}
\bibalias{micheloni2018inside}{micheloni2013inside}
\bibalias{micron2016ddr4mta18asf2g72pz}{micronddr4mta18asf2g72pz}
\bibalias{mimdram}{oliveira2024mimdram}
\bibalias{mimdramextended}{oliveira2024mimdram}
\bibalias{missingnot}{yuksel2024functionally}
\bibalias{moorelaw}{schaller1997moore}
\bibalias{MR18ABY3}{ABY3_18}
\bibalias{mutlu-emergingcomputing-2021}{mutlu2022modern}
\bibalias{mutlu-micro2007}{mutlu2007stall}
\bibalias{mutlu.aacbbtalk19}{alser2020accelerating}
\bibalias{mutlu.appttalk19}{mutlu.gwutalk19}
\bibalias{mutlu.glvlsitalk19}{mutlu.gwutalk19}
\bibalias{mutlu.iccdtalk19}{mutlu.gwutalk19}
\bibalias{mutlu.isscctalk19}{mutlu.gwutalk19}
\bibalias{mutlu.vlsi20}{mutlu.nsfpim20}
\bibalias{mutlu2019}{mutlu2019processing}
\bibalias{mutlu2020intelligent}{mutlu2021intelligent}
\bibalias{mutlu2020modern}{mutlu2022modern}
\bibalias{mutlu2020vlsi}{mutlu.nsfpim20}
\bibalias{mutlu2021intelligentdate}{mutlu2021intelligent}
\bibalias{mutlu2023tesseractretrospective}{ahn2023retrospective}
\bibalias{myers1999fast}{myers-jacm-1999}
\bibalias{MZ17SecureMLMZ}{SecureML17}
\bibalias{nai2017}{nai2017graphpim}
\bibalias{needleman1970general}{Needleman1970Ageneral}
\bibalias{nider2020}{nider2020processing}
\bibalias{niu2022184qps}{niu2022isscc}
\bibalias{nvidia2020a100}{a100}
\bibalias{oh2014ecc}{oh20143}
\bibalias{olgun2021.SLS}{olgun2021quactrng}
\bibalias{olgun2023drambender}{olgun2023dram}
\bibalias{oliveira2017samos}{oliveira2017generic}
\bibalias{oliveira2021.SLS}{oliveira2021damov}
\bibalias{oliveira2021pimbench}{oliveira2021damov}
\bibalias{oliveira2023transpimlib}{item2023transpimlibispass}
\bibalias{oliveira2025proteus}{de2025proteus}
\bibalias{ONeil07Bitmap}{oneil-ideas-2007}
\bibalias{openfhe22}{OpenFHE22}
\bibalias{Organick20DNA}{organick2020probing}
\bibalias{orosa2022spyhammer}{orosa2024spyhammer}
\bibalias{pandiyan.iiswc2014}{pandiyan2014}
\bibalias{ParaBit21}{Gao_MICRO2021}
\bibalias{park-micro-2022}{Park_MICRO2022}
\bibalias{park2021reducing}{park-asplos-2021}
\bibalias{park2022flash}{Park_MICRO2022}
\bibalias{Park20Graphene}{park2020graphene}
\bibalias{patel2024rethinking}{patel2024rethinkingieee}
\bibalias{patterson1997case}{patterson-ieeemicro-1997}
\bibalias{pattnaik.pact16}{pattnaik2016}
\bibalias{pattnaik2016scheduling}{pattnaik2016}
\bibalias{pei2019registor}{pei-tos-2019}
\bibalias{pekhimenko2012basedeltaimmediate}{pekhimenko2012base}
\bibalias{peng2023chopper}{peng_chopper_2023}
\bibalias{pimnet}{son2025pimnet}
\bibalias{puma-asplos2019}{ankit2019puma}
\bibalias{QEMU}{qemu}
\bibalias{raidr.isca2012talk}{raidr}
\bibalias{Raj20HEDatabase}{raj2020web}
\bibalias{ramanathan2020lookup}{ramanathan-micro-2020}
\bibalias{ramulator}{kim-cal-2016}
\bibalias{Regev09LWE}{regev2009lattices}
\bibalias{rhyner2024pimopt}{rhyner2024pim}
\bibalias{riedel.1998}{riedel-vldb-1998}
\bibalias{riedel2001active}{riedel-computer-2001}
\bibalias{rixner.isca00}{rixner2000memory}
\bibalias{rMat}{rmat}
\bibalias{rnsckks}{CHKKS18FullRNS}
\bibalias{RoyTuran19HPCA}{roy2019fpga}
\bibalias{Ruan19INSIDER}{ruan2019insider}
\bibalias{RVU}{santos-date-2017}
\bibalias{Saileshwar22DSAC}{saileshwar2022randomized}
\bibalias{salihoglu.ssdbm13}{salihoglu2013gps}
\bibalias{samardzic2021f1}{feldmann2021f1}
\bibalias{samsung2013k4b4g0446a}{samsung2010k4b4g0446a}
\bibalias{santos2017operand}{santos-date-2017}
\bibalias{Saxena24PTGuard}{saxena2023ptguard}
\bibalias{seaborn.2016}{seaborn.2015}
\bibalias{seshadri.arxiv16}{seshadri-arxiv-2016}
\bibalias{seshadri.bookchapter17}{seshadri2017simple}
\bibalias{seshadri.bookchapter17.arxiv}{seshadri-arxiv-2016-pum}
\bibalias{seshadri.micro17}{seshadri-micro-2017}
\bibalias{seshadri2013rowclone}{seshadri-micro-2013}
\bibalias{seshadri2014willow}{seshadri-osdi-2014}
\bibalias{seshadri2016buddy}{seshadri-arxiv-2016}
\bibalias{seshadri2016simple}{seshadri2017simple}
\bibalias{seshadri2017ambit}{seshadri-micro-2017}
\bibalias{seshadri2020indram}{seshadri-arxiv-2019}
\bibalias{Seshadri:2015:ANDOR}{seshadri-ieeecal-2015}
\bibalias{shafiee2016isaac}{Shafiee-isca-2016}
\bibalias{SHARP23ISCA}{kim2023sharp}
\bibalias{shim-ieeejetcas-2022}{shim2022gp3d}
\bibalias{Silberschatz20DBConcepts}{silberschatz1997database}
\bibalias{singh-dac-2019}{singh2019napel}
\bibalias{singh2017zen}{singh2017}
\bibalias{smith1986pipelined}{burtonsmith1978}
\bibalias{Spectre19}{kocher2018spectre}
\bibalias{stone1970logic}{stone-tc-1970}
\bibalias{strukov.nature08}{strukov2008}
\bibalias{sura.cf15}{sura2015data}
\bibalias{sutton2018reinforcement}{sutton_2018}
\bibalias{SV14FHESIMD}{smart2014fully}
\bibalias{syncron}{giannoula2021.SLS}
\bibalias{tavakkol-fast-2018}{tavakkol2018mqsim}
\bibalias{tavakkol.fast18}{tavakkol2018mqsim}
\bibalias{tavakkol.isca18}{tavakkol2018flin}
\bibalias{TFHELib}{CGGI20TFHE}
\bibalias{thornton1970design}{thornton1970}
\bibalias{TitanV}{titanv}
\bibalias{torabzadehkashi2019catalina}{torabzadehkashi-pdp-2019}
\bibalias{truong2021racer}{truong-micro-2021}
\bibalias{upmemwebsite2024}{upmem}
\bibalias{Verma19FingerprintHybrid}{verma2019hybrid}
\bibalias{vieira2018exploiting}{8645905}
\bibalias{vtune}{MemoryBo20}
\bibalias{wang-damon-2016}{wang2016ssd}
\bibalias{wang-ieeetvlsi-2018}{wang2018three}
\bibalias{wang2019project}{wang-eurosys-2019}
\bibalias{wang2020figaro}{wang-micro-2020}
\bibalias{wang2022memcore}{wang-edtm-2022}
\bibalias{Wang23HEBooster}{wang2023he}
\bibalias{wangndsearch2024}{wang2024ndsearch}
\bibalias{watkins1992q}{qlearning_ML_1992}
\bibalias{williams2009roofline}{roofline}
\bibalias{wong.procieee12}{wong2012}
\bibalias{wu1998encoded}{wu-icde-1998}
\bibalias{xin.shd.bioinformatics15}{xin-bioinformatics-2015}
\bibalias{xin2020elp2im}{xin-hpca-2020}
\bibalias{xor1}{tuyls-springer-2005}
\bibalias{yang2000push}{YangL00}
\bibalias{yasin_ispass2014}{yasin2014}
\bibalias{yu2024cambricon}{Yu2024CambriconLLM}
\bibalias{zarif2023offloading}{Zarif_2023}
\bibalias{zhang.hpdc14}{zhang-hpdc-2014}
\bibalias{zhang2014}{zhang-hpdc-2014}
\bibalias{zhang2014top}{zhang-hpdc-2014}
\bibalias{zhao-fast-2013}{zhao2013ldpc}
\bibalias{zhaoumpim2024}{zhao2024pim}
\bibalias{zhu2013accelerating}{zhu2013}
\bibalias{zhuo-micro-2019}{zhuo2019graphq}
\bibalias{zuravleff.patent97}{zuravleff1997controller}

\newif\ifshowchanges
\showchangesfalse
\providecommand{\DIFaddbegin}{}
\providecommand{\DIFaddend}{}
\providecommand{\DIFdelbegin}{}
\providecommand{\DIFdelend}{}
\providecommand{\DIFaddbeginFL}{}
\providecommand{\DIFaddendFL}{}
\providecommand{\DIFdelbeginFL}{}
\providecommand{\DIFdelendFL}{}
\ifshowchanges
  \providecommand{\DIFadd}[1]{{\color{blue}#1}}
  \providecommand{\DIFaddFL}[1]{{\color{blue}#1}}
\else
  \providecommand{\DIFadd}[1]{#1}
  \providecommand{\DIFaddFL}[1]{#1}
\fi
\providecommand{\DIFdel}[1]{}
\providecommand{\DIFdelFL}[1]{}

\title{\titlenamed:\\
An Efficient Algorithm for Enabling Fast Bulk Bitwise Operations \\ on Homomorphically Encrypted Data}

\author{
    Mayank Kabra\textsuperscript{1} \quad
    Phillip Widdowson\textsuperscript{1} \quad
    Harshita Gupta\textsuperscript{1} \quad
    Rakesh Nadig\textsuperscript{1} \quad
    Nisa Bostanc\i{}\textsuperscript{1} \\
    Eirini Tzermpou\textsuperscript{1} \quad
    Rakesh Kumar\textsuperscript{2} \quad
    Spiros Galanopoulos\textsuperscript{1} \quad
    Ataberk Olgun\textsuperscript{1} \\
    Abdullah Giray Ya\u{g}l\i{}k\c{c}\i{}\textsuperscript{3} \quad
    Mohammad Sadrosadati\textsuperscript{1} \quad
    Onur Mutlu\textsuperscript{1} \vspace{0.5em} \\
    \textsuperscript{1}\emph{ETH Z{\"u}rich} \quad
    \textsuperscript{2}\emph{NTNU} \quad
    \textsuperscript{3}\emph{CISPA}
}

\begin{document}

\maketitle
\thispagestyle{firstpage}

\sloppypar

\begin{abstract}

A wide range of important cloud applications, e.g., database analytics, \mkrev{graph analytics, and genome analysis,} \mkspkon{spend a considerable amount of time performing} bulk bitwise operations\mktches{} \nb{(e.g., AND, OR, XOR)}. These applications \mkspkon{often} operate on highly sensitive data, making data privacy a critical concern.
Homomorphic encryption (HE) enables secure computation on encrypted data without requiring decryption, ensuring privacy in untrusted and shared environments. 
\mkrevb{We analyze 13 end-to-end applications across three domains (healthcare, graph, and database analytics) and observe that bulk bitwise operations take 62--100\% of the execution time of the healthcare and graph workloads. The key problem is that existing HE approaches perform them with high-latency homomorphic operations (e.g., multiplication or lookup tables) and significantly increase the data size.}

Our goal is to 
enable fast bulk bitwise operations on homomorphically encrypted data 
by designing an efficient HE algorithm that reduces the use of 
\mkrevb{these costly operations} in end-to-end applications.
\agy{To this end, we propose \titlenamed, a new algorithm for fast bulk bitwise operations on homomorphically encrypted data.}
\mkrev{\titlename performs two key steps: (1) \mktches{it constructs bulk bitwise operations using low-cost homomorphic additions, which compute the Hamming weight of the encrypted operand bits in every slot} of a packed ciphertext without any homomorphic multiplication, and (2) it decodes the result of an AND, OR, or XOR from the Hamming weight with a single indicator polynomial, keeping the result in the standard encoding.}

\mkrev{We evaluate \titlename on 13 real-world end-to-end applications against the arithmetic approach used by prior works. Across three application domains (i.e., encrypted healthcare, graph, and database analytics), \titlename improves performance by up to $7.26\times$ (CPU) and $6.65\times$ (GPU), and on average by \mktches{$2.32\times$/$1.78\times$/$1.45\times$ (CPU) and $2.08\times$/$1.70\times$/$1.51\times$ (GPU) for healthcare/graph/database analytics}.}
\mkspkon{We further study the scalability of \titlenamed, \mktches{increasing the reduction depth} in a representative end-to-end application. We demonstrate that \titlenamed's performance scales, 
achieving up to a 10.10$\times$ speedup over the state-of-the-art GPU baseline.}

\end{abstract}

\section{Introduction}
\label{sec:introduction}

\mayankk{A wide range of \mksp{cloud} applications (e.g., data analytics~\citeDBANALYTICS,
graph processing~\citeGRAPH,
and genome analysis~\citeGENOMICS)
~\hga{often operate on} 
highly sensitive data, such as
personal health records and financial information, making data privacy a critical concern.
In conventional
systems, 
data~\hga{is generally stored} in an encrypted format (e.g., using AES~\cite{daemen1999aes}) but must~\hga{be decrypted}
before computation, 
exposing sensitive information in untrusted or shared 
environments~\citeCRYPTO.
\noindent \emph{Homomorphic encryption (HE)}~\citeHEFOUND
~addresses this limitation by enabling computations directly on encrypted data without 
decryption (\mktches{\S\ref{sec:background}}).} \mkrevb{HE performance depends on which homomorphic operations an application uses, whose latencies differ by orders of magnitude (e.g., addition vs.\ multiplication).} Prior works (e.g.,~\cite{chen2018labeled,park2019secure,iliashenko2021faster}) propose new HE algorithms \mkar{and data encodings} for improving the performance of these cloud applications.
Several~\hga{works \mktches{(e.g.,}~\cite{kabra2025ciphermatch,aziz2024secure,bonte2020homomorphic}\mktches{)} propose} 
\mkar{efficient} HE-based string matching~\hga{via improved} 
data encodings, while other works such as ArcEDB~\cite{zhang2024arcedb}, HE3DB~\cite{bian2023he3db}, \mktches{NSHEDB~\cite{jung2026nshedb},} and Engorgio~\cite{bian2025engorgio} optimize homomorphic sorting and comparison for operators such as sort/join (\S\ref{sec:key_related_works}).

\mktches{Despite these algorithmic advances,} we identify a key problem that highlights an opportunity for further performance improvement. Our analysis of \mktches{13} workloads across three application classes (database, healthcare, and graph analytics) shows that these applications \mkspkon{spend a considerable amount of time \mkrev{(62.0--100\% of the execution time of the healthcare and graph workloads)}} performing bulk bitwise operations on~\hga{homomorphically encrypted} data, which remains a major overhead. \mkrevb{A bulk bitwise operation applies one wide Boolean function (e.g., an AND, OR, XOR, or threshold over many input bits) independently to every record of a large dataset, which maps directly onto the SIMD slots of batched ciphertexts.} \mkar{\mkrevb{For example,}
genome analysis such as private k-mer based pathogen screening~\mkrevb{\cite{bed2016privately}}
encodes each DNA base in two bits, so matching a 31-base sequencing read against a
reference k-mer is one wide bulk bitwise AND over all $2\times31=62$ homomorphically
encrypted bits (AND-62); the test is repeated for each of 16 reference k-mers and the
results are combined with an OR-16. With one read per ciphertext slot\footnote{\mkrevb{A B/FV ciphertext packs $N$ values into $N$ independent \emph{slots}, and each homomorphic operation acts on all slots at once (\S\ref{sec:BFV_background}).}}, a single
evaluation screens $32{,}768$ reads against all 16 references, where the arithmetic
approach spends $991$ $\mathsf{HMul}$s (Table~\ref{tab:workload_sizes_limbs_32gib_all16}).}

\textbf{Key Problem.} Unfortunately, existing HE schemes express bitwise operations using expensive homomorphic \mkrevb{operations} (e.g., lookup table (LUT) evaluation or multiplication), resulting in high latency when executed in bulk.  \mksp{Existing} HE schemes express bitwise operations either using 
(1) the Boolean~\cite{bourse2018fast, stoian2023deep, sanyal2018tapas}, or (2) the arithmetic~\cite{xie2014crypto, juvekar2018gazelle, lagesse2021you} approach. The Boolean approach supports bit-level logic \harshita{and}
\mayank{arbitrary} Boolean circuits \harshita{via}
\mkrevb{LUT} \harshita{evaluation.}
\hga{The arithmetic approach supports homomorphic addition and multiplication, which are used to construct homomorphic bitwise operations. For example, the AND can be implemented as $a \cdot b$ and OR as $a + b - a \cdot b$, \mksp{where $\cdot$ is homomorphic multiplication and $+,-$ are homomorphic addition/subtraction} (\S\ref{sec:motivation}). \mkrevb{LUT evaluation and homomorphic multiplication} incur significant latency overheads, i.e., \pw{they are} $10^4$-$10^5\times$ slower than unencrypted~\hga{CPU} execution
\cite{samardzic2022craterlake} due to their~\hga{high} computational cost and large memory footprint (\S\ref{sec:motivation})\mkrevb{, whereas homomorphic addition is orders of magnitude cheaper}.}

\mkisca{\textbf{Our goal} 
is to enable fast bulk bitwise operations on homomorphically encrypted data by designing an efficient algorithm that reduces reliance on high-latency homomorphic operations 
in end-to-end applications.
To this end, we propose \titlenamed,\footnote{\titlename combines \underline{h}omomorphic encryption and b\underline{ulk} bitwise operations.} 
whose key idea is to construct \mktches{the bulk bitwise operations, which are prevalent in many applications, as a sequence of low-cost homomorphic additions} that reduces the need for costly homomorphic operations (e.g., homomorphic multiplications)\mktches{, thereby improving the performance of HE-based applications}.}

\mkisca{The \textbf{key insight} behind \titlename is that in many applications, bulk bitwise operations follow a sequence of the form \(a_{1} \star a_{2} \star \ldots \star a_{n}\), where $\star$ is a bitwise operation (AND/OR/XOR). This form constitutes the class of symmetric Boolean functions~\cite{canteaut2005symmetric} whose output depends only on the Hamming weight of the inputs (i.e., the number of 1s), not on their arrangement. 
As such, for bits \(a, b \in \{0, 1\}\), the sum \(a + b \in \{0, 1, 2\}\) places \(a \oplus b\) (XOR) in its low-order bit and \(a \cdot b\) (AND) in its high-order (carry) bit. For example, \(a\,\&\,b\,\&\,c\,\&\,d\) is computed as \(a + b + c + d\): the sum reaches 4 only when all four bits are 1, which for four inputs is exactly its most-significant bit (\S\ref{sec:overview}).}

\mkrev{\textbf{Key Mechanisms.} \titlename leverages the above key insight in two steps. First, \titlename encodes each bitvector using one bit per \mktches{ciphertext} slot and adds the encrypted operands with homomorphic addition, which computes the Hamming weight in every slot without any homomorphic multiplication (\S\ref{sec:encoding} and \S\ref{sec:sum_indicate}). Second, \titlename decodes the Boolean result from the Hamming weight with a single indicator polynomial, which the server evaluates under encryption, so that the result returns to the standard arithmetic encoding and supports applications with mixed bitwise and arithmetic operations (e.g., aggregation or popcount) (\S\ref{sec:server_decoding}). \mktches{This indicator is cheap because it evaluates one polynomial of the sum, which needs only $O(\sqrt{K})$ $\mathsf{HMul}$s instead of the $K{-}1$ of the product tree (e.g., 3 instead of 7 for $K{=}8$).} When revealing the Hamming weight is acceptable, the client can read the result directly
(\S\ref{sec:client_decoding}).}

\mktches{\textbf{Key Results.} We compare \titlename against the arithmetic approach implemented with Microsoft SEAL~\cite{sealcrypto} (CPU) and HEonGPU~\cite{ozcan2024heongpu} (GPU) (\S\ref{sec:methodology}). First, across 13 end-to-end applications, \titlename improves performance by up to $7.26\times$ (CPU) and $6.65\times$ (GPU), and on average by $2.32\times$/$1.78\times$/$1.45\times$ (CPU) and $2.08\times$/$1.70\times$/$1.51\times$ (GPU) for healthcare/graph/database analytics (\S\ref{sec:end-to-end}). The six database filter-and-aggregate queries gain only $1.01$--$1.31\times$ (CPU), because homomorphic comparison, which both approaches share, dominates them, whereas encrypted top-10 ranking gains $2.00\times$ (CPU). Second, when the client may learn the Hamming weight, client-side decoding removes the indicator entirely: on a microbenchmark that isolates one bulk AND, \titlename improves performance by $390.6$--$2002.2\times$ (CPU) as the reduction depth increases (\S\ref{sec:reduction}); this microbenchmark is not an end-to-end speedup. Third, a sensitivity study of \mktches{bit-sliced} PSI~\mktches{\cite{cheon2015search,kim2016efficiency}} shows that \titlenamed's speedup grows with the reduction depth, up to $10.10\times$ on the GPU (\S\ref{sec:discussion}). Fourth, two further optimizations extend these gains: memory-efficient data packing accelerates addition-only bulk bitwise kernels by $58.9$--$227.0\times$ (CPU), because it converts the bulk of the bitwise logic into a sequence of low-cost homomorphic additions, and on top of encrypted bitmap indices used in databases, \titlename accelerates database queries by a further $1.32\times$ on average (\S\ref{sec:app_optimizations}).}

This work makes the following key contributions: 

\squishlist 

\item We propose \titlenamed, which significantly accelerates applications that use \mkrevb{HE-based} bulk bitwise operations.

\item \mkrev{To our knowledge, \titlename is \mkrevb{the first work to evaluate wide bit-sliced AND, OR, and XOR reductions in packed B/FV by adding the operand bits slot-wise and decoding their Hamming weight with a single indicator, which keeps the result in the standard slot encoding and reduces} the use of homomorphic multiplication across end-to-end applications (\S\ref{sec:otherbatched} shows that our baseline, the product tree of packed arithmetic B/FV, is faster than four families of prior approaches).}

\item \titlename supports efficient conversion between its encoding and the standard arithmetic HE encoding using \mkrev{indicator evaluation (i.e.,} bit-extraction\mkrev{)}, 
enabling integration into end-to-end applications and improving overall performance.

\item \mktches{We evaluate \titlename on 13 end-to-end applications on real CPU and GPU systems, where it improves performance by up to $7.26\times$. We further show its benefit with client-side decoding, its scaling with the reduction depth in a PSI sensitivity study, and two further optimizations: memory-efficient data packing and encrypted bitmap indices.}%
\footnote{Our code is available at \url{https://github.com/CMU-SAFARI/HULK}.}

\squishend

\section{Background}
\label{sec:background}

\subsection{Homomorphic Encryption}
\label{sec:HE_background}
Homomorphic encryption (HE)~\citeHEFOUND
~is an encryption paradigm that permits computation on encrypted data without decryption.\footnote{
\hga{We use the following terminology: \textit{cleartext} for raw unencrypted input, \textit{plaintext} for HE-encoded data, and \textit{ciphertext} for encrypted plaintext.}}
\harshita{HE approaches are broadly classified into two categories based on the type of operations they enable over encrypted data: (i) Boolean HE schemes (e.g., TFHE~\cite{chillotti2016faster,chillotti2020tfhe}) and (ii) arithmetic HE schemes (e.g., B/FV~\cite{brakerski2012fully,fan2012somewhat} and CKKS~\cite{cheon2017homomorphic}).}
\mkrev{In B/FV, the secret key $s$ is a ternary polynomial in $R_Q=\mathbb{Z}_Q[X]/(X^N{+}1)$, and a message $m\in R_p=\mathbb{Z}_p[X]/(X^N{+}1)$ is encrypted as $\mathsf{ct}=\big([-a\cdot s+e+\Delta\cdot m]_Q,\ a\big)\in R_Q^2$, where $a$ is uniform in $R_Q$, $e$ is a small error polynomial (discrete Gaussian with standard deviation $\approx3.2$), and $\Delta=\lfloor Q/p\rfloor$. TFHE uses the LWE form over the torus.}

\subsubsection{Boolean Approach (TFHE)} 
\label{sec:TFHE_background} 
TFHE (Fast Fully Homomorphic Encryption over the Torus)~\citeTFHE is designed for efficient gate-by-gate evaluation of Boolean circuits.
\hgsp{In this setting, bits are encoded on the torus $[0, 1)$ ($0 \mapsto 0.0$, $1 \mapsto 0.5$), which improves noise tolerance.}

\noindent \textbf{Gate Evaluation.}
TFHE~\cite{chillotti2021programmable,chillotti2021improved} supports homomorphic evaluation of arbitrary binary gates, such as AND ($\mathsf{HomAND}$), OR ($\mathsf{HomOR}$), and XOR ($\mathsf{HomXOR}$), by encoding them as lookup tables (LUTs)~\cite{chillotti2021improved}. These LUTs are evaluated during bootstrapping~\cite{chillotti2021programmable}, which refreshes ciphertext noise after each binary gate. Bootstrapping internally requires a complex homomorphic rotation operation~\cite{chillotti2021improved}. Since each ciphertext represents a single bit, complex functions are expressed as Boolean circuits composed of basic gates and bootstrapping.

\subsubsection{Arithmetic Approach (B/FV and CKKS)}
\label{sec:BFV_background}
The Brakerski/Fan-Vercauteren (B/FV)~\citeBFV{} scheme supports exact arithmetic on integers over a modulus. 
\mkar{The Cheon-Kim-Kim-Song (CKKS)~\citeCKKS{} scheme, on the other hand, supports approximate arithmetic on floating-point numbers over a modulus.} 
Both schemes operate over polynomial rings, where ciphertexts are represented as polynomials in the ring \( R_Q = \mathbb{Z}_Q[X]/(X^N + 1) \), $Q$ being the ciphertext modulus and $N$ a power-of-two polynomial degree.\footnote{Note that for large $N$ 
these schemes use a large ciphertext modulus (e.g., $Q = 881$ bits)~\cite{sealcrypto, albrecht2022homomorphic}. To support computation on conventional CPUs, HE schemes use the RNS decomposition~\cite{halevi2019improved, bajard2016full} to decompose $Q$ into smaller moduli $\{q_1, \ldots, q_m\}$ (called limbs).}
Plaintexts are in a smaller ring \( R_p = \mathbb{Z}_p[X]/(X^N + 1) \), \( p \ll Q\) is the plaintext modulus. \mkrev{When $p$ is a prime with $p\equiv1\bmod 2N$, batching (SIMD) encoding~\cite{smart2014fully} maps a vector of $N$ values in $\mathbb{Z}_p$ to one plaintext, and each value is called a \emph{slot}: homomorphic addition and multiplication then act slot-wise. With coefficient encoding (any $p$), each polynomial coefficient holds one value; addition is coefficient-wise, but multiplication is a negacyclic convolution. $\mathsf{HMul}$ denotes a ciphertext-ciphertext multiplication with relinearization.}

\noindent \textbf{Homomorphic Operations:}  
Both schemes support both homomorphic addition ($\mathsf{HAdd}$) and homomorphic multiplication ($\mathsf{HMul}$). Given ciphertexts \( c_1 = (\mathbf{c}_0^{(1)}, \mathbf{c}_1^{(1)}) \) and \( c_2 = (\mathbf{c}_0^{(2)}, \mathbf{c}_1^{(2)}) \), $\textsf{HAdd}$ is performed component-wise as follows:
\( c_{\text{add}} = ( \mathbf{c}_0^{(1)} + \mathbf{c}_0^{(2)},\ \mathbf{c}_1^{(1)} + \mathbf{c}_1^{(2)} ) \).
The $\mathsf{HMul}$ operation produces a ciphertext with \emph{three} components \( (\mathbf{d}_0, \mathbf{d}_1, \mathbf{d}_2) \in R_Q^3 \), \phillip{requiring a reduction back to \emph{two}}. This step is \phillip{known} as \emph{relinearization}~\cite{kipnis1999cryptAnalysis, wibawa2022bfv}. 
\hgrev{Each $\mathsf{HMul}$ adds noise, requiring a larger $Q$ and bootstrapping~\cite{gentry2012better} when the budget is exhausted.}

\noindent\textbf{Homomorphic Polynomial Evaluation.}
\label{sec:ps_background}
\mkar{Arithmetic HE schemes evaluate non-linear functions, such as homomorphic comparison~\cite{iliashenko2021faster} and digit extraction for bootstrapping~\cite{chen2018homomorphic}, as a univariate polynomial of one ciphertext with plaintext coefficients. The cost of such an evaluation is dominated by its ciphertext-ciphertext $\mathsf{HMul}$s, since plaintext-ciphertext multiplications and $\mathsf{HAdd}$s are cheap. The Paterson--Stockmeyer (PS) algorithm~\cite{paterson1973number} minimizes the number of these $\mathsf{HMul}$s.}

\mkar{For a polynomial $I(t)=\sum_{k=0}^{d}c_k t^k$ of degree $d$ and a block size $B$, PS rewrites $I(t)$ as
\begin{equation}
    \label{eq:eqn6}
I(t)=\sum_{j=0}^{G-1} t^{jB}\!\left(\sum_{i=0}^{B-1} c_{jB+i} t^i\right),\quad
G=\left\lceil\frac{d+1}{B}\right\rceil
\end{equation}
PS precomputes the baby-step powers $t,\dots,t^{B-1}$ and the giant-step powers $t^{B},t^{2B},\dots$ once, forms each block as a linear combination of baby-step powers with plaintext coefficients, and combines the blocks with the giant-step powers. For $B\approx\sqrt{d}$, the number of $\mathsf{HMul}$s drops from $O(d)$ for a naive power-series evaluation to $O(\sqrt{d})$.}

\mksp{\noindent\textbf{Security.}
The security of B/FV, CKKS, and TFHE reduces to the hardness of the (Ring) Learning-with-Errors (RLWE)
problem~\cite{Regev:2005aln,lyubashevsky2010ideal,fan2012somewhat,brakerski2014leveled}.
For a fixed polynomial degree $N$, the achievable security level $\lambda$ is governed by the
ratio between the ciphertext modulus $Q$ and the error distribution: larger $N$ and smaller $Q$
yield higher $\lambda$. Throughout this work we target $\lambda = 128$-bit (classical) security
and select all parameters in accordance with the Homomorphic Encryption Security
Standard~\cite{HomomorphicEncryptionSecurityStandard}, which, for $N = 2^{15} = 32768$ and a
uniform ternary secret, permits a ciphertext modulus of up to $\log_2 Q \approx 881$~bits at
$\lambda = 128$. The plaintext modulus $p$ does not affect $\lambda$: it determines only the message space.}

\subsection{Bulk Bitwise Operations using HE}
\label{sec:bbop}

Existing HE approaches express bulk bitwise operations using either the \textit{Boolean} approach
or the \textit{arithmetic} approach.

\textbf{The Boolean (bit-by-bit) approach}
\harshita{encrypts} each bit \harshita{individually}
using Boolean
schemes (e.g., TFHE~\cite{chillotti2020tfhe, TFHE-rs}) \harshita{and applies} logic gates
directly \harshita{on}
ciphertexts via look up table (LUT) evaluations~\cite{chillotti2021improved, chillotti2021programmable}\phillip{,  enabling} arbitrary-depth computation (\S\ref{sec:HE_background}).
However, the lack of data packing results in significant
memory and computational overheads, \harshita{as each bit requires a separate ciphertext and gate evaluation.}

\textbf{The Arithmetic (packed) approach}
packs multiple bits
into a single \harshita{plaintext} polynomial, 
where each \DIFdelbegin \DIFdel{coefficient }\DIFdelend \DIFaddbegin \DIFadd{slot }\DIFaddend holds a single bit (for bitwise operations), and encrypts it using arithmetic schemes such as B/FV~\cite{fan2012somewhat}. 
Logical (Boolean) operations are expressed
arithmetically: $ \text{AND} = (a \cdot b) \mod p$, $ \text{XOR} = (a + b - 2ab) \mod p$, and $ \text{OR} = (a + b - ab) \mod p$.
\hgsp{Packed ciphertexts enable SIMD-style parallelism that amortizes memory and computational costs\mktches{, because one homomorphic operation acts on all $N$ slots at once. However, each two-input AND or OR still costs one $\mathsf{HMul}$, so a $K$-input reduction costs $K{-}1$ $\mathsf{HMul}$s}. We quantify this tradeoff in \S\ref{sec:deciding_baseline}.}

\section{Key Related Works}
\label{sec:key_related_works}
\label{sec:previous_works}

\mkrev{We describe prior HE-based works in three application domains and the constructions they use for bulk bitwise operations.}

\mkrev{\textbf{1) Encrypted Graph Analytics.} Graph analytics increasingly run in the cloud over personal relationships and identifiers (e.g., contact lists and customer graphs), so they must preserve the privacy of the underlying data~\cite{ion2017private,kales2019mobile}. Private set intersection (PSI), which finds the common elements of two parties' sets (e.g., for contact discovery~\cite{kales2019mobile} and ad-conversion measurement~\cite{ion2017private}), and private nearest-neighbor search are two representative workloads. PSI protocols (e.g.,~\cite{chen2017fast,chen2018labeled,pinkas2019efficient,chandran2022circuit}) encode each party's items as encrypted indicators and test membership element-wise. Bit-sliced constructions~\cite{cheon2015search,kim2016efficiency} compare an encrypted item with a query bit by bit and evaluate equality as the product of the matched bits, i.e., a product tree of $\mathsf{HMul}$s. Private nearest-neighbor search (e.g., Tiptoe~\cite{henzinger2023tiptoe}) retrieves results from an encrypted index. When candidates are generated by locality-sensitive hashing~\cite{indyk1998approximate,charikar2002similarity} over encrypted codes, each bucket match is a wide AND over XNOR bits, which the arithmetic approach also evaluates with $\mathsf{HMul}$s.}

\mkrev{\textbf{2) Encrypted Healthcare Analytics.} Genomic and clinical data is among the most sensitive personal data, because it permanently identifies a person and their relatives, so outsourced genome analysis must protect it~\cite{aziz2022privacy,wood2020homomorphic}. Pathogen screening with $k$-mer matching, clinical variant lookup, and genome-wide association studies are representative workloads. Prior works match encrypted genomic data against reference sequences or variants. Kim and Lauter~\cite{kim2015private} and SIG-DB~\cite{titus2018sig} evaluate private genome queries with arithmetic HE. Park et al.~\cite{park2019secure} match homomorphically encrypted $k$-mers. Yasuda et al.~\cite{yasuda2014practical} pack bits for private string matching, and Aziz et al.~\cite{aziz2024secure} use Boolean HE gates ($\mathsf{HomXOR}$ and $\mathsf{HomAND}$). CIPHERMATCH~\cite{kabra2025ciphermatch} improves exact matching with a memory-efficient data packing. Blatt et al.~\cite{blatt2020secure} perform genome-wide association studies (GWAS) under HE. In these works, each match decision is an equality or threshold over many encrypted bits, which they evaluate with $\mathsf{HMul}$ trees or lookup tables.}

\mkrev{\textbf{3) Encrypted Database Analytics.} Cloud databases store sensitive business and personal records (e.g., customer, order, and financial data), so query processing over encrypted tables is a long-standing goal~\cite{bian2023he3db,bian2025engorgio}. Filter-and-aggregate analytics, as in the TPC benchmarks~\cite{council2005transaction,nambiar2006making}, and top-$k$ ranking queries are representative workloads. HE3DB~\cite{bian2023he3db}, ArcEDB~\cite{zhang2024arcedb}, HEDA~\cite{ren2022heda}, \mktches{NSHEDB~\cite{jung2026nshedb},} and Engorgio~\cite{bian2025engorgio} evaluate relational queries on encrypted tables. They generate each predicate with homomorphic comparison~\cite{iliashenko2021faster} and compose the predicates of a query with $\mathsf{HMul}$-based AND and OR \mkrev{(or, in HE3DB\mkrevb{ and ArcEDB}, with TFHE gates)} before aggregation. Encrypted top-$k$ queries additionally rank under encryption, e.g., by counting comparison results~\cite{cetin2015depth}. \mktches{For each candidate, the server counts the candidates with a larger value and tests whether this count is below $k$, which is a THRESHOLD reduction over the comparison bits (e.g., TPC-H Q3 top-10).} Many TPC queries~\cite{council2005transaction,nambiar2006making} consist of such predicate composition followed by aggregation (see \S\ref{sec:deciding_baseline}).}

\section{\pw{The Baseline FHE Scheme}}
\label{sec:deciding_baseline}
\mktches{We select our baseline in two steps. First, we compare the Boolean (TFHE) and arithmetic (B/FV) HE approaches (\S\ref{sec:prior_approaches_motivation}). Second, we compare the product tree of packed B/FV with four other families of HE approaches that evaluate bulk bitwise operations (\S\ref{sec:otherbatched}). The product tree is $1.7$--$6{,}270\times$ faster than these families (Table~\ref{tab:prior_comparison}), so it is our baseline.}

\subsection{Analysis of Existing FHE Schemes Executing HE-based Bulk Bitwise Operations}
\label{sec:prior_approaches_motivation}

We measure the post-encryption memory footprint and execution time of $\mathsf{HomAND}$ and $\mathsf{HomXOR}$ under the Boolean approach (TFHE, via TFHE-rs~\mbox{
\cite{TFHE-rs}}\hskip0pt
) and the arithmetic approach (B/FV, via Microsoft SEAL~\mbox{
\cite{sealcrypto}}\hskip0pt
) on a real system (\S\ref{sec:methodology}), varying (1)~$K$, the number of sequential homomorphic operations on the critical path in $a_1\star\cdots\star a_K$, and (2)~the packing factor (PF), the number of bits per ciphertext. TFHE has PF$=1$ and needs one ciphertext per bit; B/FV packs up to $N{=}32768$ bits per ciphertext. For TFHE, PF remains 1, and the number of ciphertexts increases linearly~\hgsp{with} the number of bits\mktches{, so for PF$>$1, TFHE evaluates the same number of bits with PF ciphertexts}. For B/FV, a single ciphertext can pack \pw{from} 1 \pw{up} to 32768 \pw{($N$)} bits, increasing PF from 1 to 32768. 
These experiments are performed on a real system (\S\ref{sec:methodology}). 

\begin{figure}[h]
    \centering
    \includegraphics[width=\linewidth]{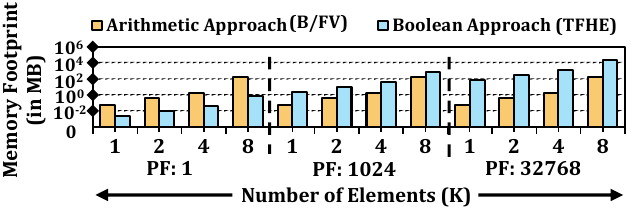}
    \caption{\DIFdelbeginFL \DIFdelFL{Comparison of memory }\DIFdelendFL \DIFaddbeginFL \DIFaddFL{Memory }\DIFaddendFL footprint after encryption for different \DIFdelbeginFL \DIFdelFL{K }\DIFdelendFL \DIFaddbeginFL \DIFaddFL{$K$ }\DIFaddendFL (sequential homomorphic operations) and three packing factors (PF).}
    \label{fig:memory_footprint_single_gates}
\end{figure}

Figure~\ref{fig:memory_footprint_single_gates} compares the memory footprint (in MB) after encryption using both approaches. We make two observations.

\DIFdelend 

\squishlist 
\item \mksp{\hgsp{\mktches{First}, for all K, packing more bits into a single \wi{B/FV} ciphertext lowers \pw{the} memory footprint per bit, such that beyond a certain PF, \wi{B/FV} uses less memory than TFHE. 
\pw{For \wi{B/FV}, increasing K requires increasing the limb count, which in turn increases the ciphertext degree $N$. Therefore, even encrypting a single bit requires 768KB at K=8, whereas the same ciphertext can accommodate 32768 data bits.}
\pw{Conversely, \mktches{at K=8, TFHE's memory footprint grows from 2.5MB at PF=1 to 81.4GB at PF=32768 (32768 ciphertexts)}. This is because TFHE requires one ciphertext per data bit, which compares poorly to \wi{B/FV}, which uses at most 192MB in total.} 
}}
\item \mksp{\mktches{Second, only for small packing factors,}\footnote{Note that the number of ciphertexts required is $2^d$ for the \wi{B/FV} scheme, and $PF\times2^d$ for the TFHE scheme, where PF is the packing factor, and d is the circuit depth. The memory footprint for a single TFHE ciphertext differs from a \wi{B/FV} ciphertext due to \phillip{different cryptographic primitives}~\cite{chillotti2020tfhe,fan2012somewhat}.} \mktches{does} the Boolean HE approach incur a significantly lower memory footprint after encryption compared to the arithmetic approach. For example, when encrypting one bit \mktches{at K=1}, TFHE uses approximately 2.5KB, \pw{compared to 32KB for \wi{B/FV}}.}

\squishend

\begin{figure}[h]
    \centering
    \includegraphics[width=\linewidth]{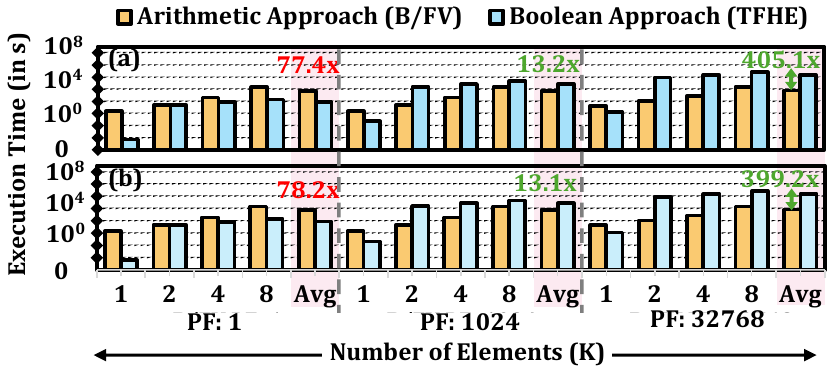}
    \caption{\DIFdelbeginFL \DIFdelFL{Comparison of execution }\DIFdelendFL \DIFaddbeginFL \DIFaddFL{Execution }\DIFaddendFL time (\DIFdelbeginFL \DIFdelFL{in }\DIFdelendFL s) \DIFdelbeginFL \DIFdelFL{for }\DIFdelendFL \DIFaddbeginFL \DIFaddFL{of }\DIFaddendFL (a) \DIFdelbeginFL \DIFdelFL{homomorphic AND ($\mathsf{homAND}$) }\DIFdelendFL \DIFaddbeginFL \DIFaddFL{$\mathsf{HomAND}$ }\DIFaddendFL and (b) \DIFdelbeginFL \DIFdelFL{homomorphic XOR ($\mathsf{homXOR}$) }\DIFdelendFL \DIFaddbeginFL \DIFaddFL{$\mathsf{HomXOR}$ }\DIFaddendFL for different \DIFdelbeginFL \DIFdelFL{K (i.e., number of sequential homomorphic operations) }\DIFdelendFL \DIFaddbeginFL \DIFaddFL{$K$ }\DIFaddendFL and three packing factors (PF).}
    \label{fig:computational_single_gates}
\end{figure}

\noindent\mayankk{Figure~\ref{fig:computational_single_gates} shows the execution time (in s\nb{econds}) for evaluating $\hgsp{\mathsf{HomAND}}$ and $\hgsp{\mathsf{HomXOR}}$ operations using both arithmetic and Boolean approaches. We make two observations.
}

\DIFdelend 

\DIFdelbegin 

\squishlist

\item \mktches{First}, \pw{the} arithmetic scheme benefits \pw{significantly} from data packing, which \pw{both} enables parallel evaluation of multiple bits within a single ciphertext, \pw{and reduces the number of total operations to perform, due to the decreased number of ciphertexts. This reduces} latency by up to 405$\times$ \pw{across different K}
for \pw{the} $\mathsf{homAND}$ operation (Fig.~\ref{fig:computational_single_gates}, PF=32768). 

\item \mktches{Second, only for small packing factors does the Boolean approach achieve} lower latency, showing a 77$\times$ average reduction compared to the arithmetic approach for \pw{the} AND operation across all circuit depths. This is because
the arithmetic approach requires $\mathsf{HMul}$, which has a higher computational complexity than the LUT evaluations used in the Boolean approach. The latency of a $\mathsf{HMul}$ operation grows with K, which requires a larger limb count (e.g., 7.6ms and 75ms for depths 2 and 8, respectively).

\squishend

\mktches{We conclude that packing makes the arithmetic approach the better fit for bulk bitwise operations over many bits: its cost per bit falls as the packing factor grows, whereas TFHE always pays one ciphertext and one bootstrapped gate per bit.}

\subsection{\mkrev{Comparison with Other Approaches}}
\label{sec:otherbatched}
\mktches{The Boolean approach of \S\ref{sec:prior_approaches_motivation} evaluates one bit per ciphertext. However, other HE approaches also batch bitwise evaluation, so they are potentially stronger baselines than packed B/FV. Packed B/FV evaluates a bulk AND-$K$ as a \emph{product tree} of $K{-}1$ $\mathsf{HMul}$s over packed ciphertexts. We compare this product tree with four families of HE approaches that target batched bitwise evaluation. We measure every family on the same CPU, each with its own parameters at 128-bit security (\S\ref{sec:methodology}), and Table~\ref{tab:prior_comparison} reports the latency per record of one bulk AND-$K$.}

\mkrev{\textbf{(1) Batched Boolean Schemes.} Batched Boolean schemes~\cite{chillotti2017faster,micciancio2018ring,liu2023amortized} pack $N$ bits into one TRLWE ciphertext or amortize bootstrapping across many LWE samples. Packing reduces the memory footprint after encryption. However, packing does not reduce the per-gate latency, because each gate extracts every operand bit into an individual LWE ciphertext and performs one bootstrapping-class operation. Hence, their per-gate latency remains on the Boolean curve of Figure~\ref{fig:computational_single_gates}. As a result, one bulk AND-$K$ is $2{,}529$--$6{,}270\times$ slower per record than the product tree for $K{=}2$--$128$ (see Table~\ref{tab:prior_comparison}).}

\mkrev{\textbf{(2) Discrete CKKS.} Discrete CKKS~\cite{bae2024bootstrapping,drucker2024bleach} evaluates Boolean gates as arithmetic over $\{0,1\}$ in the approximate CKKS scheme~\cite{cheon2017homomorphic}. One CKKS multiplication is cheaper than one B/FV multiplication (50.4\,ms vs.\ 92.5\,ms at $\log_2 Q{=}440$ and $N{=}32768$, both in Microsoft SEAL~\cite{sealcrypto}). However, discrete CKKS suffers from two limitations. First, every gate requires a cleanup polynomial that removes the accumulated approximation error~\cite{drucker2024bleach}, which adds two multiplications and two levels per gate. Hence, a depth-$d$ circuit consumes $3d$ levels. Second, these extra levels inflate the ciphertext modulus: at $d{=}7$, the required $\log_2 Q{=}960$ exceeds the 881-bit bound for 128-bit security at $N{=}32768$~\cite{HomomorphicEncryptionSecurityStandard}, which forces $N{=}65536$. As a result, discrete CKKS evaluates an AND-16 reduction $2.8\times$ slower than arithmetic B/FV (3{,}877\,ms vs.\ 1{,}388\,ms), and it is $1.7$--$3.8\times$ slower per record than the product tree for $K{=}2$--$128$ (see Table~\ref{tab:prior_comparison}).}

\mkrev{\textbf{(3) Scheme Switching.} Scheme switching~\cite{lu2021pegasus,boura2020chimera} converts between arithmetic and Boolean representations on every operation. However, every gate converts its operands to LWE ciphertexts and evaluates a lookup table, so scheme switching pays at least one bootstrapped gate per AND, i.e., at least the per-gate TFHE latency of family (1) (see Table~\ref{tab:prior_comparison}). Moreover, Engorgio~\cite{bian2025engorgio} already outperforms scheme switching on relational predicates. Hence, scheme switching does not provide a stronger baseline for bulk bitwise operations.}

\mkrev{\textbf{(4) Multi-Input Programmable Bootstrapping.} Multi-input programmable bootstrapping (PBS)~\cite{chillotti2021programmable,bon2024optimized} adds several LWE ciphertexts and evaluates a multi-input Boolean function with one PBS, i.e., one lookup table on the sum. Bon et al.~\cite{bon2024optimized} choose the encodings such that a single PBS suffices for many functions. However, multi-input PBS suffers from two limitations. First, each LWE ciphertext encrypts one value, so the approach pays at least one bootstrapping per record and cannot exploit SIMD packing. Second, the message space must hold every possible sum (e.g., 63 values for AND-62), which requires large parameter sets or a tree of PBSs (e.g., 17 PBSs per record for AND-64 with at most 5 bits per PBS). As a result, multi-input PBS is $1{,}496$--$4{,}002\times$ slower per record than the product tree for $K{=}8$--$64$ (see Table~\ref{tab:prior_comparison}).}

\begin{table}[t]
\centering
\caption{\mkrev{Latency per record (single thread, median of four runs) of one bulk AND-$K$ with four families of prior approaches and with our arithmetic baseline. B/FV rows use $N{=}2^{15}$ and their own minimum number of 57-bit limbs.}}
\label{tab:prior_comparison}
\footnotesize
\setlength{\tabcolsep}{2.5pt}
\resizebox{\columnwidth}{!}{%
\begin{tabular}{@{}llrrrrr@{}}
\toprule
\textbf{Approach} & \textbf{Ops.\ per AND-$K$} & $K{=}2$ & $K{=}8$ & $K{=}16$ & $K{=}64$ & $K{=}128$ \\
\midrule
(1) Batched Boolean~\cite{chillotti2017faster} & $K{-}1$ gates & 15.0\,ms & 120\,ms & 306\,ms & 1.21\,s & 2.28\,s \\
(2) Discrete CKKS~\cite{drucker2024bleach} & mult.\ + cleanup & 4.0\,$\mu$s & 63\,$\mu$s & 184\,$\mu$s & 1.19\,ms & 3.40\,ms \\
(3) Scheme switching~\cite{lu2021pegasus} & conv.\ + LUTs & 15.0\,ms & 120\,ms & 306\,ms & 1.21\,s & 2.28\,s \\
(4) Multi-input PBS~\cite{bon2024optimized} & PBS tree & 25.6\,ms & 91.7\,ms & 154\,ms & 534\,ms & 1.19\,s \\
\midrule
\textbf{Product tree (baseline)} & $K{-}1$ $\mathsf{HMul}$ & 2.4\,$\mu$s & 22.9\,$\mu$s & 65.8\,$\mu$s & 357\,$\mu$s & 902\,$\mu$s \\
\bottomrule
\end{tabular}%
}
\end{table}

\takeaway{\label{sec:take1}\mkrev{The product tree of packed arithmetic B/FV is the fastest existing approach for bulk bitwise operations, so we adopt it as our baseline.}}

\DIFaddend \section{Motivation and Our Goal}
\label{sec:motivation}

\begin{figure}[h]
    \centering
    \includegraphics[width=\linewidth]{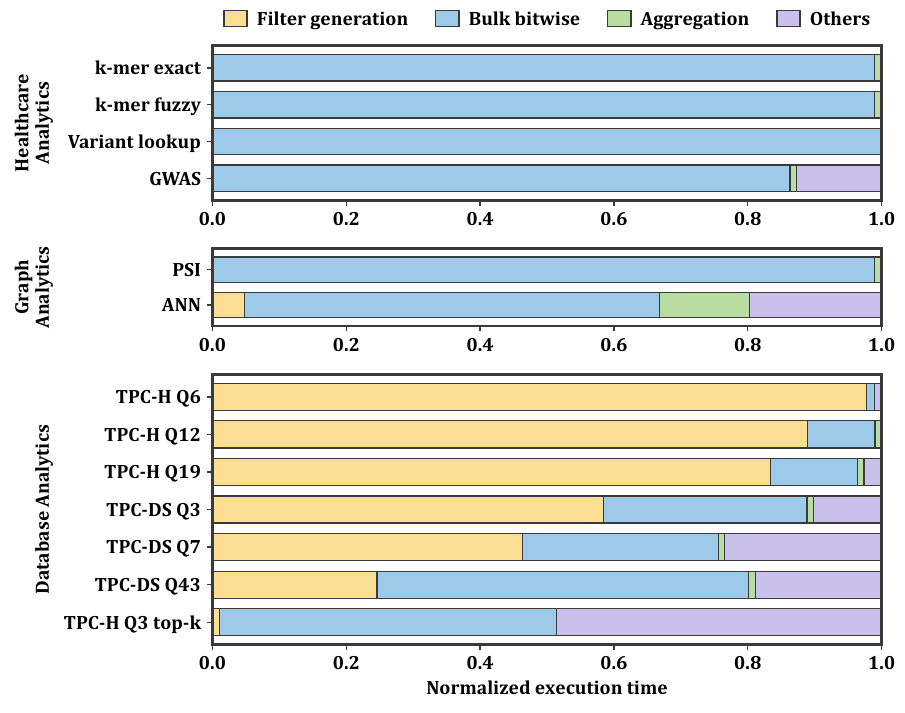}
    \caption{Contribution of HE-based Bulk Bitwise Operations in End-to-End Applications.}
    \label{fig:importance_of_bulk_bitwise}
\end{figure}

\mkar{We analyze 13 end-to-end workloads from three application classes to characterize their computational behavior and show \mktches{that bulk bitwise operations are prevalent across these application classes}:
(i)~healthcare analytics, with four genomic workloads: exact and fuzzy $k$-mer matching~\cite{bed2016privately,park2019secure}, variant lookup~\cite{fiume2019federated,titus2018sig}, and a genome-wide association study (GWAS)~\cite{blatt2020secure};
(ii)~graph analytics, with bit-sliced private set intersection (PSI)~\cite{cheon2015search,kim2016efficiency} and approximate nearest-neighbor (ANN) search~\cite{indyk1998approximate,henzinger2023tiptoe};
and (iii)~database analytics, with three TPC-H queries (Q6, Q12, Q19)~\cite{tpch}, three TPC-DS queries (Q3, Q7, Q43)~\cite{nambiar2006making}, and the TPC-H Q3 top-10 ranking query~\cite{cetin2015depth}.}
We provide more details in \S\ref{sec:workloads}.
Figure~\ref{fig:importance_of_bulk_bitwise} shows the breakdown of total execution time between HE-based bulk bitwise operations and other HE operations across all end-to-end HE-based applications.
We make \mktches{three} key observations.

\squishlist
\item \mkar{First, bulk bitwise operations dominate the execution time of the healthcare and graph workloads: they take 97.6--100\% of the time of $k$-mer matching, variant lookup, and PSI, 87.3\% of GWAS, and 62.0\% of ANN search. This is because the arithmetic approach evaluates these wide AND, OR, and threshold reductions with costly $\mathsf{HMul}$s.}
\item \mkar{Second, in TPC queries, online filter generation (i.e., homomorphic comparison) often dominates: it takes 84.3--98.1\% of the time of TPC-H Q6, Q12, and Q19. The bulk bitwise share varies from 1.3\% (TPC-H Q6) to 56.5\% (TPC-DS Q43), depending on how many predicates a query composes.}
\item \mkar{Third, ranking queries also reduce to a bulk bitwise operation. \mktches{For example, TPC-H Q3 selects its top-10 orders as follows. Within a group of 64 candidate orders, it compares each order's revenue with the revenues of the other 63 orders, which yields 63 encrypted comparison bits, and adds 10 copies of a bit that marks whether the order fails the query's filter. It then checks whether at most 9 of these 73 bits are set, i.e., whether a qualifying order ranks in the top 10. This check is a THRESHOLD reduction over 73 bits, which takes} 51.3\% of the arithmetic approach's execution time (\S\ref{sec:workloads}).}
\squishend

\takeaway{\mktches{Bulk bitwise operations are prevalent in many important HE applications and take 62--100\% of the execution time of the healthcare and graph workloads. Prior works optimize other stages of these applications (e.g., data packing, string matching, comparison, and sorting) but still evaluate bulk bitwise operations with costly $\mathsf{HMul}$-based product trees. Hence, reducing $\mathsf{HMul}$s in bulk bitwise operations accelerates these applications.}}

\hgsp{\textbf{Our goal} is to enable fast bulk bitwise operations on encrypted data for end-to-end applications by designing \pw{a} new HE algorithm that minimizes reliance on high-latency homomorphic operations (e.g., $\mathsf{HMul}$) and constructs bulk bitwise operations using low-cost homomorphic operations.}

\mksp{\textbf{Threat model and scope.}} 
\mkrev{We assume the standard semi-honest (honest-but-curious) setting~\cite{lu2021pegasus,zhang2024arcedb,bian2025engorgio}. The key holder (the client) encrypts its data under B/FV and uploads it to a server that faithfully executes the requested circuit. The structure of each circuit is public, and its execution (operations, memory accesses, ciphertext sizes, and timing) depends only on the query and the database shape, never on encrypted values. The server may also hold plaintext operands (e.g., the reference $k$-mers or the PSI server items). \mktches{Both approaches pay the same standard costs, which our measurements exclude: noise flooding for circuit privacy when the server's operands are private~\cite{chen2017fast}, and a negligible decryption-failure probability if decrypted results reach the server~\cite{cheon2024attacks}. After decryption, the key holder learns only its own per-record results or, for PSI, only the intersection cardinality.}}

\section{\titlenamed: An Algorithm for Fast Homomorphic Bulk Bitwise Operations}
\label{sec:overview}

\mkar{We propose \textbf{\titlenamed}, whose key idea is to construct bulk bitwise operations using low-cost homomorphic addition ($\mathsf{HAdd}$), which reduces costly homomorphic multiplications ($\mathsf{HMul}$s).}

\mkar{\textbf{Key Insight.} A bulk bitwise operation applies the same bitwise operator $\star$ to many operands, i.e., $a_1 \star a_2 \star \ldots \star a_K$. Such a sequence is a symmetric Boolean function~\cite{canteaut2005symmetric}, whose output depends only on the Hamming weight $\sum_{i} a_i$ of its inputs~\cite{canteaut2005symmetric}. \titlename exploits this property in two steps. First, it computes the Hamming weight with $\mathsf{HAdd}$, which operates slot-wise on binary inputs $a_i \in \{0,1\}$. \mkrev{Second, it decodes the Boolean result from the sum with one indicator polynomial, or, when revealing the sum is acceptable, reads it from a fixed bit position.} Because one $\mathsf{HAdd}$ operates on all $N$ slots of a ciphertext at once, \titlename evaluates a bulk bitwise operation over $N$ rows in parallel, using only $\mathsf{HAdd}$s and at most one indicator evaluation per reduction.}

\mkar{We present \titlename with the B/FV scheme (see \S\ref{sec:BFV_background}).}\footnote{\mkrev{\titlename targets leveled B/FV without bootstrapping: each symmetric gate (AND, OR, XOR, or THRESHOLD of any fan-in) costs one indicator, so a circuit of $\ell$ nested symmetric layers (e.g., the OR-3 inside the AND-8 of GWAS) costs $\ell$ indicators, and its total depth must fit the modulus at $N{=}32768$. Deeper circuits require B/FV bootstrapping~\cite{chen2018homomorphic}, which \titlename does not affect.}}
\label{subsec:hulk_algorithm}
\mkar{\titlename evaluates a bulk bitwise operation in three steps: (1) it encodes each bitvector into the slots of a B/FV plaintext, (2) it sums the encrypted operands with $\mathsf{HAdd}$ to compute their Hamming weight, and (3) it decodes the Boolean result from the Hamming weight, on either the client or the server. \S\ref{sec:encoding}, \S\ref{sec:sum_indicate}, and \S\ref{sec:decoding} describe these three steps, and \S\ref{sec:security_analysis} analyzes the security, correctness, and noise growth of \titlenamed.}

\begin{figure}[h]
    \centering
    \includegraphics[width=\linewidth]{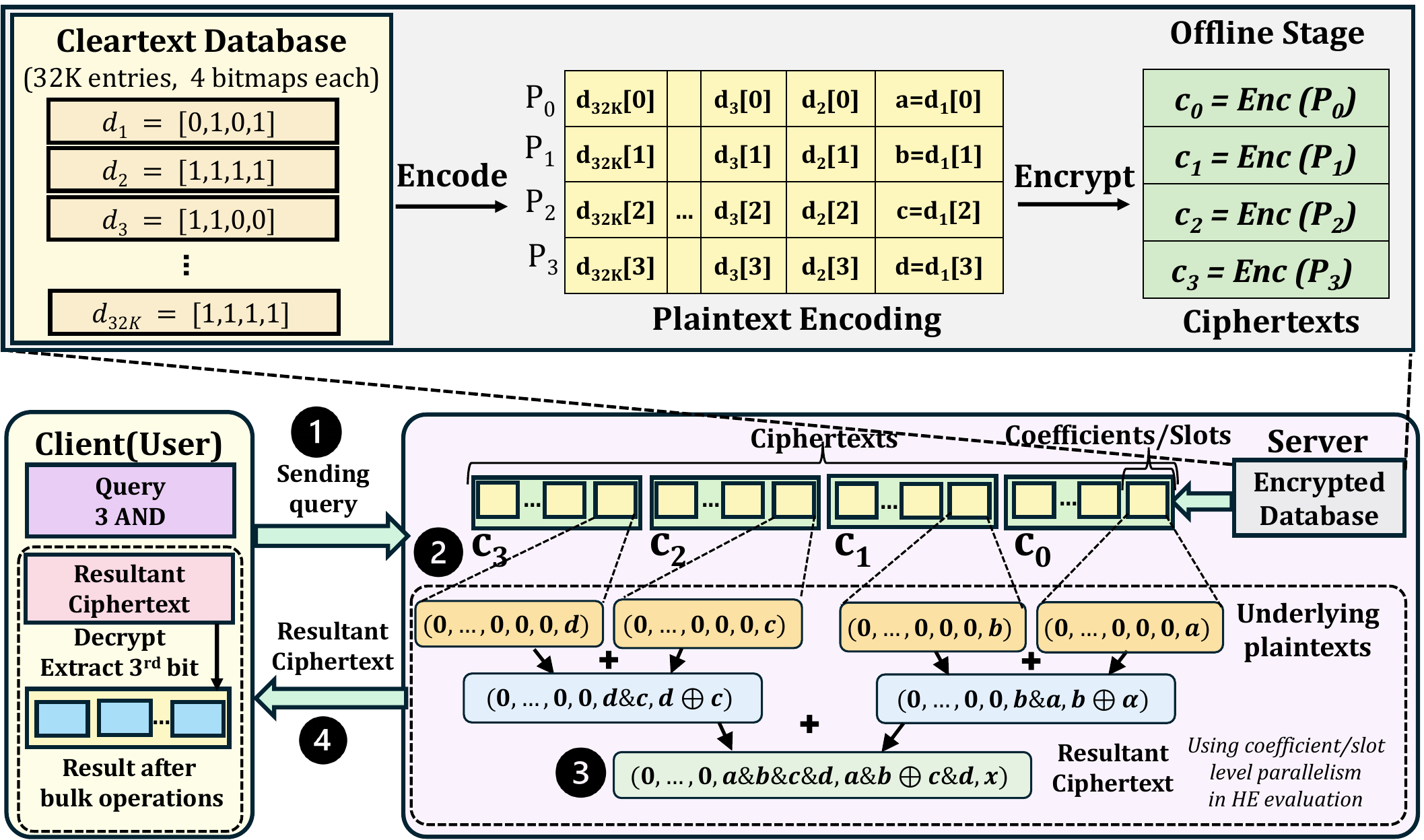}
    \caption{\titlename pipeline and encoding. Bitmaps $d_i$ encode all columns per row and are transposed so that row $i$ is in slot $i$.}
    \label{fig:packing_figure}
\end{figure}

\mkar{\textbf{End-to-End Pipeline.} Figure~\ref{fig:packing_figure} shows how \titlename evaluates a client query that ANDs four operands. Offline, the data owner computes each bitvector in plaintext, places row $i$ in slot $i$, encrypts the bitvector, and stores the resulting ciphertexts on the server. \circled{1} The client sends a query that selects a public circuit over the stored bitvectors. \circled{2} The server computes $S$ with slot-wise $\mathsf{HAdd}$s. The intermediate sums stay slot-aligned, and the number of bits reserved per slot grows only as $\lceil\log_2K\rceil+1$. \circled{3} The server adds $\beta$ (client-side decoding) or evaluates $I(S)$ (server-side decoding). \circled{4} \mktches{In client-side decoding, the client decrypts the result and extracts the result from bit $W{-}1$ of each slot.} Hence, \titlename uses only $\mathsf{HAdd}$s and one indicator per reduction.}

\subsection{Encoding}
\label{sec:encoding}

\mkar{\titlename uses the B/FV slot (SIMD) encoding~\cite{smart2014fully}. In this encoding, a plaintext is a vector of $N$ slots over $\mathbb{Z}_p$, and a $\mathsf{HAdd}$ or $\mathsf{HMul}$ adds or multiplies two ciphertexts slot-wise respectively. \titlename places row $i$ of a bitvector in slot $i$. Hence, one ciphertext holds $N$ rows of one bitvector (column-major storage format), and $K$ bitvectors occupy $K$ ciphertexts.}

\subsubsection{Sum, then Indicator}
\label{sec:sum_indicate}
\mkar{Let $a_1,\dots,a_K\in\{0,1\}$ be the $K$ encrypted operand bits \mktches{at the same logical slot position}. \titlename first computes their Hamming weight $S$ homomorphically:
\begin{equation}
S=\sum_{j=1}^{K}a_j\in\{0,\dots,K\}
\label{eq:hamming}
\end{equation}
This step uses $K{-}1$ $\mathsf{HAdd}$s and no $\mathsf{HMul}$. The output of every symmetric Boolean function of $a_1,\dots,a_K$ depends only on $S$~\cite{canteaut2005symmetric}. Hence, \titlename expresses AND, OR, and XOR as indicators on $S$:
\begin{equation}
\mathrm{AND}=[S=K],\qquad \mathrm{OR}=[S\ge1],\qquad \mathrm{XOR}=S\bmod 2
\label{eq:decode}
\end{equation}
}

\mkar{\titlename replaces the $K{-}1$ $\mathsf{HMul}$s of the arithmetic approach's product tree with $K{-}1$ $\mathsf{HAdd}$s and \emph{one} indicator on $S$. Since one $\mathsf{HAdd}$ is $\sim$300$\times$ faster than one $\mathsf{HMul}$ on our CPU system (see \S\ref{sec:reduction}), this significantly improves the performance of HE-based bitwise reduction.}

\subsection{Decoding}
\label{sec:decoding}
\mkar{\titlename provides two decoding mechanisms, and each application selects the mechanism that matches its privacy requirement: (1) client-side decoding and (2) server-side decoding. Table~\ref{tab:homadd_logic} summarizes both mechanisms for AND, OR, and XOR.}

\subsubsection{Client-Side Decoding}
\label{sec:client_decoding}
\mkar{When revealing $S$ to the decryptor is acceptable \mkrev{(its leakage is $\mathcal{L}=S$, see \S\ref{sec:security_analysis})}, the client decrypts $S$ and applies Eqn.~(\ref{eq:decode}) in plaintext to each slot, reading each operator from the bit. 
\titlename \mktches{represents $S$ with} $W=\lceil\log_2K\rceil+1$ bits \mktches{per slot}. \mkrev{To read every operator from one fixed bit, the server adds a constant $\beta$ to $S$ (Table~\ref{tab:homadd_logic}).}}

\subsubsection{Server-Side Decoding}
\label{sec:server_decoding}
\label{sec:extension}
\mkar{Addition-only workloads (e.g., fast scan~\mktches{\cite{li-sigmod-2013}} and k-clique star listing~\cite{danisch2018listing}) need no $\mathsf{HMul}$. However, end-to-end applications often need additional operations (e.g., aggregation or popcount) on the encrypted result. When the result must remain encrypted for such operations, or when $S$ leaks more than the ideal functionality (e.g., the number of matches in PSI~\cite{pinkas2019efficient,chandran2022circuit}), the server evaluates the indicator homomorphically. Since $S$ takes only the $K{+}1$ values $D=\{0,\dots,K\}$, the server evaluates the unique polynomial $I(x)$ that interpolates the target indicator on these $K{+}1$ points. Let $T\subseteq D$ be the values on which the indicator equals 1, i.e., $T=\{K\}$ for AND, $T=\{1,\dots,K\}$ for OR, and the odd values of $D$ for XOR. \mkrev{We call the evaluation of $I(x)$ \emph{indicator evaluation}; prior work calls this step bit or digit extraction~\cite{chen2018homomorphic}.} We construct $I(x)$ with Lagrange interpolation~\cite{chen2018homomorphic,huo2019note,mouchet2020lattigo}:
\begin{equation}
   \label{eq:eq3}
I(x)=\sum_{u\in T}\ell_u(x),\qquad
\ell_u(x)=\prod_{\substack{v\in D\\v\neq u}}\frac{x-v}{u-v}
\end{equation}
$I(x)$ is exact on $D$, and its degree is exactly $K$ for AND, OR, and XOR, for every $K$.
This construction handles two cases.} 

\mkar{\textbf{Example.} For AND-3, $D=\{0,1,2,3\}$ and $T=\{3\}$:
\begin{equation}
    \label{eq:eq4}
I(x)=\ell_3(x)=\frac{x(x-1)(x-2)}{6}=\frac{1}{6}x^3-\frac{1}{2}x^2+\frac{1}{3}x,
\end{equation}
which returns 1 for $S{=}3$ and 0 for $S\in\{0,1,2\}$. Both forms are identical on $D$, but the expanded form has fewer active monomials and is therefore cheaper to evaluate homomorphically once the powers of $x$ are precomputed.}

\mkar{\emph{Power-of-two $K$.} Since $S\le K=2^{W-1}$, the AND result is the carry-out bit $W{-}1$ of $S$, and $I(x)$ extracts this bit.} 

\mkar{\emph{Non-power-of-two $K$.} The AND result is \emph{not} a single bit of $S$ (e.g., for $K{=}5$, $S{=}4$ and $S{=}5$ share bit 2), so a bit extractor alone is incorrect. $I(x)$ avoids this problem, because it is defined on the values of $S$, not on its bits, and thus needs no shift $\beta$. In both cases, $I(x)$ has degree $K$ (e.g., 5 for AND-5), instead of degree $2^{W}{-}1$ (15 for AND-5) for a generic extractor over all $2^W$ values of the reserved bits. The server evaluates $I(x)$ with a root-paired factorization for AND and OR with $K\le16$, and with the Paterson--Stockmeyer algorithm~\cite{paterson1973number} otherwise (see \S\ref{sec:ps_background}). Table~\ref{tab:homadd_logic} summarizes both cases.}

\begin{table}[h]
\centering
\footnotesize
\setlength{\tabcolsep}{6pt}
\renewcommand{\arraystretch}{1.2}
\caption{\mkrev{Decoding AND, OR, and XOR from the Hamming weight $S$ of $K$ bits. $W=\lceil\log_2K\rceil+1$ bits per slot; $\beta$: plaintext shift for client-side decoding.}}

\begin{tabular}{@{}llll@{}}
\toprule
\textbf{Op} & \textbf{Indicator on $S$} & \textbf{Shift $\beta$} & \textbf{Bit read} \\
\midrule
AND & $[S=K]$   & $2^{W-1}-K$ & $W{-}1$ of $S{+}\beta$ \\
OR  & $[S\ge1]$ & $2^{W-1}-1$ & $W{-}1$ of $S{+}\beta$ \\
XOR & $S\bmod2$ & $0$         & $0$ of $S$ \\
\bottomrule
\end{tabular}
\label{tab:homadd_logic}
\end{table}

\mkar{\textbf{Evaluating $I(x)$.} A naive evaluation of $I(x)$ requires $d{=}K$ $\mathsf{HMul}$s, which is no better than the $K{-}1$ $\mathsf{HMul}$s of the arithmetic product tree. \titlename reduces this cost in two ways.}

\mkar{\textit{Optimization 1.} $I(x)$ is a univariate polynomial of one ciphertext. Hence, we evaluate it with the Paterson--Stockmeyer (PS) algorithm~\cite{paterson1973number} (see \S\ref{sec:ps_background}), which reduces its $\mathsf{HMul}$ count from $O(d)$ to $O(\sqrt{d})$. For the AND and OR indicators, whose roots are the consecutive integers $\{0,\dots,K{-}1\}$ (resp.\ $\{1,\dots,K\}$), we also use a \emph{root-paired} factorization. \mktches{For AND, whose indicator has the roots $\{0,\dots,K{-}1\}$, pairing roots $j$ and $K{-}1{-}j$ gives $(x-j)(x-(K{-}1{-}j))=y+j(K{-}1{-}j)$ with $y=x(x{-}K{+}1)$; for OR, whose indicator $1-\ell_0$ has the roots $\{1,\dots,K\}$, pairing roots $j$ and $K{+}1{-}j$ gives $y'+j(K{+}1{-}j)$ with $y'=x(x{-}K{-}1)$. \titlename multiplies the resulting $\lceil K/2\rceil$ factors, which are linear in $y$, pairwise in a balanced tree and reuses $y^2$ across pairs (e.g., for $K{=}8$, $y(y{+}12)$ and $(y{+}6)(y{+}10)$ are both linear in $y$ and $y^2$), at depth $1{+}\lceil\log_2\lceil K/2\rceil\rceil$.} This factorization beats PS up to $K{=}16$, with fewer multiplications and a shallower depth (3 vs.\ 4 $\mathsf{HMul}$s at $K{=}8$), and loses above it (20 vs.\ 11 $\mathsf{HMul}$s at $K{=}40$, at depth 6 vs.\ 9).}

\mksp{\textit{Optimization 2.} A naive \titlename implementation performs a general homomorphic multiplication ($\mathsf{HMul}$) for every indicator and recomputes every shared predicate independently. \titlename reduces this cost by employing \emph{squaring reuse.} Every indicator computes $s^2$ on a single ciphertext, which the naive approach treats as a general multiplication of two distinct operands. Instead, \titlename forward-transforms (NTT) the operand once and reuses it for both inputs, making the squaring operation $\approx$$32\%$ faster than general multiplication as measured on a CPU system.} 

\mkrev{\textit{Fermat Indicator.} $I(x)$ need not have the minimal degree. For AND and OR, a Fermat zero test on $S{-}K$ (AND) or on $S$ (OR)~\cite{cetin2015arithmetic} is also exact on $D$ and costs a fixed 16 squarings at $p{=}65537$, independent of $K$. We profile both indicators and observe that they break even at $K{\approx}128$ (188 vs.\ 192\,$\mu$s per record, single thread), and that the Fermat indicator is $1.9\times$ faster at $K{=}256$. Hence, \titlename uses the Fermat indicator for a single AND or OR with $K\ge128$ when the application fits $p{=}65537$, and the Lagrange indicator otherwise. None of our end-to-end applications meets these conditions (\S\ref{sec:workloads}).}

\mkar{\textbf{Key Insight.} Server-side decoding costs \emph{one} degree-$K$ indicator per reduction, i.e., $O(\sqrt{K})$ $\mathsf{HMul}$s, whereas the arithmetic approach's product tree costs $K{-}1$ $\mathsf{HMul}$s. For example, \titlename needs 3 instead of 7 $\mathsf{HMul}$s for $K{=}8$, and 11 instead of 39 for $K{=}40$.}

\subsection{Cost Model} \titlenamed's cost is governed by a single parameter, i.e., the sequence of bulk bitwise operations (i.e., homomorphic addition) $K$. An AND-$K$ (equivalently OR-$K$ or XOR-$K$) reduction reserves
$w=\lceil\log_2 K\rceil+1$ bits per slot and places its result at bit $\lceil\log_2 K\rceil$.
This width $w$, in turn, fixes the remaining cost in two ways. First, the
\emph{limb count}: since an AND-$K$ sum takes only $K{+}1$ distinct values, \mkrev{indicator evaluation}
interpolates a degree-$d$ polynomial with $d=K$, whose multiplicative depth is \mkrev{$d_{\mathrm{RP}}$ or $d_{\mathrm{PS}}$ (see \S\ref{sec:noise_budget}), i.e., $O(\log K)$}. Sustaining this depth fixes the ciphertext modulus $Q$, and
hence the limb count. Second, the \emph{extraction $\mathsf{HMul}$ count}: where the Paterson-Stockmeyer evaluation (\S\ref{sec:extension})
reduces the number of $\mathsf{HMuls}$ from K-1 to $\lfloor 2\sqrt{K{+}1}\rfloor-1$, i.e.\ $O(\sqrt{K})$.

\subsubsection{Client-Side Cost}
\mkar{\titlename adds no homomorphic \mktches{operations} to the client relative to standard B/FV. First, encoding uses the standard B/FV slot encoding. Second, encryption and decryption use the unmodified B/FV routines. Third, client-side decoding applies Eqn.~(\ref{eq:decode}) in plaintext in each slot, which costs $O(1)$ per slot and needs no evaluation keys. Under server-side decoding, the client receives a 0/1 mask and reads it directly. Hence, the client's cost equals that of the arithmetic baseline. 
}

\subsection{Security, Correctness, and Noise Analysis}
\label{sec:security_analysis}

\subsubsection{Security Guarantees}
\mkrev{\titlename inherits the IND-CPA security of B/FV under the RLWE assumption~\cite{fan2012somewhat,brakerski2014leveled,lyubashevsky2010ideal}: it only adds, multiplies, and rotates ciphertexts with standard B/FV operations and introduces no new assumption. We use a uniform ternary secret, a discrete Gaussian error with standard deviation $\approx3.2$, and $N{=}2^{15}$, for which the HE security standard admits $\log_2(PQ)\le881$ bits at $\lambda{=}128$~\cite{HomomorphicEncryptionSecurityStandard}; our largest modulus, including the special prime, is $13\times57=741$ bits. Every workload uses relinearization keys, and PSI and TPC-H Q3 top-10 additionally use Galois keys for slot rotations. Any plaintext shift (e.g., $m\mapsto2^{j}m\bmod p$) is applied before encryption, so a ciphertext reveals nothing about which bit position it encodes. Client-side decoding (\S\ref{sec:client_decoding}) reveals the Hamming weight $S$ of each slot to the key holder (leakage $\mathcal{L}=S$); server-side decoding reveals only the Boolean result.}

\subsubsection{Correctness Guarantees}
\label{sec:noise_budget}
\mkrev{A \titlename evaluation is correct if the decrypted and decoded result equals the intended Boolean function in every slot. This requires a structural condition (Lemma~1) and a cryptographic condition: B/FV decryption is exact when the ciphertext noise satisfies $\lVert e\rVert_\infty<\Delta/2$~\cite{fan2012somewhat}, which the noise analysis below ensures.}

\mkrev{\textbf{Lemma 1 (Decoding correctness).} Let $S=\sum_{j=1}^{K}a_j$ with $a_j\in\{0,1\}$, $W=\lceil\log_2K\rceil+1$, and plaintext modulus $p>2^{W}$. (i) \emph{Client-side:} bit $W{-}1$ of $S+\beta$ equals $[S=K]$ for $\beta=2^{W-1}-K$ and $[S\ge1]$ for $\beta=2^{W-1}-1$, and bit 0 of $S$ equals $S\bmod2$. \mktches{This also holds per lane when several lanes share a slot (\S\ref{sec:memory-efficient-dp}), since each lane has at least $W$ bits.} (ii) \emph{Server-side:} for a prime $p>K$ and any target set $T\subseteq D=\{0,\dots,K\}$, the polynomial $I(x)$ of Eqn.~(\ref{eq:eq3}) has degree at most $K$ and satisfies $I(x)=[x\in T]$ for all $x\in D$.}

\mkrev{\emph{Proof.} (i) Since $0\le S\le K\le2^{W-1}$, $S+\beta$ lies in $[\beta,\,\beta+K]$ and is below $2^{W}\le p$, so no wrap-around occurs. For AND, $S+\beta\ge2^{W-1}$ if and only if $S=K$; for OR, if and only if $S\ge1$. Bit 0 of $S$ is its parity. (ii) Each Lagrange basis polynomial $\ell_u$ is well defined because its denominators $u-v$ with $u,v\in D$ are non-zero modulo the prime $p>K$; $\ell_u$ has degree $K$ and equals $[x=u]$ on $D$, so their sum over $T$ equals $[x\in T]$ on $D$. The root-paired factorization of \S\ref{sec:server_decoding} follows from $(x-j)(x-(K{-}1{-}j))=x(x-K+1)+j(K-1-j)$ \mktches{for AND, and from $(x-j)(x-(K{+}1{-}j))=x(x-K-1)+j(K+1-j)$ for the OR indicator $1-\ell_0$}.}

\subsubsection{Noise Budget Analysis}
Let $B_{\mathrm{fresh}}$ bound fresh noise and $B_{\mathrm{ks}}$ the key-switching noise. \titlename has three noise phases.

 \mkar{\textbf{Addition.} $\mathsf{HAdd}$ adds noise: after summing $K$ fresh ciphertexts, $\lVert e_S\rVert_\infty\le K\cdot B_{\mathrm{fresh}}$, i.e., $\log_2K$ extra bits and no consumed level. Client-side decoding stops here, so its correctness requires only $K\cdot B_{\mathrm{fresh}}<\Delta/2$.}

 \textbf{Indicator evaluation.} Evaluating $I(S)$ of degree $K$ by Paterson--Stockmeyer uses multiplicative depth $d_{\mathrm{PS}}=\lceil\log_2b\rceil+\lceil\log_2g\rceil$ with $b=\lceil\sqrt{K}\rceil$ and $g=\lceil K/b\rceil$; the root-paired factorization uses $d_{\mathrm{RP}}=1+\lceil\log_2\lceil K/2\rceil\rceil$. \mkrev{\mktches{As an approximate model,} each B/FV $\mathsf{HMul}$ grows the noise by a factor of roughly $p\,\delta_R$, where $\delta_R$ is the ring expansion factor, plus a key-switching term~\cite{fan2012somewhat}, so the noise grows exponentially with the multiplicative depth $d$, starting from $K\,B_{\mathrm{fresh}}$ after the additions\mktches{; we validate the selected parameters empirically for every workload (\S\ref{sec:methodology})}.} This is \titlenamed's only source of multiplicative noise and determines the limb count. The product tree has depth $\lceil\log_2K\rceil$ over fresh operands, whereas \titlenamed's indicator has depth $d_{\mathrm{RP}}$ or $d_{\mathrm{PS}}$ over an operand whose noise is already $K\,B_{\mathrm{fresh}}$; for $K\le8$ the depths coincide and the extra $\log_2K$ bits cost at most one limb.

 \textbf{Downstream operations.} Multiplying the 0/1 mask by a value column and summing across blocks adds one level and $\log_2(\#\mathrm{blocks})$ bits, identically in both approaches. \mkrev{Table~\ref{tab:workload_sizes_limbs_32gib_all16} reports the resulting limb counts and noise margins.}

\section{Methodology}
\label{sec:methodology}

Our evaluation methodology consists of two key components: (1) a real system evaluation methodology to assess the performance improvement of \titlename algorithm (\S\ref{sec:real_System}), (2) details of the \mkrev{evaluated end-to-end applications} (\S\ref{sec:workloads}).

\subsection{Real System Evaluation Methodology}
\label{sec:real_System}

\textbf{Evaluated Systems.} We evaluate the software-based \titlename algorithm implemented using Microsoft SEAL~\cite{sealcrypto} and compare its performance against
the prior arithmetic approaches~\cite{chen2018labeled,bian2025engorgio,park2019secure} using Microsoft SEAL. We evaluate \titlename on GPU using \pw{the} HEonGPU library~\cite{ozcan2024heongpu}. For the Boolean approach, we use \mktches{Zama's TFHE-rs library (v1.7.0)}~\cite{TFHE-rs}\mktches{, whose bootstrapping is about $1.8\times$ faster than that of OpenFHE~\cite{xiao2025gpu}}. 

\mkrev{\textbf{Real CPU and GPU Systems.} Table~\ref{tab:config1} lists our CPU and GPU systems. We run all CPU experiments on an AMD EPYC 7742 CPU and GPU experiments on NVIDIA A100 GPU-80\,GB. Both systems use a PCIe 4.0 NVMe SSD.}

\mkrev{\textbf{Performance Measurements.} For the end-to-end applications, we measure the latency of the homomorphic evaluation only (excluding encoding, encryption, decryption).}

\begin{table}[h]
  \centering
  \footnotesize
  \caption{Real system configuration.}
  \setlength{\tabcolsep}{6pt}
  \renewcommand{\arraystretch}{1.25}
  \begin{tabularx}{\columnwidth}{@{}>{\arraybackslash}p{1.85cm} X@{}}
    \toprule
    \textbf{Config.} & \textbf{Details} \\
    \midrule
    \multirow{6}{*}{\makecell[l]{\textbf{CPU:}\\\textbf{AMD EPYC}\\\textbf{7742}}}
    & \textit{Microarchitecture:} AMD Zen 2~\cite{suggs2020amd} \\
    & x86-64~\cite{jowani2010x8664, gillespie2014steam}, 64~cores, out-of-order, 2.25~GHz base / 3.4~GHz boost \\
    & \emph{L1 Data + Inst. Cache:} 32~kB, 8-way, 64~B line \\
    & \emph{L2 Private Cache:} 512~kB, 4-way, 64~B line \\
    & \emph{L3 Shared Cache:} 256~MB, 16-way, 64~B line \\
    & \emph{Main Memory:} 64~GB DDR4-3200, 8~channels \\
    \midrule
    \multirow{3}{*}{\makecell[l]{\textbf{GPU:}\\\textbf{NVIDIA A100}}}
    & \textit{Architecture:} Ampere~\cite{choquette2021nvidia} \\
    & \textit{Streaming Multiprocessors:} 108 \\
    & \emph{Memory:} 80~GB \\
    \midrule
    \textbf{Storage} & Samsung 980 Pro PCIe 4.0 NVMe SSD 2~TB~\cite{samsung-980pro} \\
    \midrule
    \textbf{OS} & Ubuntu 22.04.1 LTS \\
    \bottomrule
  \end{tabularx}
  \label{tab:config1}
\end{table}

\subsection{Workloads}
\label{sec:workloads}
\mkar{We evaluate \titlename on 13 end-to-end applications from three application domains \mktches{in which bulk bitwise operations are prevalent}: healthcare analytics (four genomic workloads), graph analytics (bit-sliced private set intersection and approximate nearest-neighbor search), and database analytics (three TPC-H and three TPC-DS filter-and-aggregate queries, and the TPC-H Q3 top-10 ranking query). Every application runs end-to-end under encryption: the server receives encrypted inputs, performs every step under encryption, and returns an encrypted application result. Table~\ref{tab:workload_sizes_limbs_32gib_all16} summarizes the bulk bitwise structure, the number of encrypted input columns, the limb count, and the encrypted database size of each application. \mkrev{The database size assumes a fixed 512\,MB cleartext input with one bit per column per row, i.e., $2^{32}/N$ ciphertexts of $2N(\text{limbs}-1)\times8$\,B (the special prime is not stored); TPC-H Q3 top-10 stores each shard twice.}}

\begin{table*}[!htbp]
\centering
\footnotesize
\caption{\mkrev{End-to-end applications (A/H: arithmetic approach/\titlename). \textbf{Structure}: bulk bitwise reductions per record; \textbf{Limbs}: 57-bit primes (CPU); \textbf{$\mathsf{HMul}$s}: including squarings, per ciphertext block; \textbf{Noise}: minimum remaining budget; \textbf{Enc.\ DB}: \mktches{encrypted database} size for 512\,MB of cleartext.}}

\setlength{\tabcolsep}{4pt}
\renewcommand{\arraystretch}{1.15}
\begin{tabular}{@{}l p{5.6cm} r cc rr cc cc@{}}
\toprule
& & & \multicolumn{2}{c}{\textbf{Limbs}} & \multicolumn{2}{c}{\textbf{$\mathsf{HMul}$s}} & \multicolumn{2}{c}{\textbf{Noise (bits)}} & \multicolumn{2}{c}{\textbf{Enc.\ DB (GB)}} \\
\cmidrule(lr){4-5}
\cmidrule(lr){6-7}
\cmidrule(lr){8-9}
\cmidrule(lr){10-11}
\textbf{Workload} & \textbf{Structure} & \textbf{Enc.\ cols} & \textbf{A} & \textbf{H} & \textbf{A} & \textbf{H} & \textbf{A} & \textbf{H} & \textbf{A} & \textbf{H} \\
\midrule
\multicolumn{11}{@{}l}{\emph{Healthcare analytics}} \\
Exact $k$-mer & 16 $\times$ AND-62, then OR-16; 1 count & 62 & 9 & 10 & 991 & 264 & 26 & 31 & 550 & 618 \\
Fuzzy $k$-mer & 16 $\times$ THRESHOLD-31 ($\ge$29), then OR-16; 1 count & 124 & 8 & 9 & 1,599 & 184 & 6 & 36 & 481 & 550 \\
Variant lookup & 32 $\times$ AND-37, then plaintext-weighted label sum; 1 per-record output & 37 & 7 & 7 & 1,152 & 416 & 54 & 28 & 412 & 412 \\
GWAS & per patient and SNP: OR-3, AND-8 (case), AND-9 (control); then weighting, sum over 128 patients, $2{\times}2$ table & 1,536 & 7 & 8 & 2,052 & 1,540 & 29 & 38 & 412 & 481 \\
\midrule
\multicolumn{11}{@{}l}{\emph{Graph analytics}} \\
Bit-sliced PSI & 16 $\times$ AND-40 (one per server item), then OR-16; 1 cardinality & 40 & 9 & 10 & 639 & 216 & 27 & 54 & 550 & 618 \\
ANN search & 8 $\times$ AND-12, then OR-8 (candidate); 64 XORs (Hamming distance); $[h\le10]$; top-10 retrieval; 190 rotations & 168 & 9 & 9 & 357 & 314 & 13 & 35 & 550 & 550 \\
\midrule
\multicolumn{11}{@{}l}{\emph{Database analytics} (predicates generated online by comparison)} \\
TPC-H Q6 & 3 filters; 1 $\times$ AND-3; 1 \texttt{SUM} & 4 & 13 & 13 & 131 & 131 & 54 & 52 & 825 & 825 \\
TPC-H Q12 & 4 $\times$ AND-5; 4 \texttt{COUNT}s & 5 & 13 & 13 & 131 & 127 & 50 & 49 & 825 & 825 \\
TPC-H Q19 & 3 $\times$ AND-6; 3 \texttt{SUM}s & 7 & 9 & 9 & 113 & 107 & 25 & 22 & 550 & 550 \\
TPC-H Q3 top-10 & 1 filter; 63 comparisons per order; THRESHOLD-73 ($\le$9); per-order top-10 mask; 126 rotations & 2 & 12 & 12 & 1,577 & 780 & 42 & 43 & 1,512 & 1,512 \\
TPC-DS Q3 & 16 $\times$ AND-4; 16 \texttt{SUM}s & 5 & 11 & 11 & 148 & 132 & 19 & 16 & 687 & 687 \\
TPC-DS Q7 & 64 $\times$ AND-6; 320 outputs (\texttt{COUNT} + 4 \texttt{SUM}s per item) & 10 & 9 & 9 & 1,046 & 918 & 43 & 6 & 550 & 550 \\
TPC-DS Q43 & 21 $\times$ AND-4; 21 \texttt{SUM}s & 5 & 7 & 7 & 112 & 91 & 31 & 30 & 412 & 412 \\
\bottomrule
\end{tabular}
\label{tab:workload_sizes_limbs_32gib_all16}
\end{table*}

\mkar{\textbf{Healthcare Analytics.} Records are bit-sliced: ciphertext $j$ holds bit $j$ of all records, one record per slot. Against a plaintext query, the server takes $x_j$ or $1{-}x_j$ for each bit, so their sum $S$ counts the matching bits with $\mathsf{HAdd}$s only, and one indicator on $S$ answers the query. \mkrev{For each reduction, the arithmetic approach multiplies the matching terms in a product tree (AND) or builds a counting tree (THRESHOLD), whereas \titlename adds them and evaluates one indicator on $S$.}}
\squishlist
\item \mkar{\textbf{Exact $k$-mer matching}~\cite{bed2016privately,park2019secure,jahshan2024majork} flags the reads (2 bits per base) that equal any of 16 reference \mktches{31-base $k$-mers}: 16 AND-62, then an OR-16; the client counts the flagged reads.}
\item \mkar{\textbf{Fuzzy $k$-mer matching}\mkrev{~\cite{alser2020sneakysnake,cali2020genasm,essex2019secure,wei2022secure}} tolerates up to 2 mismatched bases (one column per nucleotide): 16 THRESHOLD-31 ($\ge$29), then an OR-16.}
\item \mkar{\textbf{Variant lookup}~\cite{fiume2019federated,titus2018sig} returns which of 32 pathogenic variants (synthetic 37-bit keys) a record carries: 32 AND-37, then a plaintext-weighted label sum.}
\item \mkar{\textbf{GWAS}~\cite{blatt2020secure} tests 32,768 SNPs for association with type-2 diabetes over 128 patients with encrypted genotypes and health-record flags: per patient and SNP, a case rule (an OR-3 inside an AND-8) and a control rule (an AND-9), then weighting, a sum over patients, and a $2{\times}2$ table per SNP.}
\squishend

\mkar{\textbf{Graph Analytics.}} \mkrev{Both workloads match encrypted items against a query: PSI tests exact set membership, and ANN search finds similar items. The arithmetic approach evaluates each equality or bucket match as a product tree over the matched bits and each OR as a product of complements, whereas \titlename adds the bits and evaluates one indicator per reduction. In ANN, the query is also encrypted, so each bit comparison costs one $\mathsf{HMul}$ ($q{+}d{-}2qd$) in the arithmetic approach and one squaring ($(q{-}d)^2$) in \titlename.}
\squishlist
\item \mkar{\textbf{Bit-sliced PSI}~\mktches{\cite{cheon2015search,kim2016efficiency}} computes the intersection cardinality of 32,768 encrypted 40-bit client items with 16 plaintext server items: 16 AND-40 (equality), an OR-16 (membership), and a server-side sum. The baseline uses the bit-sliced product-tree equality circuit~\cite{cheon2015search}.}
\item \mkar{\textbf{ANN search}~\cite{indyk1998approximate,charikar2002similarity,henzinger2023tiptoe} encodes an encrypted database and an encrypted query with SimHash~\cite{charikar2002similarity}. An item is a candidate if its 12-bit key matches the query in any of 8 LSH tables (AND-12, then OR-8); the server then computes the Hamming distance $h$ of 64-bit codes, keeps items with $h\le10$, and retrieves the top-10 payloads as in Tiptoe~\cite{henzinger2023tiptoe}.}
\squishend

\mkrev{\textbf{Database Analytics.} We evaluate TPC-H~\cite{council2005transaction} Q6, Q12, and Q19 and TPC-DS~\cite{nambiar2006making} Q3, Q7, and Q43 over the dictionary-coded TPC column domains, as in prior HE-based databases (e.g.,~\cite{bian2023he3db,zhang2024arcedb}, \mktches{\cite{jung2026nshedb}}). Each query generates its predicates online with \mktches{the comparison polynomial of Iliashenko and Zucca~\cite{iliashenko2021faster}, as in NSHEDB~\cite{jung2026nshedb}} (degree $D{-}1$ over the column's public domain of $D$ values\mktches{, so the limb count of each query depends on its column domains}), composes them with AND-3 to AND-6, weights each mask by the value column, and aggregates with \texttt{SUM} or \texttt{COUNT}; \texttt{ORDER BY} and \texttt{LIMIT} run on the decrypted aggregates. To understand ranking under encryption, we additionally evaluate the top-10 query of TPC-H Q3 (i.e., TPC-H Q3 top-k in Figures~\ref{fig:importance_of_bulk_bitwise} and~\ref{fig:evaluation}), which ranks orders by counting~\cite{cetin2015depth}. After an online filter $q_i=[\texttt{segment}_i=\texttt{BUILDING}]$, each order is compared with its 63 partners in a shard of 64 orders ($c_{ip}=[s_p>s_i]\cdot q_p$), and it is in the top 10 if and only if the sum $S_i$ of its 63 comparison bits and 10 copies of $1{-}q_i$ satisfies $[S_i\le9]$ (THRESHOLD-73). Both approaches share the comparison, weighting, and aggregation stages. We quantize revenue to 16 levels and omit Q3's date predicates, which exceed the depth budget without bootstrapping.}

\mkrev{\textbf{Arithmetic Baseline.} Our baseline is the arithmetic approach as implemented by the prior works of each application (see \S\ref{sec:key_related_works}): AND and OR as product trees of $\mathsf{HMul}$s (e.g.,~\mkrevb{\cite{cheon2015search,kim2016efficiency,kim2015private}}), equality as the product of the matched bits~\mkrevb{\cite{cheon2015search,mahdavi2022constant}}, and thresholds (THRESHOLD-$K$ and the top-$k$ rank test) as exact counting trees, as in counting-based encrypted sorting~\cite{cetin2015depth}. \mkrevb{Prior works also evaluate thresholds with an interpolation polynomial~\cite{kim2016efficiency} or a sum of Fermat tests~\cite{cetin2015arithmetic}; \S\ref{sec:server_decoding} compares \titlename with the Fermat alternative.} \mktches{Our baseline} reflects how the evaluated applications are implemented today. We implement each baseline circuit in the same library as \titlenamed, and both approaches share the same encoding, comparison filters, aggregation, HE parameters, limb search, and measurement protocol, so that only the bulk bitwise composition differs.}

\mkar{\textbf{HE Parameters and Setup.} We use B/FV with $N{=}32768$, a 25-bit batching prime as the plaintext modulus, and 57-bit coefficient primes, where the last prime is the special prime. One ciphertext block holds 32,768 records, one record per slot, and the latency scales linearly with the number of blocks. Each approach uses its own minimum limb count for which every output decrypts correctly against a plaintext oracle. All applications use server-side decoding (see \S\ref{sec:server_decoding}): the result stays encrypted until the final answer. \mkrev{For PSI, the server also reduces the result to the encrypted cardinality; for the other applications, the client sums the slots of each returned ciphertext after decryption.}}

\mkrev{\textbf{Parameter selection and validation.} For each workload, we choose the smallest limb count for which every output decrypts correctly, and validate empirically: after evaluation, we measure the remaining noise budget of every output ciphertext (SEAL's invariant-noise estimate) and compare its decryption against a plaintext oracle. \mkrev{Both approaches use the same validation, so their limb counts are comparable; a deployment that releases decrypted results to the server must add the margin needed for a negligible decryption-failure probability~\cite{cheon2024attacks}, in both approaches.} \mkrev{Table~\ref{tab:workload_sizes_limbs_32gib_all16} also reports the $\mathsf{HMul}$ count and the minimum remaining noise budget of each approach. \titlename reduces the $\mathsf{HMul}$ count in 12 of the 13 applications and matches it in TPC-H Q6\mktches{; the resulting end-to-end speedup also depends on the share of execution time spent in bulk bitwise operations and on the cost of indicator evaluation (\S\ref{sec:end-to-end})}. Both approaches retain 6--54 bits of noise budget.} Table~\ref{tab:workload_sizes_limbs_32gib_all16} reports the result. Across the 13 applications, both the arithmetic baseline and \titlename need 7--13 limbs on the CPU. \titlename needs one more limb than the baseline in 4 applications (exact and fuzzy $k$-mer matching, GWAS, and PSI), because its indicator has a higher degree, and the same number in the other 9. All configurations use $N{=}32768$ with $\log_2Q\le13\times57=741<881$ bits, so every workload retains at least 128-bit security~\cite{HomomorphicEncryptionSecurityStandard}.}
\mkrevb{\textbf{Overheads of \titlename.} Compared with the arithmetic approach, \titlename incurs four overheads. First, \titlename needs one more limb in 4 applications (exact and fuzzy $k$-mer matching, GWAS, and PSI), which enlarges every ciphertext, so its encrypted database is 12.4--16.7\% larger in these applications (Table~\ref{tab:workload_sizes_limbs_32gib_all16}); the other 9 applications need the same storage. Second, \titlename adds $\mathsf{HAdd}$s for the sums and plaintext-ciphertext multiplications for the indicator coefficients, each about $1000\times$ cheaper than an $\mathsf{HMul}$. Third, the plaintext modulus $p$ must exceed the reduction depth $K$, so that the sum does not wrap around; our 25-bit $p$ satisfies this for every workload. Fourth, client-side decoding reveals the Hamming weight $S$ to the client (\S\ref{sec:client_decoding}), so all end-to-end applications use server-side decoding, which reveals only the result. \titlename adds no client-side computation beyond standard B/FV encryption and decryption. \S\ref{sec:app_optimizations} reports the overheads of packing and bitmap indices.}

\DIFdelbegin 

\DIFdelend

\section{Evaluation}
\label{sec:evaluation}

\mkar{We evaluate \titlename in three parts. \S\ref{sec:reduction} evaluates a single homomorphic AND reduction for different reduction widths. \S\ref{sec:end-to-end} evaluates 13 end-to-end applications. \S\ref{sec:discussion} studies how \titlename scales with the reduction width.}

\subsection{Effect of Different Reduction Depth} 
\label{sec:reduction}
\begin{figure}[h]
    \centering
    \includegraphics[width=\linewidth]{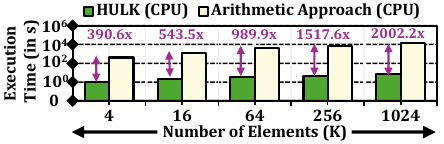}
    \caption{Execution time (in s) of AND using HULK (CPU) and the arithmetic approach for different numbers of elements ($K$).}
    \label{fig:numberofcores}
\end{figure}

Figure~\ref{fig:numberofcores} shows the execution time (in s) of the \titlename (CPU) for performing AND operation against the arithmetic approach for different number of elements in the reduction tree (i.e, \(K = 4 - 1024\)). Note that these results are only for a single AND-tree reduction across different K values. \mkrev{In this study, \titlename uses client-side decoding (see \S\ref{sec:client_decoding}), so it evaluates no indicator, whereas the arithmetic approach returns the encrypted AND result.}\footnote{\mkrev{This microbenchmark isolates the bulk bitwise operation. \titlename uses client-side decoding (\S\ref{sec:client_decoding}) and evaluates no indicator, whereas the arithmetic approach returns the encrypted AND result. \titlename also uses memory-efficient data packing with coefficient encoding (\S\ref{sec:memory-efficient-dp}) and a single 60-bit limb, whereas the end-to-end applications of \S\ref{sec:end-to-end} use one bit per slot, server-side decoding, and 57-bit limbs (\S\ref{sec:workloads}). Hence, these speedups are not end-to-end speedups.}} We make two key observations.

\begin{enumerate}[leftmargin=*, itemsep=0pt, topsep=0pt]
\item The \titlename (CPU) approach significantly outperforms the arithmetic approach for all K. 
For example, for K=4, it achieves a $390.6\times$ gain over the arithmetic approach.

\item This speedup increases with an increase in K for a single AND-tree, reaching up to $ 2002.2\times$ for K=1024 \mktches{(client-side decoding, no indicator)}. 
\cqisca{
Specifically, this large speedup arises from three compounding factors. First, $\mathsf{HAdd}$ is $\sim$300$\times$ faster than $\mathsf{HMul}$ per operation on the CPU system. Second, in this case, \titlename requires only Q={60} (1 limb) regardless of K, whereas the arithmetic approach requires \((log_2K+1)\) limbs of 60 bits 
\pw{for K} elements. Thus, the arithmetic approach operates on ciphertexts that are $log_2K$$\times$ larger than \titlenamed's (\S\ref{sec:prior_approaches_motivation}). Third, for these workloads, \titlename packs more bits per \mkrev{coefficient} (\S\ref{sec:memory-efficient-dp}), reducing the total number of operations and ciphertexts.}

\end{enumerate}

\subsection{Evaluation of End-to-End Applications} 
\label{sec:end-to-end}

\begin{figure*}[t]
    \centering
    \includegraphics[width=0.9\linewidth]{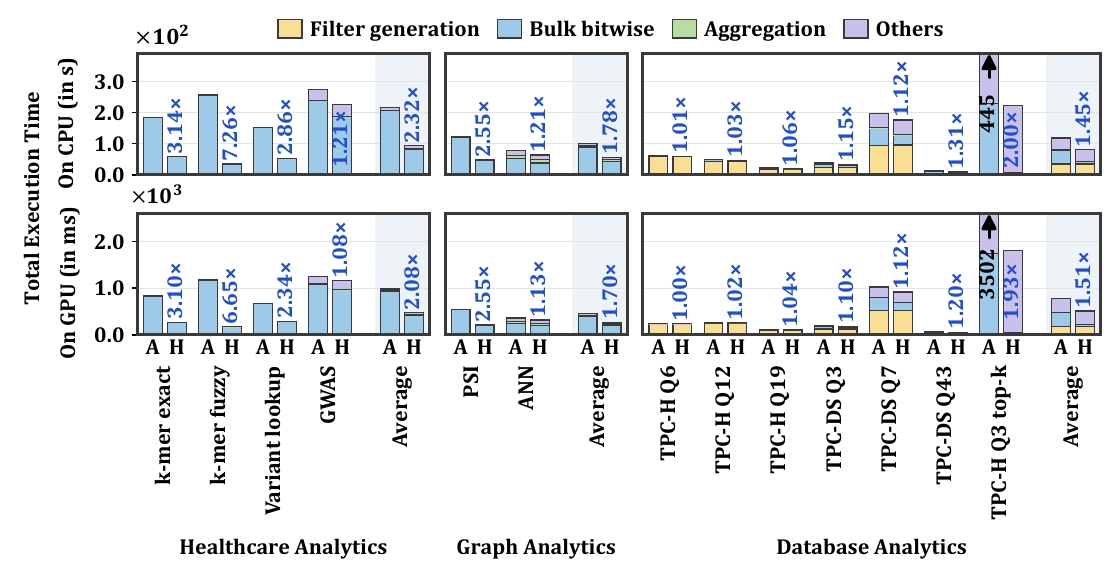}
    \caption{\mkar{Execution time of the arithmetic approach (A) and \titlename (H) on the CPU (in s) and the GPU (in ms) for all 13 end-to-end applications, broken down into filter generation, bulk bitwise operations, aggregation, and other operations. The number above each H bar is \titlenamed's speedup over the arithmetic approach. The Average bars use the total execution time of each domain.}}
    \label{fig:evaluation}
\end{figure*}
\mkar{Figure~\ref{fig:evaluation} presents the execution time of the arithmetic approach and \titlename on the CPU and GPU systems for the 13 end-to-end applications, broken down into filter generation, bulk bitwise operations, aggregation, and other operations (e.g., the compare stage of TPC-H Q3 top-10). \mkrev{The arithmetic approach is our baseline: it implements the bulk bitwise circuits of the prior works of each application (see \S\ref{sec:key_related_works} and \S\ref{sec:workloads}), and both approaches share every other stage.}}
 \mksp{Note that for these end-to-end applications, we \pw{encode} 1 bit per slot\pw{, because} \mkrev{the input range of the indicator} \pw{becomes too large to be efficient} \mkrev{(i.e., an indicator over a multi-lane slot would have to interpolate over $2^{LW}$ points, see \S\ref{sec:memory-efficient-dp})}.}
We make five key observations.

\begin{enumerate}[leftmargin=*, itemsep=0pt, topsep=0pt]
\item \mkar{\titlename matches or outperforms the arithmetic approach on all 13 applications, by $1.01$--$7.26\times$ on the CPU and $1.00$--$6.65\times$ on the GPU. On average, \titlename improves performance by $2.32\times$, $1.78\times$, and $1.45\times$ on the CPU and by $2.08\times$, $1.70\times$, and $1.51\times$ on the GPU for healthcare, graph, and database analytics, respectively. This is because \titlename replaces the $\mathsf{HMul}$-based bulk bitwise reductions with $\mathsf{HAdd}$s and one indicator per reduction.}
\item \mkar{Healthcare analytics benefits the most, because reductions are wide. For example, \titlename improves the performance of fuzzy $k$-mer matching by $7.26\times$ (CPU) and $6.65\times$ (GPU): its 16 THRESHOLD-31 reductions (one per reference) need counting trees with 1,584 $\mathsf{HMul}$s in total in the arithmetic approach, but only 176 $\mathsf{HMul}$s in \titlename (1,599 and 184 including the OR-16, Table~\ref{tab:workload_sizes_limbs_32gib_all16}). Exact $k$-mer matching (AND-62) and variant lookup (AND-37) improve by $3.14\times$ and $2.86\times$ on the CPU. GWAS improves by $1.21\times$ (CPU) and $1.08\times$ (GPU), because its case and control rules are narrower (AND-8/9), and its other stages are unchanged.}
\item \mkar{In graph analytics, \titlename improves the performance of PSI by $2.55\times$ on both systems, because each of the 16 equality tests is a wide AND-40 followed by an OR-16. ANN search improves by $1.21\times$ (CPU) and $1.13\times$ (GPU). Its query is also encrypted, so each bit comparison still needs one squaring in \titlename, and its second-round retrieval is unchanged.}
\item \mkar{In database analytics, \titlename \pw{provides the weakest speedups}, \pw{accelerating} the six filter-and-aggregate TPC queries by $1.01$--$1.31\times$ (CPU) and $1.00$--$1.20\times$ (GPU). This is because online filter generation, which is identical in both approaches, dominates execution time. Speedup grows with the share of bulk bitwise compositions, and is the largest for TPC-DS Q43 ($1.31\times$), whose 21 AND-4 trees take 56.5\% of the time. Encrypted bitmap indices remove the filter generation and \pw{provide further improvements} (see \S\ref{sec:bitmap_indices}).}
\item \mkar{\titlename also accelerates ranking under encryption. For TPC-H Q3 top-10, \titlename outperforms the arithmetic approach by $2.00\times$ on the CPU and $1.93\times$ on the GPU. This is because the rank threshold $[S_i\le9]$ over 73 bits costs 818 $\mathsf{HMul}$s in the arithmetic approach's counting tree but only 21 $\mathsf{HMul}$s in \titlename, which reduces the rank stage from 228.5\,s to 5.87\,s ($39\times$) on the CPU. The compare stage (216\,s and 756 $\mathsf{HMul}$s) is identical in both approaches and bounds the end-to-end speedup to about $2\times$. In contrast, a compare-exchange sorting network (e.g., bubble sort) contains no bulk bitwise reduction and thus gains nothing from \titlename.}
\pw{
\item \mktches{\titlenamed's speedup on the GPU is lower than or equal to that on the CPU (e.g., $6.65\times$ vs.\ $7.26\times$ for fuzzy $k$-mer matching), and PSI gains $2.55\times$ on both systems.} This is explained by the fact that the lightweight nature of $\mathsf{HAdd}$s caps the possible compute-based gains, in contrast to the large number of complex $\mathsf{HMul}$s of the arithmetic approach. This is also demonstrated in Fig.~\ref{fig:numberofcores}, where $\mathsf{HAdd}$s stop scaling beyond a few cores because of data movement.
}
\end{enumerate}

\subsection{\hgsp{Case Study: Scalability Analysis}}
\label{sec:discussion}

\begin{figure}[h]
    \centering
    \includegraphics[width=\linewidth]{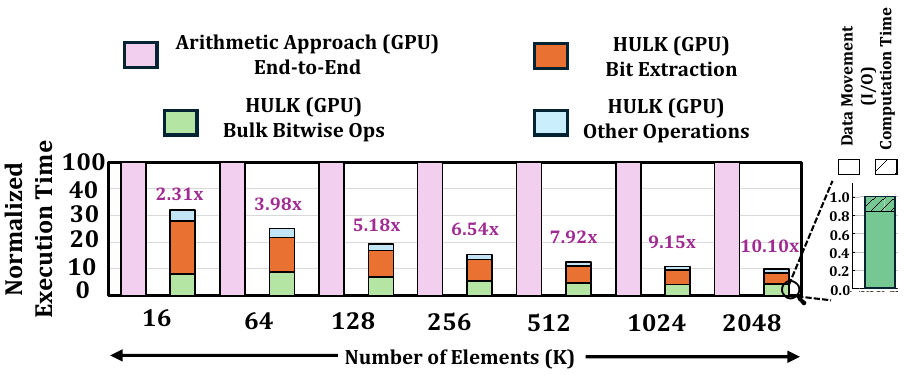}
    \caption{Sensitivity of execution time to the reduction depth $K$ (i.e., AND-$K$) for privacy-preserving set intersection (PSI) application.}
    \label{fig:discussion}
\end{figure}

\hgsp{We observe that \titlename yields greater speedups over the arithmetic approach as the bulk bitwise reductions become wider (see \S\ref{sec:end-to-end}).}
To better understand this effect, we conduct a case study of the PSI benchmark on a GPU system and quantify how \titlename improves performance as the reduction depth $K$ (i.e., the number of elements in the reduction tree) increases.
Figure~\ref{fig:discussion} presents a scalability analysis in which 
\mkrev{each of a fixed number $M=10^6$ of database entries is reduced by one AND over $K$ encrypted bits. We}
vary $K$ from 16 to 2048 to study the effect of \mkrev{the reduction depth.}
Table~\ref{tab:ops_comparison} summarizes the total number of $\mathsf{HMul}$ operations required under both approaches. \mkrev{The breakdown splits \titlenamed's execution time into bulk bitwise operations, indicator evaluation (i.e., Bit Extraction), and other operations.}\footnote{\mkrev{In this study, $K$ is the reduction depth, i.e., the number of encrypted bits that the AND reduction of each of the $M$ database entries combines (AND-$K$), and the execution time includes loading the entries from the SSD. Hence, its speedups differ from those of the end-to-end PSI application of \S\ref{sec:end-to-end}, which evaluates 16 AND-40 reductions and one OR-16 (\S\ref{sec:workloads}).}} We make two key observations.

\begin{table}[h]
\centering
\footnotesize
\setlength{\tabcolsep}{8pt}
\renewcommand{\arraystretch}{1.0}
\caption{Total number of $\mathsf{HMul}$ operations for baseline arithmetic approach and HULK.}
\begin{tabular}{|c|c|c|}
\hline
\textbf{$K$} & \textbf{Arithmetic} & \textbf{\titlename{}} \\
\hline
16   &     992 &    527 \\
64   &    3968 &   1457 \\
128  &    7936 &   2759 \\
256  &   15872 &   5394 \\
512  &   31744 &  10633 \\
1024 &   63488 &  21049 \\
2048 &  126976 &  39153 \\
\hline
\end{tabular}
\label{tab:ops_comparison}
\end{table}

\hgsp{First, \titlename achieves \pw{a performance improvement} of 2.31-10.10$\times$ over the arithmetic approach on GPU as the reduction depth $K$ increases from 16 to 2048.\footnote{\mkrev{For $K\ge128$, \titlename uses the Fermat indicator (\S\ref{sec:server_decoding}), which requires $p{=}65537$ and thus $K<65537$; \mktches{for wider values, \titlename would need digit-wise Fermat tests, which are slower than $\mathsf{HMul}$-based evaluation}.}} This is because \titlename needs fewer $\mathsf{HMul}$s than the arithmetic approach, and its $\mathsf{HMul}$ count grows more slowly with $K$ ($74\times$ vs.\ $128\times$ from $K{=}16$ to $K{=}2048$), so its $\mathsf{HMul}$ advantage grows from $1.9\times$ to $3.2\times$.}
\hgsp{Second, as 
\pw{$K$ increases}, the time spent \pw{by \titlenamed} on performing HE-based bulk bitwise operations 
grows from 25\% to 43\% of the total execution time. This is because \pw{\titlenamed's approach
is} fundamentally bottlenecked by data movement from storage to the compute unit\pw{, due to the comparative speed of $\mathsf{HAdd}$s}. \pw{Notably, more than} 80\% of $\mathsf{HAdd}$ execution time is spent moving data from the SSD to the compute unit \mkrev{(see \S\ref{sec:other_kernels})}.}

\summary{\mkrev{We conclude that \titlenamed's benefit grows with the reduction depth $K$, because its $\mathsf{HMul}$ count grows more slowly with $K$ than that of the product tree; data movement from storage then becomes its dominant cost (\S\ref{sec:ndp-discussion}).}}

\section{Optimizing \titlename for Specific Applications}
\label{sec:app_optimizations}

\mkrev{Two classes of applications allow further optimization of \titlename: addition-only workloads with client-side decoding can pack several operand bits into one slot (\S\ref{sec:memory-efficient-dp}), and database analytics can precompute predicates as encrypted bitmap indices~\cite{redis-bitmaps,chan1998bitmap} (\S\ref{sec:bitmap_indices}).}

\subsection{Memory-Efficient Data Packing}
\label{sec:memory-efficient-dp}

\mkar{The \titlename algorithm places one operand bit in each slot and \mktches{represents the Hamming weight $S$ using} $W=\lceil\log_2K\rceil+1$ bits \mktches{per slot} (see \S\ref{subsec:hulk_algorithm}). However, a $b$-bit plaintext modulus provides $b$ bits per slot, and for small $K$ most of these bits stay unused. For example, an AND-4 reduction ($W{=}3$) uses only 3 of the 46 bits of a slot when $b{=}46$. Prior work on HE-based exact string matching~\cite{kabra2025ciphermatch} packs multiple bits into each plaintext coefficient to reduce the memory footprint after encryption. Building on this idea, we propose a memory-efficient data packing scheme that uses the unused bits of each slot for applications that decode their results on the client.}

\mkar{\textbf{Lanes.} We divide each slot into \emph{lanes}. A lane is a $W$-bit field \pw{(subset)} of a slot that holds one operand bit and the Hamming weight that accumulates on that bit. Since addition is linear, values in disjoint lanes of the same slot add independently, as long as no lane overflows into its neighboring lane. A reduction of fan-in $K$ needs $W=\lceil\log_2K\rceil+1$ bits per lane. Hence, one slot of a $b$-bit plaintext modulus holds
\begin{equation}
L=\left\lfloor\frac{b}{\lceil\log_2K\rceil+1}\right\rfloor
\label{eq:lanes}
\end{equation}
lanes, and one ciphertext holds $L$ bitvectors instead of one. \mkrev{Our microbenchmarks use a power-of-two modulus $p=2^{b}$, which does not support batching, so they use the coefficient encoding: each polynomial coefficient holds $L$ lanes, and they apply only $\mathsf{HAdd}$s and plaintext shifts, which act coefficient-wise.}}

\mkar{\textbf{Lane Assignment.} \titlename assigns lanes in one of two ways. \emph{(i) At encryption.} The client encodes operand $a^{(\ell)}$, $\ell\in\{0,\dots,L{-}1\}$, as $2^{\ell W}a^{(\ell)}$ before it encrypts the operand. This assignment incurs no homomorphic cost. However, it fixes the lane width $W$, and hence the largest supported fan-in $2^{W-1}$, for the stored database. The data owner must therefore provision $W$ for the deepest reduction that the database serves, and a query of smaller fan-in leaves the extra guard bits unused. \emph{(ii) At query time.} The server multiplies a stored single-lane ciphertext by the plaintext constant $2^{\ell W}$. This assignment keeps the database fan-in-agnostic. }

\mkar{\textbf{Example.} For $b{=}46$ and $K{=}4$ ($W{=}3$), one slot holds $L{=}15$ lanes. The two packed operands are $P_1=(a^{(14)},\dots,a^{(0)})$ and $P_2=(b^{(14)},\dots,b^{(0)})$, where $a^{(\ell)}$ occupies bits $3\ell$ to $3\ell{+}2$ of the slot. Then, $P_1{+}P_2$ holds $a^{(\ell)}{+}b^{(\ell)}\in\{0,1,2\}$ in every lane, and no lane overflows.}

\mkar{\textbf{Constraints.} Memory-efficient data packing imposes three constraints. First, $p\ge2^{LW}$ must hold, so that the packed value never wraps around the plaintext modulus. Second, under lane assignment (ii), the lane shift scales the noise by up to $2^{(L-1)W}$, which the ciphertext modulus must absorb (see \S\ref{sec:noise_budget}). Hence, packing is \emph{not} parameter-agnostic: it trades $L\times$ fewer ciphertexts and $\mathsf{HAdd}$s for a larger $p$ and a noise margin. Third, packing applies only to client-side decoding. A server-side indicator would need $2^{LW}$ interpolation points.}

\subsubsection{Evaluation of Memory-Efficient Data Packing}
\label{sec:other_kernels}

\begin{figure}[h]
    \centering
    \includegraphics[width=\linewidth]{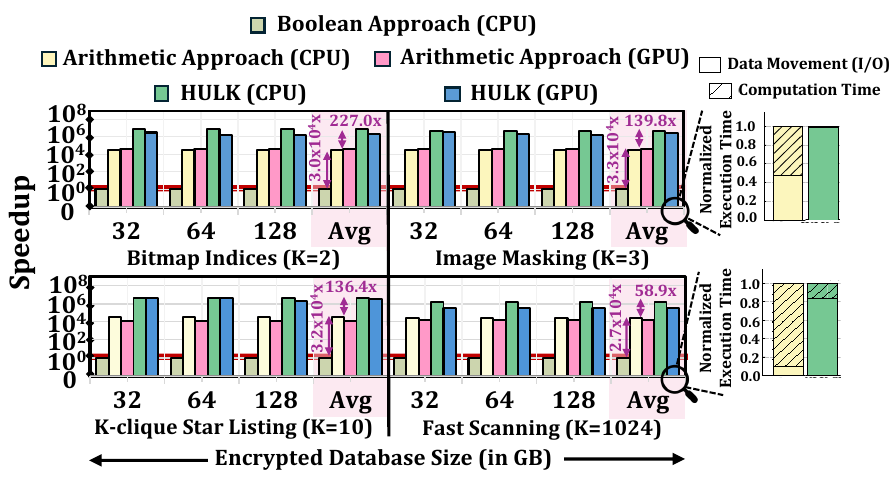}
    \caption{\mkar{Speedup (higher is better) of the arithmetic approach and \titlename, on the CPU and GPU systems, over the Boolean approach (denoted as Y=$10^0$) for different encrypted database sizes. The reduction width $K$ differs across microbenchmarks.}}
    \label{fig:effectofdatabase_sw}
\end{figure}

\mkar{Figure~\ref{fig:effectofdatabase_sw} presents the end-to-end speedup (including SSD load time) of the arithmetic approach and \titlename, each implemented on the CPU and GPU systems, over the Boolean (denoted Y=$10^0$) \mktches{and arithmetic} approaches for 32--128GB encrypted databases across four microbenchmarks: bitmap indices, image masking, k-clique star listing (KCS), and fast scanning. Each microbenchmark performs multiple AND-tree reductions with a fixed $K$, and \titlename uses the number of lanes per slot that the reduction depth of each microbenchmark allows (see Eqn.~(\ref{eq:lanes})).} We use four kernels that serve as standalone applications or building blocks of larger ones. All four use client-side decoding, lanes assigned at encryption, and one 60-bit limb.
\squishlist
\item \mkar{\textbf{Bitmap Indices}~\cite{shao2007bitmap} provide a space-efficient alternative to B-tree indices for query operations such as joins and scans. Each feature is a bitvector, and a query ANDs or ORs the bitvectors with one or two $\mathsf{HAdd}$s. \mkrev{This kernel uses $K{=}2$.} We pack 23 data bits per slot with a plaintext modulus of $2^{46}$.}
\item \mkar{\textbf{Image Masking}~\mktches{\cite{gonzales2002digital}} partitions an image into regions using masks. Each pixel is 24 bits (8 bits per RGB channel). The kernel ORs two masks and ANDs the result with the image. We reserve 2 bits for the OR and 1 bit for the AND, \mkrev{($K{=}3$)}, which packs 15 data bits per slot with $p=2^{45}$.}
\item \mkar{\textbf{K-Clique Star Listing (KCS)}~\cite{danisch2018listing,li2020ordering} finds all k-clique stars of a graph, where a k-clique star is a k-clique together with every vertex connected to all clique members. The input is the adjacency-matrix rows of the clique members, and the output is the AND of these rows. \mkrev{This kernel uses $K{=}10$, i.e., $\lceil\log_2K\rceil+1=5$ bits per lane, which packs 8 data bits per slot with a plaintext modulus of $2^{41}$.}}
\item \mkar{\textbf{Fast Scanning}~\mktches{\cite{li-sigmod-2013}} is a high-throughput filtering primitive. The input is an array of $T$ 32-bit integers, one element per ciphertext, and the kernel XORs all $T$ elements. We reserve $\lceil\log_2T\rceil+1$ bits per data bit \mkrev{($K{=}T{=}1024$)}, which packs 3 data bits per slot with a plaintext modulus of $2^{35}$. We make two key observations.}
\squishend

\squishlist 
\item \mkar{First, \titlename (CPU) outperforms the arithmetic approach by $58.9$--$227.0\times$ across the four microbenchmarks, and its speedup decreases as $K$ increases. For example, \titlename (CPU) achieves a $227.0\times$ speedup for workloads with small $K$ (e.g., bitmap indices) but a $58.9\times$ speedup for workloads with large $K$ (e.g., fast scanning). This is because a larger $K$ requires wider lanes, so each slot holds fewer lanes (see Eqn.~(\ref{eq:lanes})). As a result, \titlename needs more ciphertexts and more $\mathsf{HAdd}$s for the same amount of cleartext data.}

\item \mkar{Second, \titlename (GPU) provides a lower speedup than \titlename (CPU), despite the higher compute throughput of the GPU. This is because data movement between the SSD, main memory, and the processor dominates the execution time of the lightweight $\mathsf{HAdd}$s. Specifically, SSD-to-DRAM transfers account for more than 90\% of the total execution time on the CPU system, and the additional host-to-GPU memory transfers raise this overhead to 98\% on the GPU system.
}
\squishend

\summary{\mkrev{We conclude that memory-efficient data packing accelerates addition-only bulk bitwise workloads by $58.9$--$227.0\times$ over the arithmetic approach on the CPU, because it reduces the number of ciphertexts and $\mathsf{HAdd}$s\mktches{: compared with \titlename without packing, it needs $23\times$, $15\times$, $8\times$, and $3\times$ fewer ciphertexts for the four kernels, i.e., its gain is the number of lanes $L$, which shrinks with $\log_2K$ rather than with $K$}.}}

\subsection{Encrypted Bitmap Indices for Databases}
\label{sec:bitmap_indices}

\mkrev{\mktches{In this section, we quantify how much encrypted bitmap indices further improve database analytics with \titlename.} Relational queries are not inherently bitwise. However, a filter-and-aggregate query becomes a bulk bitwise operation when each of its predicates is precomputed as a \emph{bitmap index}~\cite{chan1998bitmap,wu-icde-1998,oneil-ideas-2007}, i.e., a bitvector whose bit $i$ indicates whether row $i$ satisfies the predicate. Prior HE-based databases (e.g., HE3DB~\cite{bian2023he3db}, ArcEDB~\cite{zhang2024arcedb}, \mktches{NSHEDB~\cite{jung2026nshedb},} and Engorgio~\cite{bian2025engorgio}) instead generate every predicate at query time with homomorphic comparison ($\mathsf{HomComp}$). We apply bitmap indices to encrypted databases: the data owner computes the bitmap of each predicate in plaintext, encrypts it once, and stores it with the relation, so a query reduces to \titlenamed's composition of the stored bitmaps and aggregation. A predicate can be precomputed if and only if the data owner knows its constant before encryption.}

\mkrev{\mktches{Precomputing bitmap indices} trades storage for the removal of $\mathsf{HomComp}$, and it reveals each predicate's constant to the server through the bitmap that the query reads. It therefore fits settings in which the data owner issues the queries or the query templates are public. An encrypted one-hot selection over the candidate bitmaps hides the constant at the cost of one \mkrevb{ciphertext-ciphertext multiplication ($\mathsf{HMul}$) per candidate and one multiplicative level}. In our setup, the query constants are plaintext to the server in both approaches.}

\mktches{\textbf{Coverage.} We call a predicate \emph{precomputable} if it compares one column with a constant, or two columns of the same row, and the constant takes at most $T{=}64$ legal values, i.e., values that the query template allows~\cite{tpch,nambiar2006making}. A query is \emph{fully precomputable} if all its predicates and group-by keys are precomputable. We classify the predicates of all 22 TPC-H and 99 TPC-DS templates, excluding foreign-key joins, which the data owner materializes when it loads the data. On average, 67.5\% and 71.6\% of the predicates of a TPC-H and a TPC-DS query are precomputable, respectively, and 7 TPC-H queries (32\%: Q1, Q5--Q7, Q12, Q14, Q19) and 14 TPC-DS queries (14\%) are fully precomputable. The remaining predicates compare against subquery results or aggregates, or filter high-cardinality columns (e.g., \texttt{i\_manufact\_id} in TPC-DS Q3, with 1,000 legal values), so they stay online. Among our six queries, TPC-H Q6, Q12, Q19 and TPC-DS Q43 are fully precomputable, TPC-DS Q3 has 50\% precomputable predicates, and TPC-DS Q7 groups by a high-cardinality key (\texttt{i\_item\_id}).}

\mkrev{\mktches{\textbf{Storage.} We consider two tiers of bitmaps, both over precomputable columns only. \emph{Tier-1} stores one bitmap per precomputable predicate of the evaluated query instances~\cite{chan1998bitmap}. \emph{Tier-2} stores one bitmap per legal value of each precomputable predicate, so that every instance of the query template runs on bitmaps~\cite{oneil-ideas-2007,wu-icde-1998}. Online comparison needs 7--13 limbs, depending on the domain size $D$ of each compared column, because its comparison polynomial has degree $D{-}1$ (\S\ref{sec:workloads}). In contrast, bitmaps need only 5 limbs in both approaches, although \titlename keeps 2--12 fewer bits of noise budget. As a result, relative to the encrypted relation of the arithmetic approach, the bitmaps of our six queries add $0.58\times$ storage for Tier-1 and $1.83\times$ for Tier-2 (Table~\ref{tab:beyond_bitmaps}). We evaluate Tier-1.}}

\subsubsection{Evaluation of Encrypted Bitmap Indices}
\label{sec:bitmap_eval}

\mkar{We evaluate the performance of encrypted bitmap indices on the CPU system (\S\ref{sec:real_System}) for Tier-1 in two steps. First, we quantify the impact of bitmaps on the prior arithmetic approach. Second, we quantify how much \titlename improves performance on top of bitmaps. Table~\ref{tab:beyond_bitmaps} reports both steps for our six TPC queries (\S\ref{sec:workloads}). In this setup, both approaches generate every predicate online with the same comparison polynomial~\cite{iliashenko2021faster}\mktches{, as in NSHEDB~\cite{jung2026nshedb}}, and only the composition differs.} 

\begin{table}[t]
\centering
\caption{\mkrevb{Encrypted bitmap indices (CPU, Tier-1). \textbf{Storage}: bitmaps (5 limbs) stored in addition to the encrypted database column (7--13 limbs). \textbf{Speedup} of A+B and H+B over the arithmetic approach with online comparison (A). \textbf{Exec.\ time reduction}: fraction of A's execution time that bitmaps remove. \textbf{All six}: storage over all six queries; means of the other columns.}}
\label{tab:beyond_bitmaps}
\footnotesize
\setlength{\tabcolsep}{3pt}
{\color{black}
\resizebox{\columnwidth}{!}{%
\begin{tabular}{@{}lrrrrrr@{}}
\toprule
& \multicolumn{2}{c}{\textbf{Storage}} & \multicolumn{2}{c}{\textbf{Speedup}} & & \textbf{Exec.\ time} \\
\cmidrule(lr){2-3}\cmidrule(lr){4-5}
\textbf{Query} & \textbf{Tier-1} & \textbf{Tier-2} & \textbf{A+B} & \textbf{H+B} & $\frac{\textbf{H+B}}{\textbf{A+B}}$ & \textbf{reduction} \\
\midrule
TPC-H Q6   & 0.25$\times$ & 1.25$\times$ & 52.55$\times$ & 55.46$\times$ & 1.06$\times$ & 98.1\% \\
TPC-H Q12  & 0.47$\times$ & 1.07$\times$ & 9.89$\times$  & 14.47$\times$ & 1.46$\times$ & 89.9\% \\
TPC-H Q19  & 1.00$\times$ & 4.50$\times$ & 6.39$\times$  & 10.31$\times$ & 1.61$\times$ & 84.4\% \\
\midrule
TPC-DS Q3  & 0.40$\times$ & 0.64$\times$ & 1.10$\times$ & 1.28$\times$ & 1.17$\times$ & 9.2\% \\
TPC-DS Q7  & 0.25$\times$ & 1.00$\times$ & 1.01$\times$ & 1.14$\times$ & 1.13$\times$ & 1.3\% \\
TPC-DS Q43 & 1.60$\times$ & 4.00$\times$ & 1.33$\times$ & 1.95$\times$ & 1.47$\times$ & 24.6\% \\
\midrule
\textbf{All six} & 0.58$\times$ & 1.83$\times$ & 12.05$\times$ & 14.10$\times$ & 1.32$\times$ & 51.2\% \\
\bottomrule
\end{tabular}%
}}
\end{table}

\mkrev{\mktches{Table~\ref{tab:beyond_bitmaps} compares the arithmetic approach with online predicates (A) against the arithmetic approach (A+B) and \titlename (H+B) with bitmaps. We make three key observations.}}

\mkrev{\mktches{First, bitmaps accelerate the arithmetic approach by $6.39$--$52.55\times$ on the three TPC-H queries but only by $1.01$--$1.33\times$ on the three TPC-DS queries ($12.05\times$ on average). This is because the TPC-H queries are fully precomputable, so bitmaps remove all online comparison, i.e., 84.4--98.1\% of their execution time. In contrast, TPC-DS Q3 and Q7 keep the comparison of their high-cardinality columns online, so bitmaps remove only 9.2\% and 1.3\% of their execution time, respectively.}}

\mkrev{\mktches{Second, on top of bitmaps, \titlename outperforms the arithmetic approach by $1.06$--$1.61\times$ on TPC-H and by $1.13$--$1.47\times$ on TPC-DS ($1.32\times$ on average). This is because bitmaps remove the comparison of the precomputable predicates, so the bulk bitwise composition, which \titlename accelerates, takes a larger share of the execution time. The gain is smallest for TPC-H Q6 ($1.06\times$), whose single AND-3 leaves little composition work, and largest for TPC-H Q19 ($1.61\times$), which evaluates three AND-6 trees. For TPC-DS Q3 and Q7, online comparison still dominates, which limits the gain to $1.17\times$ and $1.13\times$. Together, bitmaps and \titlename improve performance by $14.10\times$ on average over A, and most of this gain comes from bitmaps.}}

\mkrev{\mktches{Third, restricting bitmaps to precomputable columns keeps their storage small. Tier-1 adds $0.25$--$1.00\times$ of the encrypted relation for TPC-H and $0.25$--$1.60\times$ for TPC-DS, and Tier-2 adds $1.07$--$4.50\times$ and $0.64$--$4.00\times$, respectively. In contrast, a Tier-2 index that also covers the high-cardinality columns (e.g., 1,000 \texttt{i\_manufact\_id} values for Q3 and about 9,000 \texttt{i\_item\_id} groups for Q7) adds $90.5\times$ over our six queries, so these columns are better served by online comparison. The extra storage does not reduce performance, because the ciphertext transfer from the SSD is pipelined with computation.}}

\summary{\mkrev{\mktches{We conclude that bitmaps and \titlename are complementary: bitmaps remove the comparison of precomputable predicates ($12.05\times$ on average), and \titlename accelerates the remaining bulk bitwise composition ($1.32\times$ on top of bitmaps), at the cost of $0.58\times$ (Tier-1) to $1.83\times$ (Tier-2) extra storage.}}}

\section{Discussion and Future Work}
\label{sec:ndp-discussion}

\mkar{\textbf{1) Accelerating \titlename with Near-Data Processing.} The case study in \S\ref{sec:discussion} shows that, once \titlename replaces $\mathsf{HMul}$ with $\mathsf{HAdd}$, data movement, rather than computation, becomes the dominant cost\mkrev{, as in many data-intensive workloads~\cite{mutlu2019processing,boroumand2018google,mutlu2022modern,mutlu2023dac,mutlu2025memory}}. \titlename splits a bulk bitwise HE workload into two phases with opposite hardware demands: (1) a bulk $\mathsf{HAdd}$ phase that performs the bitwise reduction, and (2) a $\mathsf{HMul}$ phase that \titlename incurs only once per reduction for decoding (see \S\ref{sec:server_decoding}). The $\mathsf{HMul}$ phase is compute-bound and relies on complex polynomial operations (e.g., NTT~\cite{halevi2019improved}, key switching, and relinearization~\cite{bajard2016full}) that prior HE accelerators target~\cite{feldmann2021f1,samardzic2022craterlake,kim2022bts}. \mkrev{The $\mathsf{HAdd}$ phase is bound by data movement (\S\ref{sec:discussion}).}}

\mkar{This split enables a heterogeneous design: \titlename performs the $\mathsf{HAdd}$ phase \emph{where the data resides}, inside NAND flash, and offloads only the residual $\mathsf{HMul}$ phase to a compute-optimized processor. $\mathsf{HAdd}$ is element-wise addition, which reduces to bit-serial addition that \mkrev{storage-centric (near-storage and in-flash)~\citeSTORISP{} and memory-centric (near-memory~\citePIMPNM{} and in-memory~\citePIMPUM, whose security trade-offs are studied in~\cite{yuksel2026memory})} substrates support efficiently. Moreover, reducing in place collapses $K$ ciphertexts into one before they leave the device, which eliminates the bottleneck transfers. The arithmetic approach cannot exploit these substrates, because its $\mathsf{HMul}$-based bitwise operations do not map onto the addition-oriented compute capabilities of in-flash~\mkrev{\citeSTORISP{}} and in-DRAM processing~\mkrev{\citePIMPUM}. Such near-data HE accelerators remain an emerging, not yet commercially deployed, direction. To quantify the opportunity, we map \titlenamed's $\mathsf{HAdd}$ phase onto a state-of-the-art in-flash processing (IFP) substrate, CIPHERMATCH~\cite{kabra2025ciphermatch}, and run the remaining operations on the GPU (\titlename (SC-GPU)).}

\mkar{\textbf{Data Layout.} To perform element-wise addition, we use a bit-serial data layout and bit-serial addition. We map each $q$-bit polynomial coefficient along a single bitline and distribute the coefficients across multiple bitlines. Bit-serial addition then sums the bits of each bitline sequentially. To maximize throughput, we exploit the plane-, die-, chip-, and channel-level parallelism of NAND flash memory~\cite{kabra2025ciphermatch,nadig2023venice}.}

\begin{figure}[h]
    \centering
    \includegraphics[width=0.9\linewidth]{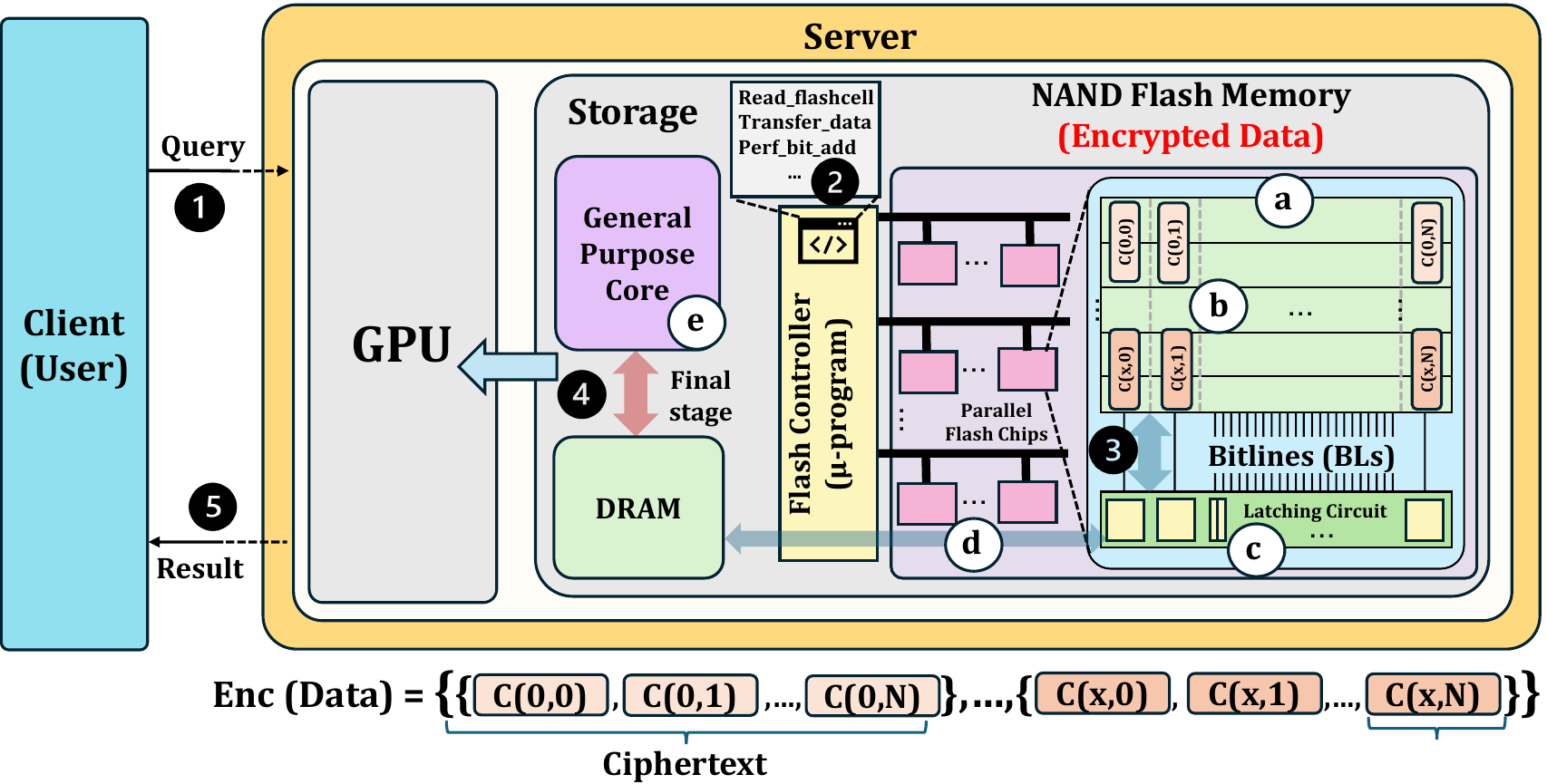}
    \caption{\mkar{Integration of \titlename with a storage-centric SSD.}}
    \label{fig:Arch_design}
\end{figure}

\mkar{\textbf{Workflow.} Figure~\ref{fig:Arch_design} extends the client-server model of Figure~\ref{fig:packing_figure} to the SSD. \circled{a} For $N{=}32768$, each ciphertext coefficient is mapped to one of 32K bitlines. \circled{b} Multiple ciphertexts are distributed across NAND flash blocks, where each plane holds approximately $(2048\text{ blocks}\times256\text{ wordlines})/q\approx8$K ciphertexts (for $q{=}60$ bits). \circled{c} Similar to prior IFP designs~\cite{chen2024ares,kabra2025ciphermatch}, we perform bit-serial additions with the data latches of the NAND flash arrays: carry bits stay in the latches, and sum bits move to the SSD-internal DRAM for the next addition. \circled{d} The results read out of multiple planes and chips are stored in the SSD-internal DRAM. \circled{e} The SSD controller cores~\cite{arm-cortexR5} perform the modular reduction, so the NAND flash arrays perform only polynomial additions. After the bulk bitwise operations, the SSD sends the results to the GPU, which performs the remaining operations (bit extraction, aggregation, popcount).}

\mkar{\textbf{Evaluation Methodology.} Table~\ref{tab:NDP_systems} presents the configuration of the simulated storage system. We develop an IFP simulator, based on CIPHERMATCH~\cite{kabra2025ciphermatch}, that models the latency of bit-serial addition, i.e., of $\mathsf{HAdd}$s. We model the modular reduction on the SSD controller cores with QEMU~\cite{qemu}.}

\begin{table}[h]
    \centering
    \footnotesize
    \setlength{\tabcolsep}{6pt}
    \renewcommand{\arraystretch}{1.25}
    \caption{\mkvvv{Simulated} system configurations.}
    \begin{tabularx}{\columnwidth}{@{}>{\arraybackslash}p{1.4cm} X@{}}
    \toprule
    \textbf{Config.} & \textbf{Details} \\
    \midrule
    \multirow{8}{*}{\textbf{\titlenamed-SC}}
    & 48-WL-layer 3D TLC NAND flash-based SSD; 2 TB\\
    & \textbf{\mkvvv{NAND-Flash Channel Bandwidth}}: 1.2-GB/s IO rate\\
    & \textbf{Controller Cores}: ARM Cortex-R5 series @1.5GHz; 5 Cores~\cite{arm-cortexR5} \\
    & \textbf{NAND Config}: 8 channels; 8 dies/channel; 2 planes/die; 2,048 blocks/plane; 196 (4$\times$48) WLs/block; 4 KiB/page\\
    & \textbf{Latency}: \texttt{T\textsubscript{read}} (SLC mode): 22.5 $\mu$s~\cite{park2022flash}; \texttt{T\textsubscript{AND/OR}}: 20 ns~\cite{gao2021parabit}; \texttt{T\textsubscript{latchtransfer}}: 20 ns~\cite{gao2021parabit}; \texttt{T\textsubscript{XOR}}: 30 ns~\cite{park2022flash}; \texttt{T\textsubscript{DMA}}: 3.3 $\mu$s\\
    \bottomrule
   \end{tabularx}
   \label{tab:NDP_systems}
\end{table}

\begin{figure}[h]
    \centering
    \includegraphics[width=\linewidth]{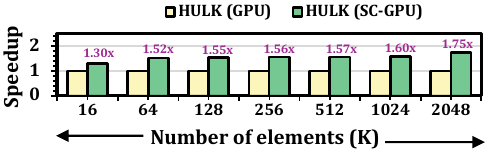}
    \caption{\mkar{Speedup (higher is better) of \titlename (SC-GPU) over \titlename (GPU) for different reduction depths $K$.}}
    \label{fig:discussion_ndp}
\end{figure}

\mkar{\textbf{Results.} Figure~\ref{fig:discussion_ndp} presents the speedup of \titlename (SC-GPU) over \titlename running only on the GPU (\titlename (GPU)). We make two key observations. First, \titlename (SC-GPU) outperforms \titlename (GPU) by $1.30$--$1.75\times$. Second, this benefit increases with $K$. Both observations have two causes: (i) NAND flash memory provides very high parallelism, which increases the throughput of $\mathsf{HAdd}$s, and (ii) performing the tree reductions inside NAND flash memory reduces the data transferred out of the SSD by $K\times$, which reduces the time spent on data transfers. Hence, \titlename maps its bulk bitwise operations, which it constructs from $\mathsf{HAdd}$s, onto near-data processing designs, which improves its performance by up to $1.75\times$.}

\textbf{\textit{2) Improving indicator evaluation.}}
\titlenamed's only homomorphic multiplication \pw{use} comes from \mkrev{indicator evaluation (i.e., bit extraction)}, which converts a
packed addition-tree result back to the standard encoding (\S\ref{sec:extension}). \titlename
implements this as a Lagrange-interpolation polynomial evaluated with the
Paterson-Stockmeyer optimization~\cite{paterson1973number}, reducing the per-reduction
cost from $O(k)$ to $O(\sqrt{k})$ $\mathsf{HMul}$s. Because \titlename collapses the bitwise
phase to near \pw{zero cost} $\mathsf{HAdd}$s, this extraction step is now the dominant remaining \pw{overhead}
in most end-to-end workloads (\S\ref{sec:evaluation}). \titlename therefore benefits
directly from any advance in homomorphic digit/bit
extraction~\cite{chen2018homomorphic,huo2019note}.

\section{Other Related Work}
\label{sec:relatedwork}

\mkrev{\mkrevb{To our knowledge, \titlename is the first work that evaluates wide bulk bitwise operations on bit-sliced data in packed B/FV by adding the operand bits slot-wise and evaluating a single indicator. Prior works add before testing in other settings: on whole words with Fermat tests~\cite{cetin2015arithmetic,jung2026nshedb}, with a randomized linear combination~\cite{bonte2020homomorphic}, or over the match indicators of many records~\cite{akavia2018secure}.} We have already quantitatively compared our arithmetic baseline with four families of prior approaches (\S\ref{sec:otherbatched}) and \titlename with this baseline (\S\ref{sec:evaluation}).}

\mkrev{\textbf{\textit{Encrypted search.}} Encrypted search composes equality tests. Akavia et al. evaluate equality as a product over GF(2)~\cite{akavia2019setup} or add match indicators across array entries to find the first match~\cite{akavia2018secure}, and Bonte and Iliashenko~\cite{bonte2020homomorphic} apply one randomized Fermat test per string. These works generate equality predicates; they do not reduce bit-sliced data. On bit-sliced data, a Fermat equality test~\cite{cetin2015arithmetic,iliashenko2021faster} packs the bits into a value, which needs one test per 16-bit digit at $p{=}65537$ and is $1.8$--$77\times$ slower than the product tree for AND-$K$ ($K{=}2$--$128$).}

\textbf{\textit{Boolean and gate-level HE.}}
A line of work evaluates Boolean circuits directly under HE\mkrev{~\citeTFHE},
enabling native support for bit-level logic. Systems such as FHEW~\cite{ducas2015fhew} and TFHE~\cite{chillotti2016faster,chillotti2020tfhe} significantly reduced the cost of bootstrapped gate
evaluation, making fine-grained encrypted Boolean computation more
practical. However, these schemes typically operate at the level of
individual gates or small Boolean words~\cite{aziz2024secure,stoian2023deep,nam2025affinity,TFHE-rs} and do not provide the packed
SIMD-style throughput that RLWE-based arithmetic HE offers. 
As a result, while Boolean HE is
well-suited to fine-grained logic, it is less appropriate for the workloads targeted in this work.

\textbf{\textit{HE acceleration.}}
Prior work on HE acceleration~\mkrev{\citeHEACCEL{}} primarily targets the dominant kernels
of arithmetic HE, such as NTT, key switching, relinearization, and
modular polynomial arithmetic, using CPUs, GPUs, FPGAs, and specialized
accelerators. In parallel, \mkrev{near-memory~\citePIMPNM{} and near-storage~\citeSTORISP{}} systems have
demonstrated efficient support for general-purpose NDP architectures\mkrev{, and prior works evaluate HE on them~\cite{mwaisela2024evaluating,zhou2025fhemem,gupta2023evaluating,gupta2021accelerating,kim2025anaheim}}. However, further performance improvements are achieved by using analog properties of memory/storage systems, \mkrev{i.e., processing-using-memory~\citePIMPUM{} and in-flash processing~\citeSTORISP}. These trends highlight a key distinction: conventional arithmetic HE remains
difficult to map efficiently onto these NDP substrates, \pw{whereas}
\titlenamed's design is better aligned with the
strengths of near-data processing architectures.

\section{Conclusion}
\label{sec:conclusion}

We introduce \titlenamed, an algorithmic approach that enables fast HE-based bulk bitwise operations for many cloud computing applications (e.g., encrypted databases and healthcare analytics). \titlename proposes an algorithm that constructs bulk bitwise operations (AND, OR, XOR\mkrev{, and THRESHOLD}) using \mksp{\emph{addition-dominant} logic that minimizes the use of} computationally expensive homomorphic multiplication and thereby reduces computational latency. \mkrev{Our evaluation on real CPU and GPU systems shows that \titlename provides significant performance gains over the arithmetic approach used by prior works, which is faster than the four other families of approaches we compare (\S\ref{sec:otherbatched}), and a simulated near-storage design further improves its performance.} \mkar{Across 13 end-to-end applications in healthcare, graph, and database analytics, \titlename improves performance by up to $7.26\times$ on the CPU and $6.65\times$ on the GPU over the state-of-the-art arithmetic approach.}


\end{document}